\documentclass[aps,twocolumn,prx,amsmath,nofootinbib,longbibliography,amssymb,superscriptaddress,10pt]{revtex4-2}

\usepackage[T1]{fontenc}  
\usepackage[utf8]{inputenc}  
\usepackage[T2A]{fontenc}    
\usepackage[russian,english]{babel}
\usepackage{graphicx}
\usepackage[normalem]{ulem}
\usepackage{soul}
\usepackage{float} 
\usepackage[svgnames]{xcolor} 
\usepackage{comment}
\usepackage[colorlinks, linkcolor=NavyBlue , citecolor= NavyBlue,urlcolor = NavyBlue, breaklinks=true]{hyperref}
\usepackage{verbatim}
\usepackage{esint}
\usepackage{marginnote}

\usepackage{soul}

\renewcommand{\vec}[1]{\boldsymbol{#1}}
\newcommand{\RNum}[1]{\uppercase\expandafter{\romannumeral #1\relax}}
\newcommand{\noga}[1]{{\color{orange}{\bf NB: #1}}}

\newcommand{\erez}[1]{{\color{purple}{\bf EB: #1}}}

\newcommand{\steve}[1]{{\color{red}{\bf SK: #1}}}

\def \k {{\vec k}}

\def \G {{\cal{G}}}

\def \t {\theta}

\def \bea {\begin{eqnarray}}
\def \eea {\end{eqnarray}}

\def \tanh {\textnormal{tanh}}
\def \coth {\textnormal{coth}}

\begin{document}

\title{Extended strange metal regime from superconducting puddles}

\author{Noga Bashan}
\thanks{These authors contributed equally to this work.}
\affiliation{Department of Condensed Matter Physics, Weizmann Institute of Science, Rehovot 76100, Israel}

\author{Evyatar Tulipman}
\thanks{These authors contributed equally to this work.}
\affiliation{Department of Condensed Matter Physics, Weizmann Institute of Science, Rehovot 76100, Israel}


\author{Steven A. Kivelson}
\affiliation{Department of Physics, Stanford University, Stanford, CA 93405}

\author{J\"{o}rg Schmalian}
\affiliation{Karlsruher Institut für Technologie, Institut f\"{u}r Theorie der Kondensierten Materie,  76049, Karlsruhe, Germany}

\affiliation{Karlsruher Institut für Technologie, Institut f\"{u}r Quantenmaterialien und Technologien,  76021, Karlsruhe, Germany}

\author{Erez Berg}
\affiliation{Department of Condensed Matter Physics, Weizmann Institute of Science, Rehovot 76100, Israel}

\date{\today}

\begin{abstract}
 We study a model of mesoscale superconducting puddles in a metal, represented as dynamical impurities interacting with a finite number of electronic channels via Andreev and normal scattering. We  identify conditions under which the collection of puddles make a $T$-linear contribution to the resistivity and a $T\ln(1/T)$ to the specific heat and thermopower. This behavior emerges in an intermediate temperature range that extends from an upper energy scale set by the renormalized charging energy of the puddles, and down to an exponentially small scale associated with a charge-Kondo crossover, provided that the number of electronic channels interacting with the puddle is large. The phenomenology of our model resembles the apparent extended strange metal regime observed in overdoped cuprates which exhibits $T$-linear resistivity at low $T$ over a finite range of doping. We also propose to engineer a strange metal from suitably designed superconducting grains in a metallic matrix.

\end{abstract}

\maketitle


\textbf{Introduction.} The interplay of strong electronic correlations and disorder is a promising route towards a theory of strange metals (SMs) \cite{ueda1977electrical,miranda1996kondo,ParcolletNFL,rosch_interplay_1999,aldape_solvable_2022,esterlis_large_2021,patel_universal_2023,BashanTunable,tulipman2024solvable,Patel_2024loc,Li_2024,patel2024strange}.
A particular form of strong disorder occurs in systems with local superconducting (SC) domains embedded in a 
metal. 
Such a
situation may arise in engineered systems, where isolated grains of a conventional SC are 
embedded in a metallic matrix~\cite{Merchant2001,Durkin2020,Bishop2022,KapInO,han2014collapse}, or in 
otherwise uniformly disordered 
unconventional SCs with a short coherence length, where pair breaking or doping disorder
leads to the 
emergence of local SC regions, often referred to as superconducting ``puddles''~\cite{Galitski2008,dodaro_generalization_2018,Kapitulnik_colloquium_2019,li_superconductor--metal_2021}.
The nature of the SC-metal transition in such 
systems has been 
extensively investigated~\cite{feigel1998quantum,Spivak2001,Feigelman2001, Spivak2008,Kapitulnik_colloquium_2019}.

High-$T_c$ cuprate SCs exhibit a striking 
form of superconductor-to-metal transition. 
The inhomogeneous nature of the superconducting state and the nearby metal are further supported by measurements of the superfluid stiffness~\cite{lemberger2011superconductor,bovzovic2016dependence}, 
scanning tunneling microscopy~\cite{howald_inherent_2001,pan_microscopic_2001,Lang_2002,tranquada_stripes_2022,gomes_visualizing_2007,tromp_puddle_2023}, 
the low-energy microwave response~\cite{mahmood2019locating,orenstein_optical_2007}, specific heat~\cite{Wen_2009,Tallon_2020,Uchida_2021}, and shot noise \cite{niu2024equivalence}. 
Moreover, in several overdoped cuprates, the ``normal state'' (i.e. in a range of $T$ above $T_c$, including in high enough magnetic fields that $T_c\to 0$) shows an apparently critical ``strange metal'' phase over an extended range of doping with $T$-linear scaling of the resistivity at the lowest temperatures~\cite{cooper_anomalous_2009,hussey_dichotomy_2011,hussey_generic_2013,giraldo2018scale,putzke2021reduced,ayres2021incoherent}.
Notably, the $T$-linear term in the dc resistivity appears to be closely linked to the proximity to 
the superconducting state; 
the coefficient of the $T$-linear term in the resistivity diminishes with doping and disappears just beyond the edge of the superconducting dome \cite{taillefer2010scattering,Yuan_2022}.

This behavior is to be distinguished from the $T$-linear resistivity observed near optimal doping that extends over a much larger range of temperatures 
with a slope that appears to be
(at most) weakly dependent on specific material parameters~\cite{zaanen_why_2004,bruin_similarity_2013,grissonnanche_linear-temperature_2021,Hartnoll2022}.
It is also different from the behavior expected in a quantum critical fan, where the range of $T$-linear resistivity shrinks to a single point as $T\to 0$~\cite{sachdev_quantum_2011}, as is seen in several heavy fermion compounds tuned near a quantum critical point~\cite{gegenwart_quantum_2008,HFreview}.  

A crucial ingredient for strange metal behavior is the presence of scatterers with a constant density of states at asymptotically low energies, as has been extensively explored in the context of ``marginal Fermi liquid'' (MFL) phenomenology~\cite{varma_phenomenology_1989}.  
Identifying the nature of the scatterers underlying strange metallicity is key to resolving this long-standing puzzle \cite{varma_phenomenology_1989,varma_singular_2002,zaanen_planckian_2019,varma_colloquium_2020,Si_review_2020,chowdhury_sachdev-ye-kitaev_2022,Hartnoll2022,phillips_stranger_2022,sachdev2025footfancupratephase}.



\begin{figure}[t]
\centering
\includegraphics[width=0.47\textwidth]{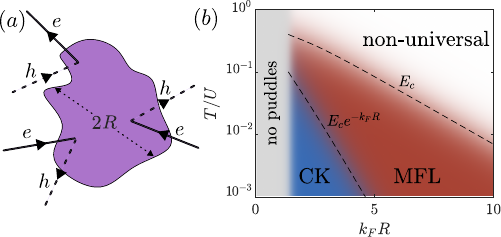} 
\caption{\textbf{(a)} Cartoon of Andreev scattering from a superconducting puddle of typical diameter $2R$. 
 \textbf{(b)} Schematic phase diagram 
 as a function of temperature $T$ and typical puddle size in units of the Fermi wavelength. For $T_{\rm 1ch}<T < E_c$  the electrons exhibit marginal-Fermi liquid 
 (MFL) behavior, where $E_c$ is the renormalized charging energy, discussed below Eq.~\ref{TLS Ham}, and $T_{\rm 1ch}\sim E_c e^{-k_F R}$ is the exponentially small temperature below which single-channel charge-Kondo (CK) screening of the puddle fluctuations becomes relevant, cutting off the MFL regime. The gray area marks $R\lesssim \xi_0$ ($\xi_0$ being the coherence length of infinitely large puddle) there is no mean-field solution for the superconducting gap due to inverse proximity effect.}
\label{fig:cartoon}
\end{figure}

In this work, we demonstrate that remnants of superconductivity in the normal state, i.e., SC puddles, can give rise to an extended regime of SM behavior. (This idea was first proposed by Eliashberg shortly after the discovery of the cuprates~\cite{Eliashberg}.) We compare our findings with measurements in overdoped cuprates and state the conditions under which the theory may apply to these systems and how to falsify it. In contrast to the traditional view where superconductivity and $T$-linear resistivity share a common mechanism as their origin, our theory is an example of a case where SC itself gives rise to $T$-linear resistivity, providing a natural explanation for why it disappears near the edge of the SC dome. We further describe the key conditions necessary to experimentally realize SM behavior in engineered granular superconductors where conventional superconducting grains are embedded in a metallic matrix.

\textbf{Summary of results.} We consider a metal with small superconducting inclusions (puddles), as illustrated in Fig.~\ref{fig:cartoon}(a). The puddles are assumed to be far apart, and the Josephson coupling between them is frustrated either by an applied magnetic field or because of the d-wave nature of the SC order parameter in each puddle~\cite{SigristDwaveReview,Kapitulnik_colloquium_2019}. We therefore neglect the contribution to transport from coherent Cooper pair tunneling between puddles; the puddles affect transport by scattering electrons in the metal, either by Andreev or normal scattering processes. 

Under these conditions, we may first consider a mesoscopic impurity problem, where the puddle is modeled by a quantum rotor 
that represents its SC phase.  
The puddle is subject to a random background charge, shifting its charging spectrum.
The two important parameters in the problem are the size of the puddle, $R$, which determines the effective number of electronic channels ($\sim k_F R$) that 
scatter off the puddle, and the
dimensionless Andreev coupling $\alpha_\perp$ between the puddle and the itinerant electrons. The problem is characterized by two energy scales: the renormalized charging energy of the puddle, $E_c$, below which the puddle can be described in terms of its two lowest charging energy states, and the charge Kondo temperature, at which the charge fluctuations of the puddle become Kondo screened by the conduction electrons \cite{taraphder1991heavy}. 

Remarkably, we find that MFL behavior emerges at temperatures below $E_c$ and down to the onset of charge-Kondo behavior [Fig.~\ref{fig:cartoon}(b)]. The MFL originates from inelastic Andreev scattering of electrons off the puddles. 
The upper limit of the MFL regime, $E_c$, is exponentially suppressed 
compared to microscopic energy scales (such as the bare Coulomb repulsion between electrons on the puddle) by a factor of $\sim e^{- c \alpha_\perp}$ where $c=\mathcal{O}(1)$. In order for this energy scale to be experimentally accessible, $\alpha_\perp$ should be at most of order unity, either because the number of electronic channels $k_F R$ is not too large, or because the Andreev reflection coefficient of each channel is small. 
The charge-Kondo scale is found to be further exponentially suppressed with the size of the puddle compared to $E_c$, opening a parametrically wide energy window between $E_c$ and the onset of charge-Kondo behavior.

In this intermediate energy window the puddle is described as a two-level system (TLS) coupled to an ohmic bath. The exact charge $U(1)$ symmetry of the problem renders the interaction between electrons and puddles to be marginally irrelevant\footnote{The marginal irrelevancy of the interactions \textit{does not} mean that the scattering of electrons off of the two-level systems should be ignored. Rather, the important implications are that (1) the spectra of the two-level systems are not strongly renormalized and (2) the strength with which they scatter the electrons is determined by the bare interaction strength. This essentially justifies the results of 2nd order perturbation theory in this intermediate energy regime.}. 
 Consequently, averaging over the random background charge on each puddle gives rise to a constant density of states of the two-level systems. These excitations lead to MFL behavior of the itinerant electrons with $T$-linear resistivity and $T\log(1/T)$ specific heat and thermopower. 

Within the MFL, the inelastic Andreev scattering rate is estimated as
\begin{equation}
\Gamma_{\rm inel}  \sim\alpha_{\perp}\frac{n_{\rm pud}E_{F}}{E_{c}}T,
\end{equation}
where $n_{\rm pud} \ll 1$ is the concentration
of puddles per unit cell in the metal, and $E_{F}$ is the Fermi energy. Assuming $\alpha_\perp\sim \mathcal{O}(1)$, $E_{F}/E_{c}$ must be large in order for the slope to be of order unity. We expect that $E_{F}/E_{c}\gg 1$ since, as mentioned above, $E_c$ is suppressed relative to microscopic scales due to the coupling of the puddle to the surrounding metal. 

The puddles also scatter elastically and contribute to the residual resistivity. At temperatures below $E_c$ the contribution to the resistivity from inelastic Andreev scattering, $\propto \Gamma_{\rm inel}$, is typically smaller than (or, at most, comparable to) the elastic contribution. Thus, our theory cannot account for situations where the dynamical range of the $T-$linear resistivity is much larger than the residual resistivity at low temperature. 




Compared to most existing studies of superconducting puddles in a metal~\cite{feigel1998quantum,Spivak2001,Feigelman2001,Spivak2008,Kapitulnik_colloquium_2019}, 
our theory is relevant when the typical puddle size is sufficiently small such that their charging 
energies are significant (despite being renormalized down by the proximity coupling to the metal), and the system is sufficiently far from the metal-to-superconductor transition 
that phase coherence between the puddles is negligible. 
These conditions may be met in short coherence length superconductors, where local superconductivity may persist in small puddles even when global superconductivity is destroyed.
See Appendix~\ref{appendix:anomalous_metals} for a more detailed comparison to existing literature and the regime of validity of our theory. 

{\bf Model of a single SC puddle in a metal.} To describe a single superconducting puddle, we consider a model of itinerant electrons coupled to a quantum rotor that resides in a finite domain of radius $R$. Hereafter we focus on two-dimensional systems. The generalization to higher dimensions is straightforward and only affects the $R$ scaling of various quantities. The electrons are coupled to the phase and charge (or number) fluctuations of the puddle, such that the Hamiltonian of the system is given by
\begin{eqnarray}
H &= H_{\rm el}+H_{\rm C}+H_{\rm int},
\end{eqnarray}
where  $H_{\rm el} = \sum_{\vec{k},s}\varepsilon_{\vec{k}} c^\dagger_{\vec{k},s}c_{\vec{k},s}$ is the kinetic energy of metallic electrons in the normal region of the system. Here $c^{\dagger}_{\vec{k},s}$ is the creation operator of an electron of wave vector $\vec{k}$ and spin $s$, $\varepsilon_{\vec{k}}$ denotes the electronic dispersion relative to the chemical potential. We assume that the electrons are ballistic, formally valid when the elastic bulk mean free path $\ell_{\rm mfp}$ is much larger than the size of the puddle. $H_{\rm C} = E_{c,0}(\hat{n}-\bar{n})^2$ describes the charging energy of a superconducting puddle, where $\hat{n}$ is the Cooper pair number operator, while  $\bar{n}$ is the background charge on the puddle.  $E_{c,0}$ is the bare charging energy associated with adding a Cooper pair to the puddle. Assuming the Coulomb repulsion is short ranged with strength $U$, the bare charging energy decreases with the puddle area $E_{c,0}\sim U/R^2$. The superconductor is characterized by the phase $\hat{\theta}$ that is canonically conjugate to $\hat{n}$, i.e. $[\hat{\theta},\hat{n}]=i$. The puddle interacts with  electrons both via Andreev scattering, in which two electrons are absorbed or emitted, with matrix elements $g_\perp(\vec{k},\vec{k}')$, and by short-range Coulomb repulsion, with matrix elements $g_z(\vec{k},\vec{k}')$:
\begin{align}
    H_{\rm int} &= \sum_{\vec{k},\vec{k}'} \left(g_\perp(\vec{k},\vec{k}') e^{i\hat{\t}} c_{\vec{k},\uparrow}c_{\vec{k}',\downarrow} +{\rm h.c.} \right)  \nonumber\\
    &+ \sum_{\vec{k},\vec{k}',s} g_z(\vec{k},\vec{k}')\hat{n}c^\dagger_{\vec{k},s}c_{\vec{k}',s}.
    \label{H rotor}
\end{align}
It is 
convenient to introduce the dimensionless couplings, defined by ($b=\perp,z$) 
\begin{equation}
      \alpha_{b} = \frac{\rho^2_F}{\pi^2} \sum_{\vec{k},\vec{k}'\in {\rm FS}} |g_{b}(\vec{k},\vec{k}')|^2, 
      \label{alphas}
\end{equation}
where $\rho_F$ is the density of states at the Fermi level and $\sum_{\k \in {\rm FS}} (\cdot) \equiv \langle (\cdot)\rangle_{\rm FS}$ denotes a Fermi surface average. 

The spectrum of a single isolated puddle is $E_n=E_{c,0}(n-\overline{n})^2$, which has excitation energies of order $E_{c0}$ for $\overline{n}\approx n_0$  with integer $n_0$. However, if $\overline{n}\approx n_0+\tfrac{1}{2}$,  two consecutive charge states are almost degenerate, and separated from other states by at least $E_{c,0}$. This near-degeneracy leads to resonant Andreev scattering at arbitrarily low energy. To  describe such puddles we can project the Hamiltonian of Eq. \eqref{H rotor} into the two-dimensional Hilbert space of the near-degenerate  states, coupled to the conduction electrons:
\begin{align}
    H_{\rm C}+H_{\rm int}
    & = \sum_{\vec{k},\vec{k}' }\left(g_\perp(\vec{k},\vec{k}')\sigma^+ c_{\vec{k}\uparrow}c_{\vec{k}'\downarrow} + {\rm h.c.} \right) \nonumber \\ &-\frac{h}{2}\sigma^z + \sum_{\vec{k},\vec{k}',s}g_z(\vec{k},\vec{k}')\sigma^z c^\dagger _{\vec{k}s}c_{\vec{k}'s} ,
    \label{TLS Ham}
\end{align}
where $\sigma^z,\sigma^\pm$ are Pauli operators acting in the space of the two charge states, while $h$ measures the deviation from perfect degeneracy and is related to the background charge as $\overline{n}=n_0+\tfrac{1}{2}+\tfrac{h}{2E_c}$. $h$ is defined such that $|h|<E_c$. 
Integrating out the higher-energy states with charge $n\neq \{n_0,n_0+1\}$ renormalizes the charging energy to $E_c=z(\alpha_\perp) E_{c,0}$, 
which serves as the upper cutoff for the theory in Eq. \eqref{TLS Ham}. The factor $z(\alpha_\perp)$ is known in the strong coupling limit $z(\alpha_\perp \gg1)\sim \alpha_\perp^2\exp(-\pi^2 \alpha_\perp)$ \cite{Zaikin1994InfluenceOC,Feigel_man_1998,feigelman_weak_2002,Lukyanov_2004}. 
As we will now show, for intermediately sized puddles $\alpha_\perp$ is not too large, $\alpha_\perp=\mathcal{O}(1)$ at most, so that we can treat $E_c$ as a well-defined energy scale and focus on energies below $E_c$. Due to the exponential renormalization in $\alpha_\perp$, statistical fluctuations in $E_c$ may 
be significant if some channels are very strongly coupled to the puddle; see Appendix~\ref{Appendix: Ec distribution}.

\textbf{Coupling matrices $g_\perp$ and $g_z$.} We now discuss the structure of the matrices $g_{\perp,z}$ and resulting dimensionless couplings $\alpha_{\perp,z}$ for mesoscopic puddles treated within a mean-field approximation. The coupling constants $\alpha_{\perp}$ and $\alpha_z$ depend on the size $R$ of the puddle and the superconducting coherence length, $\xi_0 = v_F/\Delta_0$,  that characterizes the bulk properties of a large puddle with $R \gg \xi_0$.  (Here  $\Delta_0$ is the mean-field pairing potential magnitude.) 

We consider a spatially dependent gap function $\Delta_{ij} = \Delta_{\vec{r}_i - \vec{r}_j}\left(\vec{r}\right)$, which we assume can be expressed as a product of functions of the relative coordinate and the center-of-mass coordinate $\vec{r} = (\vec{r}_i+\vec{r}_j)/2$. Its Fourier transform with respect to the relative coordinate, $\Delta_{\frac{\vec{k}-\vec{k}'}{2}}\left(\vec{r}\right)$ depends on the 
form-factor of the gap-function, $\Delta_{\vec{k}}$, e.g., $s$-wave, $d$-wave, etc.

The existence of a mean-field gap and its magnitude is set by the size of the puddle. If $R 
\ll \xi_0$, the puddle does not admit a mean-field solution due to the inverse proximity effect~\cite{Spivak2001}. 
In the crucial range for present purposes, $R\sim \xi_0$, the mean-field pairing potential $\Delta$ is smaller than 
$\Delta_0$ and the coherence length $\xi\sim v_F /\Delta$ is larger than $\xi_0$. Thus, scattering matrix elements can be evaluated within the Born approximation where $g_\perp$ is proportional to the Fourier transform of the spatially dependent pairing potential:
\begin{align}
g_\perp(\vec{k},\vec{k}')&\propto \sum_{\vec{r}} e^{i(\vec{k}+\vec{k}')\vec{r}}\Delta_{\frac{\vec{k}-\vec{k}'}{2}}\left(\vec{r}\right) \nonumber \\&= \left(\frac{R}{a}\right)^2  \Delta_{\frac{\vec{k}-\vec{k}'}{2}} f_\perp\left(\frac{|\vec{k}+\vec{k}'|}{q_0}\right).   
\label{g perp}
\end{align}
Here, $\vec{k}$ and $\vec{k}'$ lie on the Fermi surface (which we assume is isotropic), and  $q_0$ denotes the range for efficient momentum exchange on the Fermi surface. For example, circular puddles with a uniform pairing potential satisfy $q_0\sim 1/R$. 
The function $f_\perp(x)$ depends on the spatial profile of the puddle, and controls the range of scattering in momentum space. For a puddle with a sharp boundary, it oscillates in sign and satisfies $f(x\to 0) =\pi $ and $f(x\gg 1)\sim 1/x^2$. For example, a circular puddle has $f_\perp(x)=2\pi J_1(x)/x$ with $J_1(x)$ being the Bessel function.  We expect the momentum structure of $g_\perp$ to remain qualitatively similar beyond the Born approximation, i.e., when $R/\xi\gtrsim 1$. Scattering off of small puddles 
with no well-formed mean-field gap \cite{degennes,Spivak2001,feigelman_weak_2002} - i.e. typically with $R < \xi_0$, is not considered in this work.

Using the Born approximation for $g_\perp$ (Eq.~\eqref{g perp}), we find that $\alpha_\perp \sim (k_F a)^2 k_F R^3/\xi^2$.
Physically, $(R/\xi)^2$ is the scattering probability while $k_F R$ is the effective number of channels for scattering. The range of momentum exchange for puddles with gap function disorder or shape irregularities is considerably larger. We provide several more examples of $\alpha_\perp$ for non-homogeneous puddles, i.e. when $q_0$ differs from $1/R$, in Appendix \ref{appendix: puddles}.

From a mesoscopic perspective, $\alpha_\perp$ is proportional to the interface Andreev conductance between the metal and the SC puddle. The Andreev conductance is is approximately proportional to $\alpha_\perp$ provided that the conductance of the bulk metal is much larger than $\alpha_\perp$. In the absence of a tunneling barrier, this condition is satisfied in the ballistic regime, i.e. when $R\ll \ell_{\rm mfp}$, the bulk metal mean free path \cite{Beenakker1992,Skvortsov2001}.

The matrix element for normal scattering is related to Coulomb repulsion between the puddle and itinerant electrons. We therefore consider it to be of the same order as the charging energy:
    $g_z(\vec{k},\vec{k}') \propto   U (a/R)^2 f_z(|\vec{k}-\vec{k}'|/q_0)$,
where $f_z(x)$ is a function similar to $f_\perp(x)$ and the dimensionless coupling for normal scattering, $\alpha_z$, rapidly decays as $R$ increases. 
Using this form of $g_z$, we obtain $\alpha_z \sim (k_F a)^2 \rho_F^2 U^2 (a/R)^{5}$ for uniform circular puddles. 
Hence, puddles of intermediate to large size satisfy $\alpha_z < \alpha_\perp$. 
This observation will be important in the following, where we show that the SM behavior comes from (inelastic) Andreev scattering off of intermediate-size puddles, while normal scattering processes are mainly elastic and can be essentially neglected when $\alpha_z < \alpha_\perp$. 

{\bf Low-energy behavior.} Considering the low-energy Hamiltonian ~\eqref{TLS Ham}, we observe that the problem 
resembles an anisotropic multichannel charge-Kondo model~\cite{taraphder1991heavy,dzero2005superconductivity,matsushita2005evidence,garate2011charge},  where the different channels correspond to the eigenvectors of $g_\perp(\vec{k},\vec{k}')$ and $g_z(\vec{k},\vec{k}')$, i.e. the distinct fermion modes that couple to the puddle. Generically, below an emergent charge-Kondo temperature, $T_K$, the coupling to one particular channel dominates the physics, leading to Kondo-screening and a restoration of Fermi-liquid behavior.  
However, as we shall now discuss, 
the competition between 
$N\sim k_FR$ different channels with comparable couplings frustrates the Kondo screening over a wide range of energies, resulting in an exponential suppression (in $N$)  of $T_K$.

\begin{figure}[t]
\centering
\includegraphics[width=0.48\textwidth]{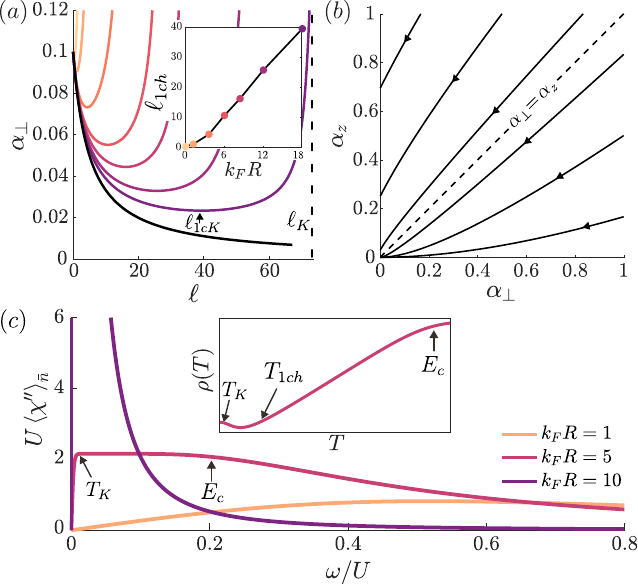} 
\caption{
\textbf{(a)} RG flow of $\alpha_\perp$ 
(see Appendix.~\ref{appendix: kondo}) for circular puddles with uniform gap of varying sizes and with $g_z(0)=0$. The black line corresponds to Eq.~\eqref{alpha of ell}. $\ell_{1{\rm ch}}$ is defined as the minimum of each curve, and $\ell_{K}$ as the point where they grow to be $\mathcal{O}(1)$.
Inset: $\ell_{1{\rm ch}}$ as a function of the puddle size $k_FR$. 
Colors correspond to $\alpha_\perp(\ell)$ curves of the main plot.
\textbf{(b)} Flow lines in the $(\alpha_\perp,\alpha_z)$-plane given by Eqs.~\eqref{RG spin-boson U1}. For $\alpha_z>\alpha_\perp$, $\alpha_z$ flows to a finite value and $\alpha_\perp$ is irrelevant, and other both are marginally irrelevant. \textbf{(c)} Schematic form of susceptibilities averaged over background charge for various sizes. Inset: Schematic resistivity curve for $k_FR=5$, with the energy scales $T_K,T_{1{\rm ch}},E_c$ marking the different features. }
\label{fig:rg and susc}
\end{figure}

To see this explicitly, in Fig.~\ref{fig:rg and susc}(a) we plot the RG flow of $\alpha_{\perp}$ for a representative case of uniform circular puddles of radius $R$. The flow of $\alpha_\perp$ is obtained from the full RG equations of the coupling matrices $g_{\perp,z}$, 
which we present in Appendix~\ref{appendix: kondo}. 
The effective number of channels $N\sim k_F R$; see Appendix~\ref{appendix: kondo} for further details. 
 Fig.~\ref{fig:rg and susc}(a) shows that $\alpha_\perp$ initially decreases under the RG flow. It reaches a minimum and then starts growing (becoming marginally relevant) only after a finite ``RG time'' which we denote $\ell_{1{\rm ch}}\sim k_FR$, indicating that the associated energy scale is suppressed exponentially in $k_F R$. This exponential suppression stems from the fact that the density of the eigenvalues of the matrix $g_{\perp}(\vec{k},\vec{k}')$ has support near the maximal eigenvalue, and scales as the number of independent fermionic modes, roughly proportional to $k_FR$ due to the finite range of scattering on the Fermi surface. This does not rely on specific structure of $g_\perp$, and happens quite generically, even for uniform circular puddles. 
For a derivation and a detailed analysis of this result for various shapes and sizes of puddles, see Appendices \ref{appendix: puddles} and \ref{appendix: kondo}.

 Consequently, the temperature below which the system exhibits single-channel charge-Kondo behavior, denoted by $T_{1{\rm ch}}$, is {\it exponentially } suppressed in the size of the puddle: $T_{1{\rm ch}} \sim E_c e^{- ck_FR}$. 
 The non-universal  prefactor $c$  depends on details of the puddle but is found to be of order unity (see Appendix \ref{appendix: puddles}).
 Note that $T_{1{\rm ch}}$ is distinct from the Kondo temperature, $T_K < T_{1{\rm ch}}$; $T_{1{\rm ch}}$ serves as an upper cutoff for the effective single-channel Kondo problem, whereas $T_{K}\sim T_{1{\rm ch}}e^{-1/\sqrt{\alpha_\perp(\ell_{1{\rm ch}})}}$ is the temperature where the Kondo screening is fully developed.  For randomly distributed coupling constants, 
the exponential suppression of $T_{1{\rm ch}}$ was previously recognized in the context of the single-electron transistor for random coupling matrices~\cite{zarand_two-channel_2000}. 

The exponential suppression of $T_{1{\rm ch}}$ opens a parametrically large low-energy window
\begin{equation}
   E_ce^{- ck_FR}\ll {\rm max}(|\omega|,T) \ll E_c. 
   \label{regime}
\end{equation}
Crucial for our analysis is that the existence of this regime does not require any fine tuning to a high symmetry point of the multichannel Kondo problem.
As discussed in App.~\ref{appendix: random mat model}, in this regime the properties of the puddle are not determined by the full matrices $g_{z,\perp}$, but rather only by the dimensionless couplings  $\alpha_{z,\perp}$ introduced in Eq.~\eqref{alphas}. They follow to leading order in $\alpha_{z,\perp}$ and $h$ the well known flow equations of an $U(1)$ symmetric spin-boson problem~\cite{belyansky_frustration-induced_2021,novais_frustration_2005,zarand_two-channel_2000}

\begin{align}
\frac{d\alpha_\perp}{d\ell} &=  - 2\alpha_z\alpha_\perp-2\alpha_\perp^2 \nonumber, \\
    \frac{d\alpha_z}{d\ell} &= -4\alpha_z\alpha_\perp \nonumber,\\
    \frac{dh}{d\ell} &= (1-2\alpha_\perp) h.
    \label{RG spin-boson U1}
\end{align}
where $\ell=\log(E_c/\Lambda)$ is the RG parameter and $\Lambda$ is the running cutoff.
As shown in Appendix~\ref{appendix: random mat model}, the 
emergence of spin-boson physics relies on the coupling of the puddle to multiple electronic channels with comparable coupling strengths; under these conditions, the electronic environment behaves effectively as a bosonic ohmic bath. This mapping is exact within a model where $g_\perp$ and $g_z$ are treated as random matrices (presented in Appendix~\ref{appendix: random mat model}), and holds to second order in $g$ for the Hamiltonian Eq.~\eqref{TLS Ham}.
The exact $U(1)$ symmetry (reflecting charge conservation of the original problem) is a vital ingredient for the SM behavior, as we will see below. In a different class of models equivalent to generic spin-boson problems (without the $U(1)$ symmetry), the resulting non-Fermi liquid exponent in the electron self-energy varies continuously as a function of the coupling strength, rather than being fixed to the MFL form~\cite{BashanTunable,tulipman2024solvable}.

 
Next we show that in the relevant regime $\alpha_\perp>\alpha_z$ the interaction is marginally irrelevant, allowing us to solve for the dynamic susceptibility of the puddle.
Considering the flow in the $(\alpha_\perp,\alpha_z)$-plane, shown in Fig.~\ref{fig:rg and susc}(b), we find two regimes: If $\alpha_z>\alpha_\perp$ then $\alpha_\perp$ flows to $0$ while $\alpha_z$ is renormalized as $\alpha_z\to \alpha_z-2\alpha_\perp + \mathcal{O}(\alpha_\perp^2)$. In this regime, the effects of inelastic Andreev scattering are suppressed. 
Conversely, in the regime of interest where $\alpha_\perp > \alpha_z$ both couplings are marginally irrelevant, with $\alpha_\perp$ decaying more slowly than $\alpha_z$.
To leading order in $\alpha_z(\ell=0)$, one obtains that
\begin{equation}
    \alpha_\perp(\ell) = \frac{\alpha_\perp(0)}{1+2\alpha_\perp(0)\ell+\alpha_z(0)\log(1+2\alpha_\perp(0)\ell)},
    \label{alpha of ell}
\end{equation}
which describes the contribution of intermediate to large puddles. Thus, for $\alpha_\perp(0) > \alpha_z(0)$ we can neglect the effect of $\alpha_z$ on the flow of $\alpha_\perp$ at low energies (large $\ell$). We henceforth assume $\alpha_z=0$, and omit the argument $(0)$ from bare coupling values.
Given the flow of $\alpha_\perp(\ell)$, we obtain the flow of the splitting $h(\ell)$:
\begin{equation}
    h(\ell)=\frac{e^\ell h(0)}{1+2\alpha_\perp(0) \ell}.
    \label{h(l)}
\end{equation}
We will be interested in the pairing susceptibility $\chi(i\omega,T,h,\alpha_\perp)=\int_\tau e^{i\omega\tau}\left<\mathcal{T}\sigma^-(\tau)\sigma^+(0)\right>$.
Since the interaction is marginally irrelevant, we can first perform the RG scaling down to a scale $\ell_* = \log(E_c/\max(h(\ell_*),\omega,T))$, at which point $\alpha_\perp(\ell_*)\ll 1$ can be treated perturbatively. The result can then be translated back to the bare values using the scaling relation coming from the field renormalization of the operators $\sigma^\pm$ \cite{belyansky_frustration-induced_2021,sheehy_quantum_2007}:
\begin{equation}
\chi(i\omega,T,\alpha_\perp,h) =\frac{e^{-\ell_*} \chi(ie^{\ell_*}\omega,e^{\ell_*} T,\alpha_\perp(\ell_*),h(\ell_*))}{1+2\alpha_\perp\ell_*} .
    \label{scaling susc}
\end{equation}
This procedure effectively sums all the contributions from the logs arising in the RG; see App.~\ref{appendix: averaged susc} for a derivation. Specifically, we approximate $\alpha_\perp(\ell_*)\approx 0$ in the r.h.s. of Eq.~\eqref{scaling susc}, and use the bare susceptibility
\begin{equation}
\chi(i\omega,T,h(\ell_*),\alpha_\perp(\ell_*)\approx0)=\frac{\tanh\left({ h(\ell_*)}/{2T}\right)}{i\omega+h(\ell_*)}.
    \label{weak coupling susc}
\end{equation}
One can go beyond the simplest approximation using an RPA-like re-summation of bubble diagrams, which accounts for broadening effects (see e.g. Ref.~\cite{belyansky_frustration-induced_2021}). Nevertheless, the zeroth order approximation correctly captures the qualitative behavior of the fully renormalized averaged susceptibility, as we show by an explicit calculation with additional corrections in App.~\ref{appendix: averaged susc}. Thus, we have solved the problem of a single puddle in the intermediate energy regime defined by Eq.~\eqref{regime}.


{\bf Averaging over puddles.} We proceed to evaluate the electronic self-energy and the transport relaxation rate due to scattering off of a collection of different puddles. We assume that electron-mediated interactions between puddles are sufficiently weak such that no global phase ordering occurs, and hence the contributions from different puddles can be summed independently.  
We further assume self-averaging and evaluate the contribution of a collection of puddles of with different sizes $R$ and background charges $\bar{n}$. 
Averaging over the different puddle sizes is rather straightforward, in particular as we are making the conservative assumption that the distribution of $R$  rapidly decays with $R$ and allows replacing $R$ by a typical value $\overline{R}$. A more careful treatment, giving the same qualitative behavior, is presented in Appendix~\ref{appendix: R average}. 
Henceforth we treat all $R$-dependent quantities (e.g. $E_c,T_{1{\rm ch}},\alpha_\perp$) as evaluated at $R=\overline{R}$.

Consider the averaging over $\overline{n}$. We expect generic distributions to have finite support around half-integer values, where the charge-gap $h$ vanishes. We focus on these contributions since they will dominate low-energy scattering processes. We therefore replace the average over arbitrary distributions of $\overline{n}$ by an average over the TLS gap $h$ which we take to be uniformly distributed between $-E_c$ and $E_c$. The non-interacting averaged TLS spectral function is therefore flat with $\chi'' \sim \text{sign}(\omega)$ at $T=0$.\footnote{Before introducing the electrons, the collection of decoupled (or weakly coupled) puddles with randomly distributed gaps is reminiscent of a disordered Bose glass \cite{FisherFisher1990}. In particular, the constant density of states of the (bare) averaged susceptibility is analogous to the one of the Bose glass phase.} However, due to the marginal interaction, it is natural to expect renormalization effects to logarithmically enhance or reduce the low-energy spectral weight.
However, using the scaling relations in Eqs.~\eqref{scaling susc} and \eqref{h(l)} we find that the susceptibility averaged over $h$ is scale invariant (see App.~\ref{appendix: averaged susc}). For $\text{max}(T,|\omega|)\ll E_c$ we can perform the averaging using  Eq.~\eqref{weak coupling susc} and obtain the imaginary part of the retarded susceptibility 
averaged over the background charge:
\begin{equation}
    \langle \chi'' (\omega,T) \rangle_{\bar{n}}= \frac{\pi}{2E_c}\tanh\left(\frac{\omega}{2T}\right).
    \label{sus averaged over h}
\end{equation}
Remarkably, we again find that the \textit{averaged} susceptibility is oblivious to the effects of the marginally irrelevant interaction and the renormalization of the gap $h$ due to cancellation between logarithmic renormalization of the field and the transverse susceptibility.
The form of Eq.~\eqref{sus averaged over h} holds qualitatively up to energies $\lesssim E_c$, after which it decays over a scale $\alpha_\perp E_c$; see App.~\ref{appendix: averaged susc} for details. For $\omega\ll T_K$, the susceptibility vanishes linearly due to Kondo screening (there is no special feature in the susceptibility at $T_{1{\rm ch}}$). In Fig.~\ref{fig:rg and susc}, we compare the gap-averaged susceptibility at $T=0$ of small, intermediate, and large puddles. We are interested in the `flat' regions of the susceptibility, i.e., $T_{1{\rm ch}}\ll \omega \ll E_c$, which give rise to strange metal behavior, as we will see below.
Small puddles have a large $T_{1{\rm ch}}$, such that there is not parametric separation from $E_c$, while large puddles have an exponentially small $E_c$, pushing this flat region to exponentially small energies. However, intermediate puddles sit on the ``sweet spot," having a not-too-small $E_c$ but a small enough $T_{1{\rm ch}}$. For these puddles, the energy regime Eq.~\eqref{regime} in which the susceptibility follows Eq.~\eqref{sus averaged over h} is therefore wide and accessible.

\textbf{Self energy and transport properties.} Finally, we determine the electronic self-energy which, due to Eq.~\eqref{sus averaged over h}, admits a marginal-Fermi-liquid form~\cite{varma_phenomenology_1989}. Indeed, we insert the fully renormalized averaged puddle susceptibility to the order $g_\perp^2$ (one-loop) expression for the imaginary part of the retarded self-energy (for details see \ref{appendix: transport}), and obtain
\begin{equation}
    \Sigma''_{\vec{k}}(\omega,T) = - \frac{\pi}{2}\lambda \eta_{\vec{k}}\omega \coth{\left(\frac{\omega}{2T}\right)},
    \label{1 loop se}
\end{equation}
where $\lambda= \pi\alpha_\perp n_{\rm pud} /(2k_Fa\rho_FE_c) $ with $n_{\rm pud}$  the puddle density per unit cell, and the form factor $\eta_{\vec{k}}=\sum_{\vec{k}'\in {\rm FS}} |g_\perp(\vec{k},\vec{k}')|^2/\sum_{\vec{k}'',\vec{k}'\in {\rm FS}} |g_\perp(\vec{k}'',\vec{k}')|^2$ is roughly proportional to $|\Delta_{\vec{k}}|^2$ ``smeared" over a scale $q_0$. For example, for $d$-wave puddles, $\eta_{\vec{k}} \approx a_{k_F R} +  b_{k_F R} \cos^2(2\theta_{\vec{k}})$ with considerable anisotropy ($a_{k_F R} \ll b_{k_F R}$) for $k_F R\gtrsim 1$.
Based on scaling considerations, higher-order contributions do not contribute more singularly than $\Sigma'' \sim |\omega|$, and we expect are subleading in powers of $(1+2\alpha_\perp\log(E_c/|\omega|))^{-1}$. In addition, in the App.~\ref{appendix: random mat model} we show that the self energy admits the same form as in Eq.~\eqref{1 loop se} in a related exactly solvable random-matrix model at arbitrary couplings. For the Sommerfeld coefficient it follows at low $T$ that $C/T=\frac{\pi^{2}}{3}\rho_{F}\left(1+\lambda \log (E_c/T)\right) $.


Consider the contribution of electron-puddle scattering to the dc resistivity. At temperatures $T_{1ch} \ll T \ll E_c$, where the form \eqref{1 loop se} holds, the inverse transport time and single particle decay rate follow the same $T$-scaling, namely, both are $T$-linear. However, the scattering momentum dependence reduces the numerical prefactor from $\alpha_\perp$ to $\alpha_{\rm tr}$, given by 
\begin{align}
  \alpha_{\rm tr} = \frac{\rho_F^2}{\pi^2} \sum_{\vec{k},\vec{k}'\in {\rm FS}} |g_\perp(\vec{k},\vec{k}')|^2 \left(1+\cos(\theta_{\vec{k}}-\theta_{\vec{k}'})\right),
  \label{transport lifetime}
\end{align}
where $\theta_{\vec{k}}$ denote the angle along the Fermi surface. Note that due to Andreev scattering the weighting is $1+\cos$ (rather than $1-\cos$ as in normal scattering). For uniform circular puddles, $\alpha_{\rm tr}/\alpha_\perp \sim 1/\left(k_FR\right)^2$, while short-wavelength puddle heterogeneities enhance it to $1/k_FR$, as shown in the App.~\ref{appendix: puddles} for several microscopic examples. The resulting resistivity is then
\begin{equation}
    \rho(T)=\rho_0+  \frac{m_{\rm tr}}{e^2n_{\rm el}} 2\pi\lambda_{\rm tr} T
    \label{resistivity}
\end{equation}
where $\lambda_{\rm tr}=\tfrac{\alpha_{\rm tr}}{\alpha_\perp}\lambda$, while
$m_{\rm tr}$ is the transport mass
, $n_{\rm el}$ the electron density and $\rho_0$ the resistivity at $T=0$ due to elastic processes. Anisotropy in the scattering rate, e.g., due to $d$-wave puddles, may result in an intermediate window with linear-in-$B$ magnetoresistance \cite{grissonnanche_linear-temperature_2021,hinlopen_b2_2022,kim_linear_2024}. 

We also estimate the contribution to the resistivity due to elastic scattering off of the puddles. Note that this contribution corresponds to the mean charge on the puddle, i.e., $n_0$ (see definition below Eq.~\eqref{TLS Ham}), rather than the fluctuations in $\sigma^z$. Assuming that the puddle scatter electrons elastically with a cross-section $\sim k_F R$, the elastic scattering rate is roughly 
$
\sim n_{\rm pud }\left(k_{F}R\right)E_{F}
$.
In the case where $\alpha_\perp \sim k_F R$, this contribution to the residual resistivity is comparable to the inelastic contribution at temperatures near the upper cutoff $E_c$.



At energies below $T_{1ch}$, the interaction becomes marginally \textit{relevant}, and higher-order processes must be included in the scattering. In this regime, the resistivity exhibits an upturn due to charge-Kondo screening of the SC puddles that saturates at $T_K$~\cite{hewson_kondo_1993}; see the inset of Fig.~\ref{fig:rg and susc}(c). 


\begin{figure}[t!]
\centering
\includegraphics[width=0.46\textwidth]{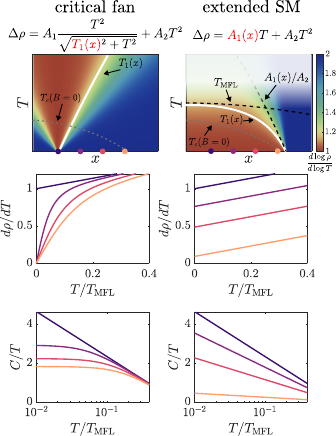} 
\caption{Schematic comparison of the expected singular $T$ dependence from a critical fan (left) and an extended strange metal (right), with an additional Fermi-liquid contribution $A_2T^2$. We assume that both descriptions are valid below an upper cutoff, denoted by $T_{\rm MFL}$.  For illustrative purposes, in the critical fan, we assume that $T_{\rm MFL}$ is sufficiently large and set $\nu z =1$.
\textbf{Top:} Map of $\tfrac{d\log \Delta \rho}{d\log T}$ in the
$(x,T)$-plane where $x$ is a non-thermal control parameter and we assume superconductivity is suppressed by a magnetic field $B$. Here $\Delta\rho\equiv\rho(T)-\rho(0)$. The dashed gray line denotes $T_c(B=0)$, and the white line $T_1(x)$ is the crossover temperature from $T^2$- to $T$-linear resistivity. For a critical fan, $T_1$ sets the lower cutoff, while in an extended SM scenario, $T_1(x) \sim \min\left[T_{\rm MFL},A_1(x)/A_2\right]$, where 
$A_1(x)$ is the slope of the $T$-linear contribution. The shaded top region of the right panel marks the limit of validity for our discussion. A very low temperature crossover or transition that is expected to bound the SM regime from below has been omitted for graphical simplicity.
\textbf{Center} and \textbf{bottom:}
$T_c(x,B=0)$ Comparison of the expected plots of $d\rho/dT$  and Sommerfeld coefficient $C/T$ for the two cases (assuming superconductivity is suppressed with a magnetic field), as a result of the variation of either the lower cutoff or the slope. The temperature is normalized by the upper cutoff of the MFL behavior  $T_{\rm MFL}$ ($=E_c$ in our case). Colored curves correspond to values of $x$ marked by the dots in the top panels.}
\label{fig:discussion}
\end{figure}

{\bf Discussion.} We have shown 
that inelastic Andreev scattering of electrons from quantum fluctuating superconducting puddles 
can give rise to a parametrically wide energy window with marginal Fermi liquid behavior and $T$-linear resistivity. This behavior occurs for generic distribution functions of the puddle sizes, which results in an extended strange metal regime, characterized by MFL behavior. 

It is important to contrast the predictions of our theory 
with the behavior expected in the vicinity of an isolated quantum critical point. To this end, Fig.~\ref{fig:discussion} shows the variation of 
the slope 
of the electrical resistivity $d \rho/dT$ and the Sommerfeld coefficient $C/T$ in our theory and that expected near a quantum critical point. 
As was pointed out in Refs.~\cite{paul_thermoelectric_2001,georges2021skewed,maebashi2022quantum}, the thermopower $S/T$ is expected to display a similar behavior to that of the Sommerfeld coefficient, 
but with a coefficient that depends on the particle-hole asymmetry and can have either sign.

 The behavior shown in the right panel of Fig.~\ref{fig:discussion} is  similar to 
 that reported for the electrical resistivity in three classes of overdoped cuprates: La$_{2-x}$Sr$_{x}$CuO$_{4}$\cite{cooper_anomalous_2009,hussey_dichotomy_2011,hussey_generic_2013,giraldo2018scale}, Tl$_2$Ba$_2$CuO$_{6+\delta}$ (Tl2201)~\cite{hussey_generic_2013,putzke2021reduced,ayres2021incoherent}, and Bi$_2$Sr$_2$CuO$_{6+\delta}$ (Bi2201)~\cite{putzke2021reduced,ayres2021incoherent}. 
 The typical size of the superconducting puddles in these systems are expected to be of size $\sim\xi_0$ (since much larger puddles are statistically rare and the pairing in puddles with $R\ll \xi_0$ is destroyed due to the inverse-proximity effect). The coherence length is presumably of the order of a few lattice constants, as is  observed in STM~\cite{gomes_visualizing_2007}. 
 This yields $\alpha_\perp \sim \mathcal{O}(1)$. The charging energy in such circumstances is only moderately renormalized compared to the bare one (roughly $E_c/E_{c,0} \sim 10^{-1}{\text - }10^{-2})$, which could account for a substantial temperature window of SM behavior for large enough $E_{c,0}$.
 
 We stress that our theory cannot explain the linear resistivity observed in many cuprates near optimal doping, which extends to very high temperatures (typically far above $300$K~\cite{Martin1990,Takagi1992}), as local superconductivity is unlikely to persists to such high temperatures, and the dynamical range of the $T$-linear resistivity is very large. Thus, within our theory, the $T$-linear behaviors in the low-$T$ overdoped and the high-$T$ optimally doped regimes are assumed to have different origins, and our mechanism has no bearing on the latter \cite{sachdev2025footfancupratephase}.

There are several implications of our theory that 
could be used to confirm or to falsify it. The most striking feature of the normal state resistivity in the overdoped cuprates is that, upon suppressing $T_c$ using a magnetic field, the resistivity is linear down to very low temperatures with a nearly field-independent slope~\cite{cooper_anomalous_2009,Boebinger2018}. Within our theory, this implies that the superconducting puddles survive in the presence of a high field; moreover, that the concentration of the puddles is not strongly field-dependent. This requires the puddles to have a very high local critical magnetic field -- 
larger even than the macroscopically determined critical field of the optimally doped bulk superconductor\footnote{Some experiments have suggested the existence of superconductivity at high fields, beyond the apparent $H_{c2}$~\cite{wang2006nernst,li2010diamagnetism,hsu2021unconventional}.}.
The presence of local superconducting gaps at high magnetic fields and low temperatures in the overdoped regime could be detected using STM (
ideally near the edge of the dome where the resistive $H_{c2}$ occurs in an accessible range of fields), or by observation of weak diamagnetism. Conversely, the absence of such local superconductivity would rule out our theory as an explanation for the extended strange metal regime. 


Our analysis also implies that future experiments should observe a logarithmic behavior of the Sommerfeld coefficient and the thermopower in the entire temperature and doping regime where $\rho \propto T$. Not finding such behavior, or observing a strong field variation of the coefficient of this logarithmic behavior (in contrast to the resistivity slope), would invalidate our theory. The observation of a logarithmic Sommerfeld coefficient for La$_{1.8-x}$Eu$_{0.2}$Sr$_x$CuO$_4$ (Eu-LSCO) and La$_{1.6-x}$Nd$_{0.4}$Sr$_x$CuO$_4$ (Nd-LSCO)\cite{Michon2019} with a slope that changes with doping,  similar to what is shown in the right panel of Fig.~\ref{fig:discussion}, and related behavior of the thermopower in Nd-LSCO~\cite{gourgout2022seebeck} are encouraging in this regard. 
Lastly, within our theory, strange metal behavior is a consequence of randomness, namely, a system without disorder or intrinsic randomness \cite{schmalian_stripe_2000} should be a homogeneous superconductor or a Fermi liquid metal.

Our proposed mechanism 
needs to be considered in the context of broader circumstances 
in which SM behavior could be observed. For example, the observation of SM behavior in conventional granular superconductors 
is not expected,
given their large coherence lengths. In that case, puddles of size $\xi_0$ or larger should have $\alpha_\perp \gg 1$, resulting in a strong suppression of their charging energy (as e.g. depicted in Fig.~\ref{fig:rg and susc}(c)) which restricts the potential MFL regime to exponentially low energies which might be practically unobservable. Nevertheless, it is interesting to ask whether these obstacles could be circumvented in engineered platforms, e.g., by embedding conventional SC grains in a metallic matrix. Observing SM behavior in these systems would require a significant reduction of $\alpha_\perp$. One possible means to this end is an application of an insulating barrier between the grains and the metallic matrix in which they are embedded. This would both reduce the Andreev tunnelling between the two (thus reducing the magnitude of $\alpha_\perp$), and additionally weaken the inverse proximity effect, allowing for grains smaller than $\xi_0$ to retain local superconductivity. Observing MFL behavior in such an engineered system would confirm the viability of our mechanism. 



\noindent{\bf Acknowledgements.}  We are  grateful to  
M. P. Allan, J. C. Davis, 
B. Gout\'eraux, E. Van Heumen, N. E. Hussey, H. Y. Hwang, M. Kiselev, P. A. Lee, M. Metlitski, Y. Oreg, A. Pandey, B. Ramshaw, S. Sachdev, T. Senthil, B. Spivak, L. Taillefer, J. M. Tranquada, A. M. Tremblay, and C. M. Varma for helpful discussions, and P. Nosov for referring us to Ref.~\cite{Eliashberg}. SAK and EB wish to acknowledge useful critiques by P. A. Lee and C. M. Varma of an earlier dynamical impurity based theory of strange metals. This work was supported in part by NSF-BSF Grant DMR-2000987. E.B. acknowledges support by the European Research Council (ERC) under grant HQMAT (Grant Agreement No. 817799) and the Simons Foundation Collaboration on New Frontiers in Superconductivity (Grant SFI-MPS-NFS-00006741-03). JS was supported by the German Research Foundation (DFG) through CRC TRR 288 “ElastoQMat,” project B01,  a grant from the Simons Foundation (SFI-MPS-NFS-00006741-05), and a Weston Visiting Professorship at the Weizmann Institute of Science. This research was supported in part by grant NSF PHY-2309135 to the Kavli Institute for Theoretical Physics (KITP).

\bibliography{TLSs.bib}

\begin{thebibliography}{109}%
\makeatletter
\providecommand \@ifxundefined [1]{%
 \@ifx{#1\undefined}
}%
\providecommand \@ifnum [1]{%
 \ifnum #1\expandafter \@firstoftwo
 \else \expandafter \@secondoftwo
 \fi
}%
\providecommand \@ifx [1]{%
 \ifx #1\expandafter \@firstoftwo
 \else \expandafter \@secondoftwo
 \fi
}%
\providecommand \natexlab [1]{#1}%
\providecommand \enquote  [1]{``#1''}%
\providecommand \bibnamefont  [1]{#1}%
\providecommand \bibfnamefont [1]{#1}%
\providecommand \citenamefont [1]{#1}%
\providecommand \href@noop [0]{\@secondoftwo}%
\providecommand \href [0]{\begingroup \@sanitize@url \@href}%
\providecommand \@href[1]{\@@startlink{#1}\@@href}%
\providecommand \@@href[1]{\endgroup#1\@@endlink}%
\providecommand \@sanitize@url [0]{\catcode `\\12\catcode `\$12\catcode `\&12\catcode `\#12\catcode `\^12\catcode `\_12\catcode `\%12\relax}%
\providecommand \@@startlink[1]{}%
\providecommand \@@endlink[0]{}%
\providecommand \url  [0]{\begingroup\@sanitize@url \@url }%
\providecommand \@url [1]{\endgroup\@href {#1}{\urlprefix }}%
\providecommand \urlprefix  [0]{URL }%
\providecommand \Eprint [0]{\href }%
\providecommand \doibase [0]{https://doi.org/}%
\providecommand \selectlanguage [0]{\@gobble}%
\providecommand \bibinfo  [0]{\@secondoftwo}%
\providecommand \bibfield  [0]{\@secondoftwo}%
\providecommand \translation [1]{[#1]}%
\providecommand \BibitemOpen [0]{}%
\providecommand \bibitemStop [0]{}%
\providecommand \bibitemNoStop [0]{.\EOS\space}%
\providecommand \EOS [0]{\spacefactor3000\relax}%
\providecommand \BibitemShut  [1]{\csname bibitem#1\endcsname}%
\let\auto@bib@innerbib\@empty
\bibitem [{\citenamefont {Ueda}(1977)}]{ueda1977electrical}%
  \BibitemOpen
  \bibfield  {author} {\bibinfo {author} {\bibfnamefont {K.}~\bibnamefont {Ueda}},\ }\bibfield  {title} {\bibinfo {title} {Electrical resistivity of antiferromagnetic metals},\ }\href {https://journals.jps.jp/doi/10.1143/JPSJ.43.1497?mobileUi=0} {\bibfield  {journal} {\bibinfo  {journal} {Journal of the Physical Society of Japan}\ }\textbf {\bibinfo {volume} {43}},\ \bibinfo {pages} {1497} (\bibinfo {year} {1977})}\BibitemShut {NoStop}%
\bibitem [{\citenamefont {Miranda}\ \emph {et~al.}(1996)\citenamefont {Miranda}, \citenamefont {Dobrosavljevic},\ and\ \citenamefont {Kotliar}}]{miranda1996kondo}%
  \BibitemOpen
  \bibfield  {author} {\bibinfo {author} {\bibfnamefont {E.}~\bibnamefont {Miranda}}, \bibinfo {author} {\bibfnamefont {V.}~\bibnamefont {Dobrosavljevic}},\ and\ \bibinfo {author} {\bibfnamefont {G.}~\bibnamefont {Kotliar}},\ }\bibfield  {title} {\bibinfo {title} {Kondo disorder: a possible route towards non-fermi-liquid behaviour},\ }\href {https://iopscience.iop.org/article/10.1088/0953-8984/8/48/014} {\bibfield  {journal} {\bibinfo  {journal} {Journal of Physics: Condensed Matter}\ }\textbf {\bibinfo {volume} {8}},\ \bibinfo {pages} {9871} (\bibinfo {year} {1996})}\BibitemShut {NoStop}%
\bibitem [{\citenamefont {Parcollet}\ and\ \citenamefont {Georges}(1999)}]{ParcolletNFL}%
  \BibitemOpen
  \bibfield  {author} {\bibinfo {author} {\bibfnamefont {O.}~\bibnamefont {Parcollet}}\ and\ \bibinfo {author} {\bibfnamefont {A.}~\bibnamefont {Georges}},\ }\bibfield  {title} {\bibinfo {title} {Non-fermi-liquid regime of a doped mott insulator},\ }\href {https://doi.org/10.1103/PhysRevB.59.5341} {\bibfield  {journal} {\bibinfo  {journal} {Phys. Rev. B}\ }\textbf {\bibinfo {volume} {59}},\ \bibinfo {pages} {5341} (\bibinfo {year} {1999})}\BibitemShut {NoStop}%
\bibitem [{\citenamefont {Rosch}(1999)}]{rosch_interplay_1999}%
  \BibitemOpen
  \bibfield  {author} {\bibinfo {author} {\bibfnamefont {A.}~\bibnamefont {Rosch}},\ }\bibfield  {title} {\bibinfo {title} {Interplay of {Disorder} and {Spin} {Fluctuations} in the {Resistivity} near a {Quantum} {Critical} {Point}},\ }\href {https://doi.org/10.1103/PhysRevLett.82.4280} {\bibfield  {journal} {\bibinfo  {journal} {Physical Review Letters}\ }\textbf {\bibinfo {volume} {82}},\ \bibinfo {pages} {4280} (\bibinfo {year} {1999})}\BibitemShut {NoStop}%
\bibitem [{\citenamefont {Aldape}\ \emph {et~al.}(2022)\citenamefont {Aldape}, \citenamefont {Cookmeyer}, \citenamefont {Patel},\ and\ \citenamefont {Altman}}]{aldape_solvable_2022}%
  \BibitemOpen
  \bibfield  {author} {\bibinfo {author} {\bibfnamefont {E.~E.}\ \bibnamefont {Aldape}}, \bibinfo {author} {\bibfnamefont {T.}~\bibnamefont {Cookmeyer}}, \bibinfo {author} {\bibfnamefont {A.~A.}\ \bibnamefont {Patel}},\ and\ \bibinfo {author} {\bibfnamefont {E.}~\bibnamefont {Altman}},\ }\bibfield  {title} {\bibinfo {title} {Solvable {Theory} of a {Strange} {Metal} at the {Breakdown} of a {Heavy} {Fermi} {Liquid}},\ }\href {https://doi.org/10.1103/PhysRevB.105.235111} {\bibfield  {journal} {\bibinfo  {journal} {Physical Review B}\ }\textbf {\bibinfo {volume} {105}},\ \bibinfo {pages} {235111} (\bibinfo {year} {2022})}\BibitemShut {NoStop}%
\bibitem [{\citenamefont {Esterlis}\ \emph {et~al.}(2021)\citenamefont {Esterlis}, \citenamefont {Guo}, \citenamefont {Patel},\ and\ \citenamefont {Sachdev}}]{esterlis_large_2021}%
  \BibitemOpen
  \bibfield  {author} {\bibinfo {author} {\bibfnamefont {I.}~\bibnamefont {Esterlis}}, \bibinfo {author} {\bibfnamefont {H.}~\bibnamefont {Guo}}, \bibinfo {author} {\bibfnamefont {A.~A.}\ \bibnamefont {Patel}},\ and\ \bibinfo {author} {\bibfnamefont {S.}~\bibnamefont {Sachdev}},\ }\bibfield  {title} {\bibinfo {title} {Large ${N}$ theory of critical {Fermi} surfaces},\ }\href {https://doi.org/10.1103/PhysRevB.103.235129} {\bibfield  {journal} {\bibinfo  {journal} {Physical Review B}\ }\textbf {\bibinfo {volume} {103}},\ \bibinfo {pages} {235129} (\bibinfo {year} {2021})}\BibitemShut {NoStop}%
\bibitem [{\citenamefont {Patel}\ \emph {et~al.}(2023)\citenamefont {Patel}, \citenamefont {Guo}, \citenamefont {Esterlis},\ and\ \citenamefont {Sachdev}}]{patel_universal_2023}%
  \BibitemOpen
  \bibfield  {author} {\bibinfo {author} {\bibfnamefont {A.~A.}\ \bibnamefont {Patel}}, \bibinfo {author} {\bibfnamefont {H.}~\bibnamefont {Guo}}, \bibinfo {author} {\bibfnamefont {I.}~\bibnamefont {Esterlis}},\ and\ \bibinfo {author} {\bibfnamefont {S.}~\bibnamefont {Sachdev}},\ }\href {http://arxiv.org/abs/2203.04990} {\bibinfo {title} {Universal theory of strange metals from spatially random interactions}} (\bibinfo {year} {2023})\BibitemShut {NoStop}%
\bibitem [{\citenamefont {Bashan}\ \emph {et~al.}(2024)\citenamefont {Bashan}, \citenamefont {Tulipman}, \citenamefont {Schmalian},\ and\ \citenamefont {Berg}}]{BashanTunable}%
  \BibitemOpen
  \bibfield  {author} {\bibinfo {author} {\bibfnamefont {N.}~\bibnamefont {Bashan}}, \bibinfo {author} {\bibfnamefont {E.}~\bibnamefont {Tulipman}}, \bibinfo {author} {\bibfnamefont {J.}~\bibnamefont {Schmalian}},\ and\ \bibinfo {author} {\bibfnamefont {E.}~\bibnamefont {Berg}},\ }\bibfield  {title} {\bibinfo {title} {Tunable non-fermi liquid phase from coupling to two-level systems},\ }\href {https://doi.org/10.1103/PhysRevLett.132.236501} {\bibfield  {journal} {\bibinfo  {journal} {Phys. Rev. Lett.}\ }\textbf {\bibinfo {volume} {132}},\ \bibinfo {pages} {236501} (\bibinfo {year} {2024})}\BibitemShut {NoStop}%
\bibitem [{\citenamefont {Tulipman}\ \emph {et~al.}(2024)\citenamefont {Tulipman}, \citenamefont {Bashan}, \citenamefont {Schmalian},\ and\ \citenamefont {Berg}}]{tulipman2024solvable}%
  \BibitemOpen
  \bibfield  {author} {\bibinfo {author} {\bibfnamefont {E.}~\bibnamefont {Tulipman}}, \bibinfo {author} {\bibfnamefont {N.}~\bibnamefont {Bashan}}, \bibinfo {author} {\bibfnamefont {J.}~\bibnamefont {Schmalian}},\ and\ \bibinfo {author} {\bibfnamefont {E.}~\bibnamefont {Berg}},\ }\bibfield  {title} {\bibinfo {title} {Solvable models of two-level systems coupled to itinerant electrons: Robust non-fermi liquid and quantum critical pairing},\ }\href {https://doi.org/10.1103/PhysRevB.110.155118} {\bibfield  {journal} {\bibinfo  {journal} {Phys. Rev. B}\ }\textbf {\bibinfo {volume} {110}},\ \bibinfo {pages} {155118} (\bibinfo {year} {2024})}\BibitemShut {NoStop}%
\bibitem [{\citenamefont {Patel}\ \emph {et~al.}(2024{\natexlab{a}})\citenamefont {Patel}, \citenamefont {Lunts},\ and\ \citenamefont {Sachdev}}]{Patel_2024loc}%
  \BibitemOpen
  \bibfield  {author} {\bibinfo {author} {\bibfnamefont {A.~A.}\ \bibnamefont {Patel}}, \bibinfo {author} {\bibfnamefont {P.}~\bibnamefont {Lunts}},\ and\ \bibinfo {author} {\bibfnamefont {S.}~\bibnamefont {Sachdev}},\ }\bibfield  {title} {\bibinfo {title} {Localization of overdamped bosonic modes and transport in strange metals},\ }\bibfield  {journal} {\bibinfo  {journal} {Proceedings of the National Academy of Sciences}\ }\textbf {\bibinfo {volume} {121}},\ \href {https://doi.org/10.1073/pnas.2402052121} {10.1073/pnas.2402052121} (\bibinfo {year} {2024}{\natexlab{a}})\BibitemShut {NoStop}%
\bibitem [{\citenamefont {Li}\ \emph {et~al.}(2024)\citenamefont {Li}, \citenamefont {Valentinis}, \citenamefont {Patel}, \citenamefont {Guo}, \citenamefont {Schmalian}, \citenamefont {Sachdev},\ and\ \citenamefont {Esterlis}}]{Li_2024}%
  \BibitemOpen
  \bibfield  {author} {\bibinfo {author} {\bibfnamefont {C.}~\bibnamefont {Li}}, \bibinfo {author} {\bibfnamefont {D.}~\bibnamefont {Valentinis}}, \bibinfo {author} {\bibfnamefont {A.~A.}\ \bibnamefont {Patel}}, \bibinfo {author} {\bibfnamefont {H.}~\bibnamefont {Guo}}, \bibinfo {author} {\bibfnamefont {J.}~\bibnamefont {Schmalian}}, \bibinfo {author} {\bibfnamefont {S.}~\bibnamefont {Sachdev}},\ and\ \bibinfo {author} {\bibfnamefont {I.}~\bibnamefont {Esterlis}},\ }\bibfield  {title} {\bibinfo {title} {{Strange} {Metal} and {Superconductor} in the {Two}-{Dimensional Yukawa-Sachdev-Ye-Kitaev Model}},\ }\href {https://doi.org/10.1103/PhysRevLett.133.186502} {\bibfield  {journal} {\bibinfo  {journal} {Phys. Rev. Lett.}\ }\textbf {\bibinfo {volume} {133}},\ \bibinfo {pages} {186502} (\bibinfo {year} {2024})}\BibitemShut {NoStop}%
\bibitem [{\citenamefont {Patel}\ \emph {et~al.}(2024{\natexlab{b}})\citenamefont {Patel}, \citenamefont {Lunts},\ and\ \citenamefont {Albergo}}]{patel2024strange}%
  \BibitemOpen
  \bibfield  {author} {\bibinfo {author} {\bibfnamefont {A.~A.}\ \bibnamefont {Patel}}, \bibinfo {author} {\bibfnamefont {P.}~\bibnamefont {Lunts}},\ and\ \bibinfo {author} {\bibfnamefont {M.~S.}\ \bibnamefont {Albergo}},\ }\href@noop {} {\bibinfo {title} {Strange metals and planckian transport in a gapless phase from spatially random interactions}} (\bibinfo {year} {2024}{\natexlab{b}}),\ \Eprint {https://arxiv.org/abs/2410.05365} {arXiv:2410.05365} \BibitemShut {NoStop}%
\bibitem [{\citenamefont {Merchant}\ \emph {et~al.}(2001)\citenamefont {Merchant}, \citenamefont {Ostrick}, \citenamefont {Barber},\ and\ \citenamefont {Dynes}}]{Merchant2001}%
  \BibitemOpen
  \bibfield  {author} {\bibinfo {author} {\bibfnamefont {L.}~\bibnamefont {Merchant}}, \bibinfo {author} {\bibfnamefont {J.}~\bibnamefont {Ostrick}}, \bibinfo {author} {\bibfnamefont {R.~P.}\ \bibnamefont {Barber}},\ and\ \bibinfo {author} {\bibfnamefont {R.~C.}\ \bibnamefont {Dynes}},\ }\bibfield  {title} {\bibinfo {title} {Crossover from phase fluctuation to amplitude-dominated superconductivity: A model system},\ }\href {https://doi.org/10.1103/PhysRevB.63.134508} {\bibfield  {journal} {\bibinfo  {journal} {Phys. Rev. B}\ }\textbf {\bibinfo {volume} {63}},\ \bibinfo {pages} {134508} (\bibinfo {year} {2001})}\BibitemShut {NoStop}%
\bibitem [{\citenamefont {Durkin}\ \emph {et~al.}(2020)\citenamefont {Durkin}, \citenamefont {Garrido-Menacho}, \citenamefont {Gopalakrishnan}, \citenamefont {Jaggi}, \citenamefont {Kwon}, \citenamefont {Zuo},\ and\ \citenamefont {Mason}}]{Durkin2020}%
  \BibitemOpen
  \bibfield  {author} {\bibinfo {author} {\bibfnamefont {M.}~\bibnamefont {Durkin}}, \bibinfo {author} {\bibfnamefont {R.}~\bibnamefont {Garrido-Menacho}}, \bibinfo {author} {\bibfnamefont {S.}~\bibnamefont {Gopalakrishnan}}, \bibinfo {author} {\bibfnamefont {N.~K.}\ \bibnamefont {Jaggi}}, \bibinfo {author} {\bibfnamefont {J.-H.}\ \bibnamefont {Kwon}}, \bibinfo {author} {\bibfnamefont {J.-M.}\ \bibnamefont {Zuo}},\ and\ \bibinfo {author} {\bibfnamefont {N.}~\bibnamefont {Mason}},\ }\bibfield  {title} {\bibinfo {title} {Rare-region onset of superconductivity in niobium nanoislands},\ }\href {https://doi.org/10.1103/PhysRevB.101.035409} {\bibfield  {journal} {\bibinfo  {journal} {Phys. Rev. B}\ }\textbf {\bibinfo {volume} {101}},\ \bibinfo {pages} {035409} (\bibinfo {year} {2020})}\BibitemShut {NoStop}%
\bibitem [{\citenamefont {Bishop-Van~Horn}\ \emph {et~al.}(2022)\citenamefont {Bishop-Van~Horn}, \citenamefont {Zhang}, \citenamefont {Waite}, \citenamefont {Mondragon-Shem}, \citenamefont {Jensen}, \citenamefont {Oh}, \citenamefont {Lippman}, \citenamefont {Durkin}, \citenamefont {Hughes}, \citenamefont {Mason}, \citenamefont {Moler},\ and\ \citenamefont {Sochnikov}}]{Bishop2022}%
  \BibitemOpen
  \bibfield  {author} {\bibinfo {author} {\bibfnamefont {L.}~\bibnamefont {Bishop-Van~Horn}}, \bibinfo {author} {\bibfnamefont {I.~P.}\ \bibnamefont {Zhang}}, \bibinfo {author} {\bibfnamefont {E.~N.}\ \bibnamefont {Waite}}, \bibinfo {author} {\bibfnamefont {I.}~\bibnamefont {Mondragon-Shem}}, \bibinfo {author} {\bibfnamefont {S.}~\bibnamefont {Jensen}}, \bibinfo {author} {\bibfnamefont {J.}~\bibnamefont {Oh}}, \bibinfo {author} {\bibfnamefont {T.}~\bibnamefont {Lippman}}, \bibinfo {author} {\bibfnamefont {M.}~\bibnamefont {Durkin}}, \bibinfo {author} {\bibfnamefont {T.~L.}\ \bibnamefont {Hughes}}, \bibinfo {author} {\bibfnamefont {N.}~\bibnamefont {Mason}}, \bibinfo {author} {\bibfnamefont {K.~A.}\ \bibnamefont {Moler}},\ and\ \bibinfo {author} {\bibfnamefont {I.}~\bibnamefont {Sochnikov}},\ }\bibfield  {title} {\bibinfo {title} {Local imaging of diamagnetism in proximity-coupled niobium nanoisland arrays on gold thin films},\ }\href {https://doi.org/10.1103/PhysRevB.106.054521} {\bibfield  {journal}
  {\bibinfo  {journal} {Phys. Rev. B}\ }\textbf {\bibinfo {volume} {106}},\ \bibinfo {pages} {054521} (\bibinfo {year} {2022})}\BibitemShut {NoStop}%
\bibitem [{\citenamefont {Zhang}\ \emph {et~al.}(2021)\citenamefont {Zhang}, \citenamefont {Hen}, \citenamefont {Palevski},\ and\ \citenamefont {Kapitulnik}}]{KapInO}%
  \BibitemOpen
  \bibfield  {author} {\bibinfo {author} {\bibfnamefont {X.}~\bibnamefont {Zhang}}, \bibinfo {author} {\bibfnamefont {B.}~\bibnamefont {Hen}}, \bibinfo {author} {\bibfnamefont {A.}~\bibnamefont {Palevski}},\ and\ \bibinfo {author} {\bibfnamefont {A.}~\bibnamefont {Kapitulnik}},\ }\bibfield  {title} {\bibinfo {title} {Robust anomalous metallic states and vestiges of self-duality in two-dimensional granular in-ino$_x$ composites},\ }\href {https://doi.org/10.1038/s41535-021-00329-2} {\bibfield  {journal} {\bibinfo  {journal} {npj Quantum Mater}\ }\textbf {\bibinfo {volume} {6}},\ \bibinfo {pages} {30} (\bibinfo {year} {2021})}\BibitemShut {NoStop}%
\bibitem [{\citenamefont {Han}\ \emph {et~al.}(2014)\citenamefont {Han}, \citenamefont {Allain}, \citenamefont {Arjmandi-Tash}, \citenamefont {Tikhonov}, \citenamefont {Feigel’Man}, \citenamefont {Sac{\'e}p{\'e}},\ and\ \citenamefont {Bouchiat}}]{han2014collapse}%
  \BibitemOpen
  \bibfield  {author} {\bibinfo {author} {\bibfnamefont {Z.}~\bibnamefont {Han}}, \bibinfo {author} {\bibfnamefont {A.}~\bibnamefont {Allain}}, \bibinfo {author} {\bibfnamefont {H.}~\bibnamefont {Arjmandi-Tash}}, \bibinfo {author} {\bibfnamefont {K.}~\bibnamefont {Tikhonov}}, \bibinfo {author} {\bibfnamefont {M.}~\bibnamefont {Feigel’Man}}, \bibinfo {author} {\bibfnamefont {B.}~\bibnamefont {Sac{\'e}p{\'e}}},\ and\ \bibinfo {author} {\bibfnamefont {V.}~\bibnamefont {Bouchiat}},\ }\bibfield  {title} {\bibinfo {title} {Collapse of superconductivity in a hybrid tin--graphene josephson junction array},\ }\href {https://www.nature.com/articles/nphys2929} {\bibfield  {journal} {\bibinfo  {journal} {Nature Physics}\ }\textbf {\bibinfo {volume} {10}},\ \bibinfo {pages} {380} (\bibinfo {year} {2014})}\BibitemShut {NoStop}%
\bibitem [{\citenamefont {Galitski}(2008)}]{Galitski2008}%
  \BibitemOpen
  \bibfield  {author} {\bibinfo {author} {\bibfnamefont {V.}~\bibnamefont {Galitski}},\ }\bibfield  {title} {\bibinfo {title} {Mesoscopic gap fluctuations in an unconventional superconductor},\ }\href {https://doi.org/10.1103/PhysRevB.77.100502} {\bibfield  {journal} {\bibinfo  {journal} {Phys. Rev. B}\ }\textbf {\bibinfo {volume} {77}},\ \bibinfo {pages} {100502} (\bibinfo {year} {2008})}\BibitemShut {NoStop}%
\bibitem [{\citenamefont {Dodaro}\ and\ \citenamefont {Kivelson}(2018)}]{dodaro_generalization_2018}%
  \BibitemOpen
  \bibfield  {author} {\bibinfo {author} {\bibfnamefont {J.~F.}\ \bibnamefont {Dodaro}}\ and\ \bibinfo {author} {\bibfnamefont {S.~A.}\ \bibnamefont {Kivelson}},\ }\bibfield  {title} {\bibinfo {title} {Generalization of anderson's theorem for disordered superconductors},\ }\href {https://doi.org/10.1103/PhysRevB.98.174503} {\bibfield  {journal} {\bibinfo  {journal} {Phys. Rev. B}\ }\textbf {\bibinfo {volume} {98}},\ \bibinfo {pages} {174503} (\bibinfo {year} {2018})}\BibitemShut {NoStop}%
\bibitem [{\citenamefont {Kapitulnik}\ \emph {et~al.}(2019)\citenamefont {Kapitulnik}, \citenamefont {Kivelson},\ and\ \citenamefont {Spivak}}]{Kapitulnik_colloquium_2019}%
  \BibitemOpen
  \bibfield  {author} {\bibinfo {author} {\bibfnamefont {A.}~\bibnamefont {Kapitulnik}}, \bibinfo {author} {\bibfnamefont {S.~A.}\ \bibnamefont {Kivelson}},\ and\ \bibinfo {author} {\bibfnamefont {B.}~\bibnamefont {Spivak}},\ }\bibfield  {title} {\bibinfo {title} {Colloquium: Anomalous metals: Failed superconductors},\ }\href {https://doi.org/10.1103/RevModPhys.91.011002} {\bibfield  {journal} {\bibinfo  {journal} {Rev. Mod. Phys.}\ }\textbf {\bibinfo {volume} {91}},\ \bibinfo {pages} {011002} (\bibinfo {year} {2019})}\BibitemShut {NoStop}%
\bibitem [{\citenamefont {Li}\ \emph {et~al.}(2021)\citenamefont {Li}, \citenamefont {Kivelson},\ and\ \citenamefont {Lee}}]{li_superconductor--metal_2021}%
  \BibitemOpen
  \bibfield  {author} {\bibinfo {author} {\bibfnamefont {Z.-X.}\ \bibnamefont {Li}}, \bibinfo {author} {\bibfnamefont {S.~A.}\ \bibnamefont {Kivelson}},\ and\ \bibinfo {author} {\bibfnamefont {D.-H.}\ \bibnamefont {Lee}},\ }\bibfield  {title} {{\selectlanguage {english}\bibinfo {title} {Superconductor-to-metal transition in overdoped cuprates}},\ }\href {https://doi.org/10.1038/s41535-021-00335-4} {\bibfield  {journal} {\bibinfo  {journal} {npj Quantum Materials}\ }\textbf {\bibinfo {volume} {6}},\ \bibinfo {pages} {1} (\bibinfo {year} {2021})}\BibitemShut {NoStop}%
\bibitem [{\citenamefont {{Feigel'man, M. V. and Larkin, A. I.}}(1998)}]{feigel1998quantum}%
  \BibitemOpen
  \bibfield  {author} {\bibinfo {author} {\bibnamefont {{Feigel'man, M. V. and Larkin, A. I.}}},\ }\bibfield  {title} {\bibinfo {title} {Quantum superconductor--metal transition in a 2d proximity-coupled array},\ }\href {https://www.sciencedirect.com/science/article/abs/pii/S0301010498000755?via%3Dihub} {\bibfield  {journal} {\bibinfo  {journal} {Chemical physics}\ }\textbf {\bibinfo {volume} {235}},\ \bibinfo {pages} {107} (\bibinfo {year} {1998})}\BibitemShut {NoStop}%
\bibitem [{\citenamefont {Spivak}\ \emph {et~al.}(2001)\citenamefont {Spivak}, \citenamefont {Zyuzin},\ and\ \citenamefont {Hruska}}]{Spivak2001}%
  \BibitemOpen
  \bibfield  {author} {\bibinfo {author} {\bibfnamefont {B.}~\bibnamefont {Spivak}}, \bibinfo {author} {\bibfnamefont {A.}~\bibnamefont {Zyuzin}},\ and\ \bibinfo {author} {\bibfnamefont {M.}~\bibnamefont {Hruska}},\ }\bibfield  {title} {\bibinfo {title} {Quantum superconductor-metal transition},\ }\href {https://doi.org/10.1103/PhysRevB.64.132502} {\bibfield  {journal} {\bibinfo  {journal} {Phys. Rev. B}\ }\textbf {\bibinfo {volume} {64}},\ \bibinfo {pages} {132502} (\bibinfo {year} {2001})}\BibitemShut {NoStop}%
\bibitem [{\citenamefont {Feigel'man}\ \emph {et~al.}(2001)\citenamefont {Feigel'man}, \citenamefont {Larkin},\ and\ \citenamefont {Skvortsov}}]{Feigelman2001}%
  \BibitemOpen
  \bibfield  {author} {\bibinfo {author} {\bibfnamefont {M.~V.}\ \bibnamefont {Feigel'man}}, \bibinfo {author} {\bibfnamefont {A.~I.}\ \bibnamefont {Larkin}},\ and\ \bibinfo {author} {\bibfnamefont {M.~A.}\ \bibnamefont {Skvortsov}},\ }\bibfield  {title} {\bibinfo {title} {Quantum superconductor-metal transition in a proximity array},\ }\href {https://doi.org/10.1103/PhysRevLett.86.1869} {\bibfield  {journal} {\bibinfo  {journal} {Phys. Rev. Lett.}\ }\textbf {\bibinfo {volume} {86}},\ \bibinfo {pages} {1869} (\bibinfo {year} {2001})}\BibitemShut {NoStop}%
\bibitem [{\citenamefont {Spivak}\ \emph {et~al.}(2008)\citenamefont {Spivak}, \citenamefont {Oreto},\ and\ \citenamefont {Kivelson}}]{Spivak2008}%
  \BibitemOpen
  \bibfield  {author} {\bibinfo {author} {\bibfnamefont {B.}~\bibnamefont {Spivak}}, \bibinfo {author} {\bibfnamefont {P.}~\bibnamefont {Oreto}},\ and\ \bibinfo {author} {\bibfnamefont {S.~A.}\ \bibnamefont {Kivelson}},\ }\bibfield  {title} {\bibinfo {title} {Theory of quantum metal to superconductor transitions in highly conducting systems},\ }\href {https://doi.org/10.1103/PhysRevB.77.214523} {\bibfield  {journal} {\bibinfo  {journal} {Phys. Rev. B}\ }\textbf {\bibinfo {volume} {77}},\ \bibinfo {pages} {214523} (\bibinfo {year} {2008})}\BibitemShut {NoStop}%
\bibitem [{\citenamefont {Lemberger}\ \emph {et~al.}(2011)\citenamefont {Lemberger}, \citenamefont {Hetel}, \citenamefont {Tsukada}, \citenamefont {Naito},\ and\ \citenamefont {Randeria}}]{lemberger2011superconductor}%
  \BibitemOpen
  \bibfield  {author} {\bibinfo {author} {\bibfnamefont {T.~R.}\ \bibnamefont {Lemberger}}, \bibinfo {author} {\bibfnamefont {I.}~\bibnamefont {Hetel}}, \bibinfo {author} {\bibfnamefont {A.}~\bibnamefont {Tsukada}}, \bibinfo {author} {\bibfnamefont {M.}~\bibnamefont {Naito}},\ and\ \bibinfo {author} {\bibfnamefont {M.}~\bibnamefont {Randeria}},\ }\bibfield  {title} {\bibinfo {title} {Superconductor-to-metal quantum phase transition in overdoped la${}_{2\ensuremath{-}x}$sr${}_{x}$cuo${}_{4}$},\ }\href {https://doi.org/10.1103/PhysRevB.83.140507} {\bibfield  {journal} {\bibinfo  {journal} {Phys. Rev. B}\ }\textbf {\bibinfo {volume} {83}},\ \bibinfo {pages} {140507} (\bibinfo {year} {2011})}\BibitemShut {NoStop}%
\bibitem [{\citenamefont {Bo{\v{z}}ovi{\'c}}\ \emph {et~al.}(2016)\citenamefont {Bo{\v{z}}ovi{\'c}}, \citenamefont {He}, \citenamefont {Wu},\ and\ \citenamefont {Bollinger}}]{bovzovic2016dependence}%
  \BibitemOpen
  \bibfield  {author} {\bibinfo {author} {\bibfnamefont {I.}~\bibnamefont {Bo{\v{z}}ovi{\'c}}}, \bibinfo {author} {\bibfnamefont {X.}~\bibnamefont {He}}, \bibinfo {author} {\bibfnamefont {J.}~\bibnamefont {Wu}},\ and\ \bibinfo {author} {\bibfnamefont {A.}~\bibnamefont {Bollinger}},\ }\bibfield  {title} {\bibinfo {title} {Dependence of the critical temperature in overdoped copper oxides on superfluid density},\ }\href {https://doi.org/10.1038/nature19061} {\bibfield  {journal} {\bibinfo  {journal} {Nature}\ }\textbf {\bibinfo {volume} {536}},\ \bibinfo {pages} {309} (\bibinfo {year} {2016})}\BibitemShut {NoStop}%
\bibitem [{\citenamefont {Howald}\ \emph {et~al.}(2001)\citenamefont {Howald}, \citenamefont {Fournier},\ and\ \citenamefont {Kapitulnik}}]{howald_inherent_2001}%
  \BibitemOpen
  \bibfield  {author} {\bibinfo {author} {\bibfnamefont {C.}~\bibnamefont {Howald}}, \bibinfo {author} {\bibfnamefont {P.}~\bibnamefont {Fournier}},\ and\ \bibinfo {author} {\bibfnamefont {A.}~\bibnamefont {Kapitulnik}},\ }\bibfield  {title} {\bibinfo {title} {{Inherent inhomogeneities in tunneling spectra of ${\textrm{Bi}}_{2}{\textrm{Sr}}_{2}{\textrm{CaCu}}_{2}{\textrm{O}}_{8\ensuremath{-}x}$ crystals in the superconducting state}},\ }\href {https://doi.org/10.1103/PhysRevB.64.100504} {\bibfield  {journal} {\bibinfo  {journal} {Phys. Rev. B}\ }\textbf {\bibinfo {volume} {64}},\ \bibinfo {pages} {100504} (\bibinfo {year} {2001})}\BibitemShut {NoStop}%
\bibitem [{\citenamefont {Pan}\ \emph {et~al.}(2001)\citenamefont {Pan}, \citenamefont {O'Neal}, \citenamefont {Badzey}, \citenamefont {Chamon}, \citenamefont {Ding}, \citenamefont {Engelbrecht}, \citenamefont {Wang}, \citenamefont {Eisaki}, \citenamefont {Uchida}, \citenamefont {Gupta}, \citenamefont {Ng}, \citenamefont {Hudson}, \citenamefont {Lang},\ and\ \citenamefont {Davis}}]{pan_microscopic_2001}%
  \BibitemOpen
  \bibfield  {author} {\bibinfo {author} {\bibfnamefont {S.~H.}\ \bibnamefont {Pan}}, \bibinfo {author} {\bibfnamefont {J.~P.}\ \bibnamefont {O'Neal}}, \bibinfo {author} {\bibfnamefont {R.~L.}\ \bibnamefont {Badzey}}, \bibinfo {author} {\bibfnamefont {C.}~\bibnamefont {Chamon}}, \bibinfo {author} {\bibfnamefont {H.}~\bibnamefont {Ding}}, \bibinfo {author} {\bibfnamefont {J.~R.}\ \bibnamefont {Engelbrecht}}, \bibinfo {author} {\bibfnamefont {Z.}~\bibnamefont {Wang}}, \bibinfo {author} {\bibfnamefont {H.}~\bibnamefont {Eisaki}}, \bibinfo {author} {\bibfnamefont {S.}~\bibnamefont {Uchida}}, \bibinfo {author} {\bibfnamefont {A.~K.}\ \bibnamefont {Gupta}}, \bibinfo {author} {\bibfnamefont {K.-W.}\ \bibnamefont {Ng}}, \bibinfo {author} {\bibfnamefont {E.~W.}\ \bibnamefont {Hudson}}, \bibinfo {author} {\bibfnamefont {K.~M.}\ \bibnamefont {Lang}},\ and\ \bibinfo {author} {\bibfnamefont {J.~C.}\ \bibnamefont {Davis}},\ }\bibfield  {title} {{\selectlanguage {english}\bibinfo {title} {Microscopic electronic
  inhomogeneity in the high-\textrm{T}$_c$ superconductor \textrm{Bi}$_2$\textrm{Sr}$_2$\textrm{CaCu}$_2$\textrm{O}$_{8+x}$}},\ }\href {https://doi.org/10.1038/35095012} {\bibfield  {journal} {\bibinfo  {journal} {Nature}\ }\textbf {\bibinfo {volume} {413}},\ \bibinfo {pages} {282} (\bibinfo {year} {2001})}\BibitemShut {NoStop}%
\bibitem [{\citenamefont {Lang}\ \emph {et~al.}(2002)\citenamefont {Lang}, \citenamefont {Madhavan}, \citenamefont {Hoffman}, \citenamefont {Hudson}, \citenamefont {Eisaki}, \citenamefont {Uchida},\ and\ \citenamefont {Davis}}]{Lang_2002}%
  \BibitemOpen
  \bibfield  {author} {\bibinfo {author} {\bibfnamefont {K.~M.}\ \bibnamefont {Lang}}, \bibinfo {author} {\bibfnamefont {V.}~\bibnamefont {Madhavan}}, \bibinfo {author} {\bibfnamefont {J.~E.}\ \bibnamefont {Hoffman}}, \bibinfo {author} {\bibfnamefont {E.~W.}\ \bibnamefont {Hudson}}, \bibinfo {author} {\bibfnamefont {H.}~\bibnamefont {Eisaki}}, \bibinfo {author} {\bibfnamefont {S.}~\bibnamefont {Uchida}},\ and\ \bibinfo {author} {\bibfnamefont {J.~C.}\ \bibnamefont {Davis}},\ }\bibfield  {title} {\bibinfo {title} {Imaging the granular structure of high-\textrm{T}$_c$ superconductivity in underdoped \textrm{Bi}$_2$\textrm{Sr}$_2$\textrm{CaCu}$_2$\textrm{O}$_{8+\delta}$},\ }\href {https://doi.org/10.1038/415412a} {\bibfield  {journal} {\bibinfo  {journal} {Nature}\ }\textbf {\bibinfo {volume} {415}},\ \bibinfo {pages} {412–416} (\bibinfo {year} {2002})}\BibitemShut {NoStop}%
\bibitem [{\citenamefont {Li}\ \emph {et~al.}(2022)\citenamefont {Li}, \citenamefont {Sapkota}, \citenamefont {Lozano}, \citenamefont {Du}, \citenamefont {Li}, \citenamefont {Wu}, \citenamefont {Kundu}, \citenamefont {Koch}, \citenamefont {Wu}, \citenamefont {Winn}, \citenamefont {Chi}, \citenamefont {Matsuda}, \citenamefont {Frontzek}, \citenamefont {Bo\ifmmode~\check{z}\else \v{z}\fi{}in}, \citenamefont {Zhu}, \citenamefont {Bo\ifmmode \check{z}\else \v{z}\fi{}ovi\ifmmode~\acute{c}\else \'{c}\fi{}}, \citenamefont {Pasupathy}, \citenamefont {Drozdov}, \citenamefont {Fujita}, \citenamefont {Gu}, \citenamefont {Zaliznyak}, \citenamefont {Li},\ and\ \citenamefont {Tranquada}}]{tranquada_stripes_2022}%
  \BibitemOpen
  \bibfield  {author} {\bibinfo {author} {\bibfnamefont {Y.}~\bibnamefont {Li}}, \bibinfo {author} {\bibfnamefont {A.}~\bibnamefont {Sapkota}}, \bibinfo {author} {\bibfnamefont {P.~M.}\ \bibnamefont {Lozano}}, \bibinfo {author} {\bibfnamefont {Z.}~\bibnamefont {Du}}, \bibinfo {author} {\bibfnamefont {H.}~\bibnamefont {Li}}, \bibinfo {author} {\bibfnamefont {Z.}~\bibnamefont {Wu}}, \bibinfo {author} {\bibfnamefont {A.~K.}\ \bibnamefont {Kundu}}, \bibinfo {author} {\bibfnamefont {R.~J.}\ \bibnamefont {Koch}}, \bibinfo {author} {\bibfnamefont {L.}~\bibnamefont {Wu}}, \bibinfo {author} {\bibfnamefont {B.~L.}\ \bibnamefont {Winn}}, \bibinfo {author} {\bibfnamefont {S.}~\bibnamefont {Chi}}, \bibinfo {author} {\bibfnamefont {M.}~\bibnamefont {Matsuda}}, \bibinfo {author} {\bibfnamefont {M.}~\bibnamefont {Frontzek}}, \bibinfo {author} {\bibfnamefont {E.~S.}\ \bibnamefont {Bo\ifmmode~\check{z}\else \v{z}\fi{}in}}, \bibinfo {author} {\bibfnamefont {Y.}~\bibnamefont {Zhu}}, \bibinfo {author} {\bibfnamefont
  {I.}~\bibnamefont {Bo\ifmmode \check{z}\else \v{z}\fi{}ovi\ifmmode~\acute{c}\else \'{c}\fi{}}}, \bibinfo {author} {\bibfnamefont {A.~N.}\ \bibnamefont {Pasupathy}}, \bibinfo {author} {\bibfnamefont {I.~K.}\ \bibnamefont {Drozdov}}, \bibinfo {author} {\bibfnamefont {K.}~\bibnamefont {Fujita}}, \bibinfo {author} {\bibfnamefont {G.~D.}\ \bibnamefont {Gu}}, \bibinfo {author} {\bibfnamefont {I.~A.}\ \bibnamefont {Zaliznyak}}, \bibinfo {author} {\bibfnamefont {Q.}~\bibnamefont {Li}},\ and\ \bibinfo {author} {\bibfnamefont {J.~M.}\ \bibnamefont {Tranquada}},\ }\bibfield  {title} {\bibinfo {title} {Strongly overdoped ${\textrm{la}}_{2\ensuremath{-}x}{\textrm{sr}}_{x}{\textrm{cuo}}_{4}$: Evidence for josephson-coupled grains of strongly correlated superconductor},\ }\href {https://doi.org/10.1103/PhysRevB.106.224515} {\bibfield  {journal} {\bibinfo  {journal} {Phys. Rev. B}\ }\textbf {\bibinfo {volume} {106}},\ \bibinfo {pages} {224515} (\bibinfo {year} {2022})}\BibitemShut {NoStop}%
\bibitem [{\citenamefont {Gomes}\ \emph {et~al.}(2007)\citenamefont {Gomes}, \citenamefont {Pasupathy}, \citenamefont {Pushp}, \citenamefont {Ono}, \citenamefont {Ando},\ and\ \citenamefont {Yazdani}}]{gomes_visualizing_2007}%
  \BibitemOpen
  \bibfield  {author} {\bibinfo {author} {\bibfnamefont {K.~K.}\ \bibnamefont {Gomes}}, \bibinfo {author} {\bibfnamefont {A.~N.}\ \bibnamefont {Pasupathy}}, \bibinfo {author} {\bibfnamefont {A.}~\bibnamefont {Pushp}}, \bibinfo {author} {\bibfnamefont {S.}~\bibnamefont {Ono}}, \bibinfo {author} {\bibfnamefont {Y.}~\bibnamefont {Ando}},\ and\ \bibinfo {author} {\bibfnamefont {A.}~\bibnamefont {Yazdani}},\ }\bibfield  {title} {{\selectlanguage {english}\bibinfo {title} {Visualizing pair formation on the atomic scale in the high-{Tc} superconductor $\textrm{Bi}_2\textrm{Sr}_2\textrm{Ca}\textrm{Cu}_2\textrm{O}_{8+\delta}$}},\ }\href {https://doi.org/10.1038/nature05881} {\bibfield  {journal} {\bibinfo  {journal} {Nature}\ }\textbf {\bibinfo {volume} {447}},\ \bibinfo {pages} {569} (\bibinfo {year} {2007})}\BibitemShut {NoStop}%
\bibitem [{\citenamefont {Tromp}\ \emph {et~al.}(2023)\citenamefont {Tromp}, \citenamefont {Benschop}, \citenamefont {Ge}, \citenamefont {Battisti}, \citenamefont {Bastiaans}, \citenamefont {Chatzopoulos}, \citenamefont {Vervloet}, \citenamefont {Smit}, \citenamefont {van Heumen}, \citenamefont {Golden}, \citenamefont {Huang}, \citenamefont {Kondo}, \citenamefont {Takeuchi}, \citenamefont {Yin}, \citenamefont {Hoffman}, \citenamefont {Sulangi}, \citenamefont {Zaanen},\ and\ \citenamefont {Allan}}]{tromp_puddle_2023}%
  \BibitemOpen
  \bibfield  {author} {\bibinfo {author} {\bibfnamefont {W.~O.}\ \bibnamefont {Tromp}}, \bibinfo {author} {\bibfnamefont {T.}~\bibnamefont {Benschop}}, \bibinfo {author} {\bibfnamefont {J.-F.}\ \bibnamefont {Ge}}, \bibinfo {author} {\bibfnamefont {I.}~\bibnamefont {Battisti}}, \bibinfo {author} {\bibfnamefont {K.~M.}\ \bibnamefont {Bastiaans}}, \bibinfo {author} {\bibfnamefont {D.}~\bibnamefont {Chatzopoulos}}, \bibinfo {author} {\bibfnamefont {A.~H.~M.}\ \bibnamefont {Vervloet}}, \bibinfo {author} {\bibfnamefont {S.}~\bibnamefont {Smit}}, \bibinfo {author} {\bibfnamefont {E.}~\bibnamefont {van Heumen}}, \bibinfo {author} {\bibfnamefont {M.~S.}\ \bibnamefont {Golden}}, \bibinfo {author} {\bibfnamefont {Y.}~\bibnamefont {Huang}}, \bibinfo {author} {\bibfnamefont {T.}~\bibnamefont {Kondo}}, \bibinfo {author} {\bibfnamefont {T.}~\bibnamefont {Takeuchi}}, \bibinfo {author} {\bibfnamefont {Y.}~\bibnamefont {Yin}}, \bibinfo {author} {\bibfnamefont {J.~E.}\ \bibnamefont {Hoffman}}, \bibinfo {author} {\bibfnamefont
  {M.~A.}\ \bibnamefont {Sulangi}}, \bibinfo {author} {\bibfnamefont {J.}~\bibnamefont {Zaanen}},\ and\ \bibinfo {author} {\bibfnamefont {M.~P.}\ \bibnamefont {Allan}},\ }\bibfield  {title} {{\selectlanguage {english}\bibinfo {title} {Puddle formation and persistent gaps across the non-mean-field breakdown of superconductivity in overdoped ({Pb},{Bi}){2Sr2CuO6}+$\delta$}},\ }\href {https://doi.org/10.1038/s41563-023-01497-1} {\bibfield  {journal} {\bibinfo  {journal} {Nature Materials}\ }\textbf {\bibinfo {volume} {22}},\ \bibinfo {pages} {703} (\bibinfo {year} {2023})}\BibitemShut {NoStop}%
\bibitem [{\citenamefont {Mahmood}\ \emph {et~al.}(2019)\citenamefont {Mahmood}, \citenamefont {He}, \citenamefont {Bo\ifmmode \check{z}\else \v{z}\fi{}ovi\ifmmode~\acute{c}\else \'{c}\fi{}},\ and\ \citenamefont {Armitage}}]{mahmood2019locating}%
  \BibitemOpen
  \bibfield  {author} {\bibinfo {author} {\bibfnamefont {F.}~\bibnamefont {Mahmood}}, \bibinfo {author} {\bibfnamefont {X.}~\bibnamefont {He}}, \bibinfo {author} {\bibfnamefont {I.}~\bibnamefont {Bo\ifmmode \check{z}\else \v{z}\fi{}ovi\ifmmode~\acute{c}\else \'{c}\fi{}}},\ and\ \bibinfo {author} {\bibfnamefont {N.~P.}\ \bibnamefont {Armitage}},\ }\bibfield  {title} {\bibinfo {title} {Locating the missing superconducting electrons in the overdoped cuprates $\textrm{La}_{2-x}\textrm{Sr}_x\textrm{CuO}_4$},\ }\href {https://doi.org/10.1103/PhysRevLett.122.027003} {\bibfield  {journal} {\bibinfo  {journal} {Phys. Rev. Lett.}\ }\textbf {\bibinfo {volume} {122}},\ \bibinfo {pages} {027003} (\bibinfo {year} {2019})}\BibitemShut {NoStop}%
\bibitem [{\citenamefont {Orenstein}(2007)}]{orenstein_optical_2007}%
  \BibitemOpen
  \bibfield  {author} {\bibinfo {author} {\bibfnamefont {J.}~\bibnamefont {Orenstein}},\ }\bibfield  {title} {{\selectlanguage {english}\bibinfo {title} {Optical {Conductivity} and {Spatial} {Inhomogeneity} in {Cuprate} {Superconductors}}},\ }in\ \href {https://doi.org/10.1007/978-0-387-68734-6_7} {{\selectlanguage {english}\emph {\bibinfo {booktitle} {Handbook of {High}-{Temperature} {Superconductivity}: {Theory} and {Experiment}}}}},\ \bibinfo {editor} {edited by\ \bibinfo {editor} {\bibfnamefont {J.~R.}\ \bibnamefont {Schrieffer}}\ and\ \bibinfo {editor} {\bibfnamefont {J.~S.}\ \bibnamefont {Brooks}}}\ (\bibinfo  {publisher} {Springer},\ \bibinfo {address} {New York, NY},\ \bibinfo {year} {2007})\ pp.\ \bibinfo {pages} {299--324}\BibitemShut {NoStop}%
\bibitem [{\citenamefont {Wen}\ \emph {et~al.}(2009)\citenamefont {Wen}, \citenamefont {Mu}, \citenamefont {Luo}, \citenamefont {Yang}, \citenamefont {Shan}, \citenamefont {Ren}, \citenamefont {Cheng}, \citenamefont {Yan},\ and\ \citenamefont {Fang}}]{Wen_2009}%
  \BibitemOpen
  \bibfield  {author} {\bibinfo {author} {\bibfnamefont {H.-H.}\ \bibnamefont {Wen}}, \bibinfo {author} {\bibfnamefont {G.}~\bibnamefont {Mu}}, \bibinfo {author} {\bibfnamefont {H.}~\bibnamefont {Luo}}, \bibinfo {author} {\bibfnamefont {H.}~\bibnamefont {Yang}}, \bibinfo {author} {\bibfnamefont {L.}~\bibnamefont {Shan}}, \bibinfo {author} {\bibfnamefont {C.}~\bibnamefont {Ren}}, \bibinfo {author} {\bibfnamefont {P.}~\bibnamefont {Cheng}}, \bibinfo {author} {\bibfnamefont {J.}~\bibnamefont {Yan}},\ and\ \bibinfo {author} {\bibfnamefont {L.}~\bibnamefont {Fang}},\ }\bibfield  {title} {\bibinfo {title} {Specific-heat measurement of a residual superconducting state in the normal state of underdoped $\textrm{Bi}_2\textrm{Sr}_{2-x}\textrm{La}_x\textrm{CuO}_{6+\delta}$ cuprate superconductors},\ }\href {https://doi.org/10.1103/PhysRevLett.103.067002} {\bibfield  {journal} {\bibinfo  {journal} {Phys. Rev. Lett.}\ }\textbf {\bibinfo {volume} {103}},\ \bibinfo {pages} {067002} (\bibinfo {year} {2009})}\BibitemShut
  {NoStop}%
\bibitem [{\citenamefont {Tallon}\ and\ \citenamefont {Loram}(2020)}]{Tallon_2020}%
  \BibitemOpen
  \bibfield  {author} {\bibinfo {author} {\bibfnamefont {J.~L.}\ \bibnamefont {Tallon}}\ and\ \bibinfo {author} {\bibfnamefont {J.~W.}\ \bibnamefont {Loram}},\ }\bibfield  {title} {\bibinfo {title} {Field-dependent specific heat of the canonical underdoped cuprate superconductor $\hbox {YBa}_2\hbox {Cu}_4\hbox {O}_8$},\ }\bibfield  {journal} {\bibinfo  {journal} {Scientific Reports}\ }\textbf {\bibinfo {volume} {10}},\ \href {https://doi.org/10.1038/s41598-020-79017-3} {10.1038/s41598-020-79017-3} (\bibinfo {year} {2020})\BibitemShut {NoStop}%
\bibitem [{\citenamefont {Uchida}(2021)}]{Uchida_2021}%
  \BibitemOpen
  \bibfield  {author} {\bibinfo {author} {\bibfnamefont {S.-i.}\ \bibnamefont {Uchida}},\ }\bibfield  {title} {\bibinfo {title} {Ubiquitous charge order correlations in high-temperature superconducting cuprates},\ }\href {https://doi.org/10.7566/jpsj.90.111001} {\bibfield  {journal} {\bibinfo  {journal} {Journal of the Physical Society of Japan}\ }\textbf {\bibinfo {volume} {90}},\ \bibinfo {pages} {111001} (\bibinfo {year} {2021})}\BibitemShut {NoStop}%
\bibitem [{\citenamefont {Niu}\ \emph {et~al.}(2024)\citenamefont {Niu}, \citenamefont {Larrazabal}, \citenamefont {Gozlinski}, \citenamefont {Sato}, \citenamefont {Bastiaans}, \citenamefont {Benschop}, \citenamefont {Ge}, \citenamefont {Blanter}, \citenamefont {Gu}, \citenamefont {Swart},\ and\ \citenamefont {Allan}}]{niu2024equivalence}%
  \BibitemOpen
  \bibfield  {author} {\bibinfo {author} {\bibfnamefont {J.}~\bibnamefont {Niu}}, \bibinfo {author} {\bibfnamefont {M.~O.}\ \bibnamefont {Larrazabal}}, \bibinfo {author} {\bibfnamefont {T.}~\bibnamefont {Gozlinski}}, \bibinfo {author} {\bibfnamefont {Y.}~\bibnamefont {Sato}}, \bibinfo {author} {\bibfnamefont {K.~M.}\ \bibnamefont {Bastiaans}}, \bibinfo {author} {\bibfnamefont {T.}~\bibnamefont {Benschop}}, \bibinfo {author} {\bibfnamefont {J.-F.}\ \bibnamefont {Ge}}, \bibinfo {author} {\bibfnamefont {Y.~M.}\ \bibnamefont {Blanter}}, \bibinfo {author} {\bibfnamefont {G.}~\bibnamefont {Gu}}, \bibinfo {author} {\bibfnamefont {I.}~\bibnamefont {Swart}},\ and\ \bibinfo {author} {\bibfnamefont {M.~P.}\ \bibnamefont {Allan}},\ }\href@noop {} {\bibinfo {title} {Equivalence of pseudogap and pairing energy in a cuprate high-temperature superconductor}} (\bibinfo {year} {2024}),\ \Eprint {https://arxiv.org/abs/2409.15928} {arXiv:2409.15928} \BibitemShut {NoStop}%
\bibitem [{\citenamefont {Cooper}\ \emph {et~al.}(2009)\citenamefont {Cooper}, \citenamefont {Wang}, \citenamefont {Vignolle}, \citenamefont {Lipscombe}, \citenamefont {Hayden}, \citenamefont {Tanabe}, \citenamefont {Adachi}, \citenamefont {Koike}, \citenamefont {Nohara}, \citenamefont {Takagi}, \citenamefont {Proust},\ and\ \citenamefont {Hussey}}]{cooper_anomalous_2009}%
  \BibitemOpen
  \bibfield  {author} {\bibinfo {author} {\bibfnamefont {R.~A.}\ \bibnamefont {Cooper}}, \bibinfo {author} {\bibfnamefont {Y.}~\bibnamefont {Wang}}, \bibinfo {author} {\bibfnamefont {B.}~\bibnamefont {Vignolle}}, \bibinfo {author} {\bibfnamefont {O.~J.}\ \bibnamefont {Lipscombe}}, \bibinfo {author} {\bibfnamefont {S.~M.}\ \bibnamefont {Hayden}}, \bibinfo {author} {\bibfnamefont {Y.}~\bibnamefont {Tanabe}}, \bibinfo {author} {\bibfnamefont {T.}~\bibnamefont {Adachi}}, \bibinfo {author} {\bibfnamefont {Y.}~\bibnamefont {Koike}}, \bibinfo {author} {\bibfnamefont {M.}~\bibnamefont {Nohara}}, \bibinfo {author} {\bibfnamefont {H.}~\bibnamefont {Takagi}}, \bibinfo {author} {\bibfnamefont {C.}~\bibnamefont {Proust}},\ and\ \bibinfo {author} {\bibfnamefont {N.~E.}\ \bibnamefont {Hussey}},\ }\bibfield  {title} {\bibinfo {title} {Anomalous {Criticality} in the {Electrical} {Resistivity} of $\textrm{La}_2\textrm{Sr}_x\textrm{CuO}_4$},\ }\href {https://doi.org/10.1126/science.1165015} {\bibfield  {journal} {\bibinfo
  {journal} {Science}\ }\textbf {\bibinfo {volume} {323}},\ \bibinfo {pages} {603} (\bibinfo {year} {2009})}\BibitemShut {NoStop}%
\bibitem [{\citenamefont {Hussey}\ \emph {et~al.}(2011)\citenamefont {Hussey}, \citenamefont {Cooper}, \citenamefont {Xu}, \citenamefont {Wang}, \citenamefont {Mouzopoulou}, \citenamefont {Vignolle},\ and\ \citenamefont {Proust}}]{hussey_dichotomy_2011}%
  \BibitemOpen
  \bibfield  {author} {\bibinfo {author} {\bibfnamefont {N.~E.}\ \bibnamefont {Hussey}}, \bibinfo {author} {\bibfnamefont {R.~A.}\ \bibnamefont {Cooper}}, \bibinfo {author} {\bibfnamefont {X.}~\bibnamefont {Xu}}, \bibinfo {author} {\bibfnamefont {Y.}~\bibnamefont {Wang}}, \bibinfo {author} {\bibfnamefont {I.}~\bibnamefont {Mouzopoulou}}, \bibinfo {author} {\bibfnamefont {B.}~\bibnamefont {Vignolle}},\ and\ \bibinfo {author} {\bibfnamefont {C.}~\bibnamefont {Proust}},\ }\bibfield  {title} {\bibinfo {title} {Dichotomy in the {T}-linear resistivity in hole-doped cuprates},\ }\href {https://doi.org/10.1098/rsta.2010.0196} {\bibfield  {journal} {\bibinfo  {journal} {Philosophical Transactions of the Royal Society A: Mathematical, Physical and Engineering Sciences}\ }\textbf {\bibinfo {volume} {369}},\ \bibinfo {pages} {1626} (\bibinfo {year} {2011})}\BibitemShut {NoStop}%
\bibitem [{\citenamefont {Hussey}\ \emph {et~al.}(2013)\citenamefont {Hussey}, \citenamefont {Gordon-Moys}, \citenamefont {Kokalj},\ and\ \citenamefont {McKenzie}}]{hussey_generic_2013}%
  \BibitemOpen
  \bibfield  {author} {\bibinfo {author} {\bibfnamefont {N.~E.}\ \bibnamefont {Hussey}}, \bibinfo {author} {\bibfnamefont {H.}~\bibnamefont {Gordon-Moys}}, \bibinfo {author} {\bibfnamefont {J.}~\bibnamefont {Kokalj}},\ and\ \bibinfo {author} {\bibfnamefont {R.~H.}\ \bibnamefont {McKenzie}},\ }\bibfield  {title} {{\selectlanguage {english}\bibinfo {title} {Generic strange-metal behaviour of overdoped cuprates}},\ }\href {https://doi.org/10.1088/1742-6596/449/1/012004} {\bibfield  {journal} {\bibinfo  {journal} {Journal of Physics: Conference Series}\ }\textbf {\bibinfo {volume} {449}},\ \bibinfo {pages} {012004} (\bibinfo {year} {2013})}\BibitemShut {NoStop}%
\bibitem [{\citenamefont {Giraldo-Gallo}\ \emph {et~al.}(2018{\natexlab{a}})\citenamefont {Giraldo-Gallo}, \citenamefont {Galvis}, \citenamefont {Stegen}, \citenamefont {Modic}, \citenamefont {Balakirev}, \citenamefont {Betts}, \citenamefont {Lian}, \citenamefont {Moir}, \citenamefont {Riggs}, \citenamefont {Wu} \emph {et~al.}}]{giraldo2018scale}%
  \BibitemOpen
  \bibfield  {author} {\bibinfo {author} {\bibfnamefont {P.}~\bibnamefont {Giraldo-Gallo}}, \bibinfo {author} {\bibfnamefont {J.}~\bibnamefont {Galvis}}, \bibinfo {author} {\bibfnamefont {Z.}~\bibnamefont {Stegen}}, \bibinfo {author} {\bibfnamefont {K.~A.}\ \bibnamefont {Modic}}, \bibinfo {author} {\bibfnamefont {F.}~\bibnamefont {Balakirev}}, \bibinfo {author} {\bibfnamefont {J.}~\bibnamefont {Betts}}, \bibinfo {author} {\bibfnamefont {X.}~\bibnamefont {Lian}}, \bibinfo {author} {\bibfnamefont {C.}~\bibnamefont {Moir}}, \bibinfo {author} {\bibfnamefont {S.}~\bibnamefont {Riggs}}, \bibinfo {author} {\bibfnamefont {J.}~\bibnamefont {Wu}}, \emph {et~al.},\ }\bibfield  {title} {\bibinfo {title} {Scale-invariant magnetoresistance in a cuprate superconductor},\ }\href {https://www.science.org/doi/10.1126/science.aan3178} {\bibfield  {journal} {\bibinfo  {journal} {Science}\ }\textbf {\bibinfo {volume} {361}},\ \bibinfo {pages} {479} (\bibinfo {year} {2018}{\natexlab{a}})}\BibitemShut {NoStop}%
\bibitem [{\citenamefont {Putzke}\ \emph {et~al.}(2021)\citenamefont {Putzke}, \citenamefont {Benhabib}, \citenamefont {Tabis}, \citenamefont {Ayres}, \citenamefont {Wang}, \citenamefont {Malone}, \citenamefont {Licciardello}, \citenamefont {Lu}, \citenamefont {Kondo}, \citenamefont {Takeuchi} \emph {et~al.}}]{putzke2021reduced}%
  \BibitemOpen
  \bibfield  {author} {\bibinfo {author} {\bibfnamefont {C.}~\bibnamefont {Putzke}}, \bibinfo {author} {\bibfnamefont {S.}~\bibnamefont {Benhabib}}, \bibinfo {author} {\bibfnamefont {W.}~\bibnamefont {Tabis}}, \bibinfo {author} {\bibfnamefont {J.}~\bibnamefont {Ayres}}, \bibinfo {author} {\bibfnamefont {Z.}~\bibnamefont {Wang}}, \bibinfo {author} {\bibfnamefont {L.}~\bibnamefont {Malone}}, \bibinfo {author} {\bibfnamefont {S.}~\bibnamefont {Licciardello}}, \bibinfo {author} {\bibfnamefont {J.}~\bibnamefont {Lu}}, \bibinfo {author} {\bibfnamefont {T.}~\bibnamefont {Kondo}}, \bibinfo {author} {\bibfnamefont {T.}~\bibnamefont {Takeuchi}}, \emph {et~al.},\ }\bibfield  {title} {\bibinfo {title} {Reduced hall carrier density in the overdoped strange metal regime of cuprate superconductors},\ }\href {https://doi.org/10.1038/s41567-021-01197-0} {\bibfield  {journal} {\bibinfo  {journal} {Nature Physics}\ }\textbf {\bibinfo {volume} {17}},\ \bibinfo {pages} {826} (\bibinfo {year} {2021})}\BibitemShut {NoStop}%
\bibitem [{\citenamefont {Ayres}\ \emph {et~al.}(2021)\citenamefont {Ayres}, \citenamefont {Berben}, \citenamefont {{\v{C}}ulo}, \citenamefont {Hsu}, \citenamefont {van Heumen}, \citenamefont {Huang}, \citenamefont {Zaanen}, \citenamefont {Kondo}, \citenamefont {Takeuchi}, \citenamefont {Cooper} \emph {et~al.}}]{ayres2021incoherent}%
  \BibitemOpen
  \bibfield  {author} {\bibinfo {author} {\bibfnamefont {J.}~\bibnamefont {Ayres}}, \bibinfo {author} {\bibfnamefont {M.}~\bibnamefont {Berben}}, \bibinfo {author} {\bibfnamefont {M.}~\bibnamefont {{\v{C}}ulo}}, \bibinfo {author} {\bibfnamefont {Y.-T.}\ \bibnamefont {Hsu}}, \bibinfo {author} {\bibfnamefont {E.}~\bibnamefont {van Heumen}}, \bibinfo {author} {\bibfnamefont {Y.}~\bibnamefont {Huang}}, \bibinfo {author} {\bibfnamefont {J.}~\bibnamefont {Zaanen}}, \bibinfo {author} {\bibfnamefont {T.}~\bibnamefont {Kondo}}, \bibinfo {author} {\bibfnamefont {T.}~\bibnamefont {Takeuchi}}, \bibinfo {author} {\bibfnamefont {J.}~\bibnamefont {Cooper}}, \emph {et~al.},\ }\bibfield  {title} {\bibinfo {title} {Incoherent transport across the strange-metal regime of overdoped cuprates},\ }\href {https://www.nature.com/articles/s41586-021-03622-z} {\bibfield  {journal} {\bibinfo  {journal} {Nature}\ }\textbf {\bibinfo {volume} {595}},\ \bibinfo {pages} {661} (\bibinfo {year} {2021})}\BibitemShut {NoStop}%
\bibitem [{\citenamefont {Taillefer}(2010)}]{taillefer2010scattering}%
  \BibitemOpen
  \bibfield  {author} {\bibinfo {author} {\bibfnamefont {L.}~\bibnamefont {Taillefer}},\ }\bibfield  {title} {\bibinfo {title} {Scattering and pairing in cuprate superconductors},\ }\href {https://www.annualreviews.org/content/journals/10.1146/annurev-conmatphys-070909-104117} {\bibfield  {journal} {\bibinfo  {journal} {Annu. Rev. Condens. Matter Phys.}\ }\textbf {\bibinfo {volume} {1}},\ \bibinfo {pages} {51} (\bibinfo {year} {2010})}\BibitemShut {NoStop}%
\bibitem [{\citenamefont {Yuan}\ \emph {et~al.}(2022)\citenamefont {Yuan}, \citenamefont {Chen}, \citenamefont {Jiang}, \citenamefont {Feng}, \citenamefont {Lin}, \citenamefont {Yu}, \citenamefont {He}, \citenamefont {Zhang}, \citenamefont {Jiang}, \citenamefont {Zhang}, \citenamefont {Shi}, \citenamefont {Zhang}, \citenamefont {Qin}, \citenamefont {Cheng}, \citenamefont {Tamura}, \citenamefont {Yang}, \citenamefont {Xiang}, \citenamefont {Hu}, \citenamefont {Takeuchi}, \citenamefont {Jin},\ and\ \citenamefont {Zhao}}]{Yuan_2022}%
  \BibitemOpen
  \bibfield  {author} {\bibinfo {author} {\bibfnamefont {J.}~\bibnamefont {Yuan}}, \bibinfo {author} {\bibfnamefont {Q.}~\bibnamefont {Chen}}, \bibinfo {author} {\bibfnamefont {K.}~\bibnamefont {Jiang}}, \bibinfo {author} {\bibfnamefont {Z.}~\bibnamefont {Feng}}, \bibinfo {author} {\bibfnamefont {Z.}~\bibnamefont {Lin}}, \bibinfo {author} {\bibfnamefont {H.}~\bibnamefont {Yu}}, \bibinfo {author} {\bibfnamefont {G.}~\bibnamefont {He}}, \bibinfo {author} {\bibfnamefont {J.}~\bibnamefont {Zhang}}, \bibinfo {author} {\bibfnamefont {X.}~\bibnamefont {Jiang}}, \bibinfo {author} {\bibfnamefont {X.}~\bibnamefont {Zhang}}, \bibinfo {author} {\bibfnamefont {Y.}~\bibnamefont {Shi}}, \bibinfo {author} {\bibfnamefont {Y.}~\bibnamefont {Zhang}}, \bibinfo {author} {\bibfnamefont {M.}~\bibnamefont {Qin}}, \bibinfo {author} {\bibfnamefont {Z.~G.}\ \bibnamefont {Cheng}}, \bibinfo {author} {\bibfnamefont {N.}~\bibnamefont {Tamura}}, \bibinfo {author} {\bibfnamefont {Y.-f.}\ \bibnamefont {Yang}}, \bibinfo {author} {\bibfnamefont
  {T.}~\bibnamefont {Xiang}}, \bibinfo {author} {\bibfnamefont {J.}~\bibnamefont {Hu}}, \bibinfo {author} {\bibfnamefont {I.}~\bibnamefont {Takeuchi}}, \bibinfo {author} {\bibfnamefont {K.}~\bibnamefont {Jin}},\ and\ \bibinfo {author} {\bibfnamefont {Z.}~\bibnamefont {Zhao}},\ }\bibfield  {title} {\bibinfo {title} {Scaling of the strange-metal scattering in unconventional superconductors},\ }\href {https://doi.org/10.1038/s41586-021-04305-5} {\bibfield  {journal} {\bibinfo  {journal} {Nature}\ }\textbf {\bibinfo {volume} {602}},\ \bibinfo {pages} {431–436} (\bibinfo {year} {2022})}\BibitemShut {NoStop}%
\bibitem [{\citenamefont {Zaanen}(2004)}]{zaanen_why_2004}%
  \BibitemOpen
  \bibfield  {author} {\bibinfo {author} {\bibfnamefont {J.}~\bibnamefont {Zaanen}},\ }\bibfield  {title} {{\selectlanguage {english}\bibinfo {title} {Why the temperature is high}},\ }\href {https://doi.org/10.1038/430512a} {\bibfield  {journal} {\bibinfo  {journal} {Nature}\ }\textbf {\bibinfo {volume} {430}},\ \bibinfo {pages} {512} (\bibinfo {year} {2004})}\BibitemShut {NoStop}%
\bibitem [{\citenamefont {Bruin}\ \emph {et~al.}(2013)\citenamefont {Bruin}, \citenamefont {Sakai}, \citenamefont {Perry},\ and\ \citenamefont {Mackenzie}}]{bruin_similarity_2013}%
  \BibitemOpen
  \bibfield  {author} {\bibinfo {author} {\bibfnamefont {J.~a.~N.}\ \bibnamefont {Bruin}}, \bibinfo {author} {\bibfnamefont {H.}~\bibnamefont {Sakai}}, \bibinfo {author} {\bibfnamefont {R.~S.}\ \bibnamefont {Perry}},\ and\ \bibinfo {author} {\bibfnamefont {A.~P.}\ \bibnamefont {Mackenzie}},\ }\bibfield  {title} {{\selectlanguage {english}\bibinfo {title} {Similarity of {Scattering} {Rates} in {Metals} {Showing} {T}-{Linear} {Resistivity}}},\ }\href {https://doi.org/10.1126/science.1227612} {\bibfield  {journal} {\bibinfo  {journal} {Science}\ }\textbf {\bibinfo {volume} {339}},\ \bibinfo {pages} {804} (\bibinfo {year} {2013})}\BibitemShut {NoStop}%
\bibitem [{\citenamefont {Grissonnanche}\ \emph {et~al.}(2021)\citenamefont {Grissonnanche}, \citenamefont {Fang}, \citenamefont {Legros}, \citenamefont {Verret}, \citenamefont {Laliberté}, \citenamefont {Collignon}, \citenamefont {Zhou}, \citenamefont {Graf}, \citenamefont {Goddard}, \citenamefont {Taillefer},\ and\ \citenamefont {Ramshaw}}]{grissonnanche_linear-temperature_2021}%
  \BibitemOpen
  \bibfield  {author} {\bibinfo {author} {\bibfnamefont {G.}~\bibnamefont {Grissonnanche}}, \bibinfo {author} {\bibfnamefont {Y.}~\bibnamefont {Fang}}, \bibinfo {author} {\bibfnamefont {A.}~\bibnamefont {Legros}}, \bibinfo {author} {\bibfnamefont {S.}~\bibnamefont {Verret}}, \bibinfo {author} {\bibfnamefont {F.}~\bibnamefont {Laliberté}}, \bibinfo {author} {\bibfnamefont {C.}~\bibnamefont {Collignon}}, \bibinfo {author} {\bibfnamefont {J.}~\bibnamefont {Zhou}}, \bibinfo {author} {\bibfnamefont {D.}~\bibnamefont {Graf}}, \bibinfo {author} {\bibfnamefont {P.~A.}\ \bibnamefont {Goddard}}, \bibinfo {author} {\bibfnamefont {L.}~\bibnamefont {Taillefer}},\ and\ \bibinfo {author} {\bibfnamefont {B.~J.}\ \bibnamefont {Ramshaw}},\ }\bibfield  {title} {\bibinfo {title} {Linear-in temperature resistivity from an isotropic planckian scattering rate},\ }\href {https://doi.org/10.1038/s41586-021-03697-8} {\bibfield  {journal} {\bibinfo  {journal} {Nature}\ }\textbf {\bibinfo {volume} {595}},\ \bibinfo {pages} {667}
  (\bibinfo {year} {2021})}\BibitemShut {NoStop}%
\bibitem [{\citenamefont {Hartnoll}\ and\ \citenamefont {Mackenzie}(2022)}]{Hartnoll2022}%
  \BibitemOpen
  \bibfield  {author} {\bibinfo {author} {\bibfnamefont {S.~A.}\ \bibnamefont {Hartnoll}}\ and\ \bibinfo {author} {\bibfnamefont {A.~P.}\ \bibnamefont {Mackenzie}},\ }\bibfield  {title} {\bibinfo {title} {Colloquium: Planckian dissipation in metals},\ }\href {https://doi.org/10.1103/RevModPhys.94.041002} {\bibfield  {journal} {\bibinfo  {journal} {Rev. Mod. Phys.}\ }\textbf {\bibinfo {volume} {94}},\ \bibinfo {pages} {041002} (\bibinfo {year} {2022})}\BibitemShut {NoStop}%
\bibitem [{\citenamefont {Sachdev}(2011)}]{sachdev_quantum_2011}%
  \BibitemOpen
  \bibfield  {author} {\bibinfo {author} {\bibfnamefont {S.}~\bibnamefont {Sachdev}},\ }\href {https://doi.org/10.1017/CBO9780511973765} {\emph {\bibinfo {title} {Quantum Phase Transitions}}},\ \bibinfo {edition} {2nd}\ ed.\ (\bibinfo  {publisher} {Cambridge University Press},\ \bibinfo {year} {2011})\BibitemShut {NoStop}%
\bibitem [{\citenamefont {Gegenwart}\ \emph {et~al.}(2008)\citenamefont {Gegenwart}, \citenamefont {Si},\ and\ \citenamefont {Steglich}}]{gegenwart_quantum_2008}%
  \BibitemOpen
  \bibfield  {author} {\bibinfo {author} {\bibfnamefont {P.}~\bibnamefont {Gegenwart}}, \bibinfo {author} {\bibfnamefont {Q.}~\bibnamefont {Si}},\ and\ \bibinfo {author} {\bibfnamefont {F.}~\bibnamefont {Steglich}},\ }\bibfield  {title} {{\selectlanguage {english}\bibinfo {title} {Quantum criticality in heavy-fermion metals}},\ }\href {https://doi.org/10.1038/nphys892} {\bibfield  {journal} {\bibinfo  {journal} {Nature Physics}\ }\textbf {\bibinfo {volume} {4}},\ \bibinfo {pages} {186} (\bibinfo {year} {2008})}\BibitemShut {NoStop}%
\bibitem [{\citenamefont {Kirchner}\ \emph {et~al.}(2020{\natexlab{a}})\citenamefont {Kirchner}, \citenamefont {Paschen}, \citenamefont {Chen}, \citenamefont {Wirth}, \citenamefont {Feng}, \citenamefont {Thompson},\ and\ \citenamefont {Si}}]{HFreview}%
  \BibitemOpen
  \bibfield  {author} {\bibinfo {author} {\bibfnamefont {S.}~\bibnamefont {Kirchner}}, \bibinfo {author} {\bibfnamefont {S.}~\bibnamefont {Paschen}}, \bibinfo {author} {\bibfnamefont {Q.}~\bibnamefont {Chen}}, \bibinfo {author} {\bibfnamefont {S.}~\bibnamefont {Wirth}}, \bibinfo {author} {\bibfnamefont {D.}~\bibnamefont {Feng}}, \bibinfo {author} {\bibfnamefont {J.~D.}\ \bibnamefont {Thompson}},\ and\ \bibinfo {author} {\bibfnamefont {Q.}~\bibnamefont {Si}},\ }\bibfield  {title} {\bibinfo {title} {Colloquium: Heavy-electron quantum criticality and single-particle spectroscopy},\ }\href {https://doi.org/10.1103/RevModPhys.92.011002} {\bibfield  {journal} {\bibinfo  {journal} {Rev. Mod. Phys.}\ }\textbf {\bibinfo {volume} {92}},\ \bibinfo {pages} {011002} (\bibinfo {year} {2020}{\natexlab{a}})}\BibitemShut {NoStop}%
\bibitem [{\citenamefont {Varma}\ \emph {et~al.}(1989)\citenamefont {Varma}, \citenamefont {Littlewood}, \citenamefont {Schmitt-Rink}, \citenamefont {Abrahams},\ and\ \citenamefont {Ruckenstein}}]{varma_phenomenology_1989}%
  \BibitemOpen
  \bibfield  {author} {\bibinfo {author} {\bibfnamefont {C.~M.}\ \bibnamefont {Varma}}, \bibinfo {author} {\bibfnamefont {P.~B.}\ \bibnamefont {Littlewood}}, \bibinfo {author} {\bibfnamefont {S.}~\bibnamefont {Schmitt-Rink}}, \bibinfo {author} {\bibfnamefont {E.}~\bibnamefont {Abrahams}},\ and\ \bibinfo {author} {\bibfnamefont {A.~E.}\ \bibnamefont {Ruckenstein}},\ }\bibfield  {title} {\bibinfo {title} {Phenomenology of the normal state of {Cu}-{O} high-temperature superconductors},\ }\href {https://doi.org/10.1103/PhysRevLett.63.1996} {\bibfield  {journal} {\bibinfo  {journal} {Physical Review Letters}\ }\textbf {\bibinfo {volume} {63}},\ \bibinfo {pages} {1996} (\bibinfo {year} {1989})}\BibitemShut {NoStop}%
\bibitem [{\citenamefont {Varma}\ \emph {et~al.}(2002)\citenamefont {Varma}, \citenamefont {Nussinov},\ and\ \citenamefont {van Saarloos}}]{varma_singular_2002}%
  \BibitemOpen
  \bibfield  {author} {\bibinfo {author} {\bibfnamefont {C.~M.}\ \bibnamefont {Varma}}, \bibinfo {author} {\bibfnamefont {Z.}~\bibnamefont {Nussinov}},\ and\ \bibinfo {author} {\bibfnamefont {W.}~\bibnamefont {van Saarloos}},\ }\bibfield  {title} {{\selectlanguage {english}\bibinfo {title} {Singular or non-{Fermi} liquids}},\ }\href {https://doi.org/10.1016/S0370-1573(01)00060-6} {\bibfield  {journal} {\bibinfo  {journal} {Physics Reports}\ }\textbf {\bibinfo {volume} {361}},\ \bibinfo {pages} {267} (\bibinfo {year} {2002})}\BibitemShut {NoStop}%
\bibitem [{\citenamefont {Zaanen}(2019)}]{zaanen_planckian_2019}%
  \BibitemOpen
  \bibfield  {author} {\bibinfo {author} {\bibfnamefont {J.}~\bibnamefont {Zaanen}},\ }\bibfield  {title} {\bibinfo {title} {Planckian dissipation, minimal viscosity and the transport in cuprate strange metals},\ }\href {https://doi.org/10.21468/SciPostPhys.6.5.061} {\bibfield  {journal} {\bibinfo  {journal} {SciPost Physics}\ ,\ \bibinfo {pages} {061}} (\bibinfo {year} {2019})}\BibitemShut {NoStop}%
\bibitem [{\citenamefont {Varma}(2020)}]{varma_colloquium_2020}%
  \BibitemOpen
  \bibfield  {author} {\bibinfo {author} {\bibfnamefont {C.~M.}\ \bibnamefont {Varma}},\ }\bibfield  {title} {\bibinfo {title} {Colloquium: {Linear} in temperature resistivity and associated mysteries including high temperature superconductivity},\ }\href {https://doi.org/10.1103/RevModPhys.92.031001} {\bibfield  {journal} {\bibinfo  {journal} {Reviews of Modern Physics}\ }\textbf {\bibinfo {volume} {92}},\ \bibinfo {pages} {031001} (\bibinfo {year} {2020})}\BibitemShut {NoStop}%
\bibitem [{\citenamefont {Kirchner}\ \emph {et~al.}(2020{\natexlab{b}})\citenamefont {Kirchner}, \citenamefont {Paschen}, \citenamefont {Chen}, \citenamefont {Wirth}, \citenamefont {Feng}, \citenamefont {Thompson},\ and\ \citenamefont {Si}}]{Si_review_2020}%
  \BibitemOpen
  \bibfield  {author} {\bibinfo {author} {\bibfnamefont {S.}~\bibnamefont {Kirchner}}, \bibinfo {author} {\bibfnamefont {S.}~\bibnamefont {Paschen}}, \bibinfo {author} {\bibfnamefont {Q.}~\bibnamefont {Chen}}, \bibinfo {author} {\bibfnamefont {S.}~\bibnamefont {Wirth}}, \bibinfo {author} {\bibfnamefont {D.}~\bibnamefont {Feng}}, \bibinfo {author} {\bibfnamefont {J.~D.}\ \bibnamefont {Thompson}},\ and\ \bibinfo {author} {\bibfnamefont {Q.}~\bibnamefont {Si}},\ }\bibfield  {title} {\bibinfo {title} {Colloquium: Heavy-electron quantum criticality and single-particle spectroscopy},\ }\href {https://doi.org/10.1103/RevModPhys.92.011002} {\bibfield  {journal} {\bibinfo  {journal} {Rev. Mod. Phys.}\ }\textbf {\bibinfo {volume} {92}},\ \bibinfo {pages} {011002} (\bibinfo {year} {2020}{\natexlab{b}})}\BibitemShut {NoStop}%
\bibitem [{\citenamefont {Chowdhury}\ \emph {et~al.}(2022)\citenamefont {Chowdhury}, \citenamefont {Georges}, \citenamefont {Parcollet},\ and\ \citenamefont {Sachdev}}]{chowdhury_sachdev-ye-kitaev_2022}%
  \BibitemOpen
  \bibfield  {author} {\bibinfo {author} {\bibfnamefont {D.}~\bibnamefont {Chowdhury}}, \bibinfo {author} {\bibfnamefont {A.}~\bibnamefont {Georges}}, \bibinfo {author} {\bibfnamefont {O.}~\bibnamefont {Parcollet}},\ and\ \bibinfo {author} {\bibfnamefont {S.}~\bibnamefont {Sachdev}},\ }\bibfield  {title} {\bibinfo {title} {Sachdev-{Ye}-{Kitaev} models and beyond: {Window} into non-{Fermi} liquids},\ }\href {https://doi.org/10.1103/RevModPhys.94.035004} {\bibfield  {journal} {\bibinfo  {journal} {Reviews of Modern Physics}\ }\textbf {\bibinfo {volume} {94}},\ \bibinfo {pages} {035004} (\bibinfo {year} {2022})}\BibitemShut {NoStop}%
\bibitem [{\citenamefont {Phillips}\ \emph {et~al.}(2022)\citenamefont {Phillips}, \citenamefont {Hussey},\ and\ \citenamefont {Abbamonte}}]{phillips_stranger_2022}%
  \BibitemOpen
  \bibfield  {author} {\bibinfo {author} {\bibfnamefont {P.~W.}\ \bibnamefont {Phillips}}, \bibinfo {author} {\bibfnamefont {N.~E.}\ \bibnamefont {Hussey}},\ and\ \bibinfo {author} {\bibfnamefont {P.}~\bibnamefont {Abbamonte}},\ }\bibfield  {title} {\bibinfo {title} {Stranger than metals},\ }\href {https://doi.org/10.1126/science.abh4273} {\bibfield  {journal} {\bibinfo  {journal} {Science}\ }\textbf {\bibinfo {volume} {377}},\ \bibinfo {pages} {eabh4273} (\bibinfo {year} {2022})}\BibitemShut {NoStop}%
\bibitem [{\citenamefont {Sachdev}(2025)}]{sachdev2025footfancupratephase}%
  \BibitemOpen
  \bibfield  {author} {\bibinfo {author} {\bibfnamefont {S.}~\bibnamefont {Sachdev}},\ }\href {https://arxiv.org/abs/2501.16417} {\bibinfo {title} {The foot, the fan, and the cuprate phase diagram: Fermi-volume-changing quantum phase transitions}} (\bibinfo {year} {2025})\BibitemShut {NoStop}%
\bibitem [{\citenamefont {Eliashberg}(1987)}]{Eliashberg}%
  \BibitemOpen
  \bibfield  {author} {\bibinfo {author} {\bibfnamefont {G.}~\bibnamefont {Eliashberg}},\ }\href@noop {} {}\bibinfo {howpublished} {О возможном механизме сверхпроводимости и линейного по т сопротивления (translated from Russian: On a Possible Mechanism for Superconductivity and Linear Resistance), JETP Letters, 46, Appendix} (\bibinfo {year} {1987})\BibitemShut {NoStop}%
\bibitem [{\citenamefont {Sigrist}\ and\ \citenamefont {Rice}(1995)}]{SigristDwaveReview}%
  \BibitemOpen
  \bibfield  {author} {\bibinfo {author} {\bibfnamefont {M.}~\bibnamefont {Sigrist}}\ and\ \bibinfo {author} {\bibfnamefont {T.~M.}\ \bibnamefont {Rice}},\ }\bibfield  {title} {\bibinfo {title} {Unusual paramagnetic phenomena in granular high-temperature superconductors---a consequence of $d$- wave pairing?},\ }\href {https://doi.org/10.1103/RevModPhys.67.503} {\bibfield  {journal} {\bibinfo  {journal} {Rev. Mod. Phys.}\ }\textbf {\bibinfo {volume} {67}},\ \bibinfo {pages} {503} (\bibinfo {year} {1995})}\BibitemShut {NoStop}%
\bibitem [{\citenamefont {Taraphder}\ and\ \citenamefont {Coleman}(1991)}]{taraphder1991heavy}%
  \BibitemOpen
  \bibfield  {author} {\bibinfo {author} {\bibfnamefont {A.}~\bibnamefont {Taraphder}}\ and\ \bibinfo {author} {\bibfnamefont {P.}~\bibnamefont {Coleman}},\ }\bibfield  {title} {\bibinfo {title} {Heavy-fermion behavior in a negative-u anderson model},\ }\href {https://doi.org/10.1103/PhysRevLett.66.2814} {\bibfield  {journal} {\bibinfo  {journal} {Phys. Rev. Lett.}\ }\textbf {\bibinfo {volume} {66}},\ \bibinfo {pages} {2814} (\bibinfo {year} {1991})}\BibitemShut {NoStop}%
\bibitem [{\citenamefont {Zaikin}(1994)}]{Zaikin1994InfluenceOC}%
  \BibitemOpen
  \bibfield  {author} {\bibinfo {author} {\bibfnamefont {A.~D.}\ \bibnamefont {Zaikin}},\ }\bibfield  {title} {\bibinfo {title} {Influence of coulomb and proximity effects on electron tunneling through normal metal-superconductor interfaces},\ }\href {https://api.semanticscholar.org/CorpusID:121810198} {\bibfield  {journal} {\bibinfo  {journal} {Physica B-condensed Matter}\ }\textbf {\bibinfo {volume} {203}},\ \bibinfo {pages} {255} (\bibinfo {year} {1994})}\BibitemShut {NoStop}%
\bibitem [{\citenamefont {Feigel’man}\ and\ \citenamefont {Larkin}(1998)}]{Feigel_man_1998}%
  \BibitemOpen
  \bibfield  {author} {\bibinfo {author} {\bibfnamefont {M.}~\bibnamefont {Feigel’man}}\ and\ \bibinfo {author} {\bibfnamefont {A.}~\bibnamefont {Larkin}},\ }\bibfield  {title} {\bibinfo {title} {Quantum superconductor–metal transition in a 2d proximity-coupled array},\ }\href {https://doi.org/10.1016/s0301-0104(98)00075-5} {\bibfield  {journal} {\bibinfo  {journal} {Chemical Physics}\ }\textbf {\bibinfo {volume} {235}},\ \bibinfo {pages} {107–114} (\bibinfo {year} {1998})}\BibitemShut {NoStop}%
\bibitem [{\citenamefont {Feigelman}\ \emph {et~al.}(2002)\citenamefont {Feigelman}, \citenamefont {Kamenev}, \citenamefont {Larkin},\ and\ \citenamefont {Skvortsov}}]{feigelman_weak_2002}%
  \BibitemOpen
  \bibfield  {author} {\bibinfo {author} {\bibfnamefont {M.~V.}\ \bibnamefont {Feigelman}}, \bibinfo {author} {\bibfnamefont {A.}~\bibnamefont {Kamenev}}, \bibinfo {author} {\bibfnamefont {A.~I.}\ \bibnamefont {Larkin}},\ and\ \bibinfo {author} {\bibfnamefont {M.~A.}\ \bibnamefont {Skvortsov}},\ }\bibfield  {title} {\bibinfo {title} {Weak charge quantization on a superconducting island},\ }\href {https://doi.org/10.1103/PhysRevB.66.054502} {\bibfield  {journal} {\bibinfo  {journal} {Physical Review B}\ }\textbf {\bibinfo {volume} {66}},\ \bibinfo {pages} {054502} (\bibinfo {year} {2002})}\BibitemShut {NoStop}%
\bibitem [{\citenamefont {Lukyanov}\ and\ \citenamefont {Zamolodchikov}(2004)}]{Lukyanov_2004}%
  \BibitemOpen
  \bibfield  {author} {\bibinfo {author} {\bibfnamefont {S.~L.}\ \bibnamefont {Lukyanov}}\ and\ \bibinfo {author} {\bibfnamefont {A.~B.}\ \bibnamefont {Zamolodchikov}},\ }\bibfield  {title} {\bibinfo {title} {Integrable circular brane model and coulomb charging at large conduction},\ }\href {https://doi.org/10.1088/1742-5468/2004/05/p05003} {\bibfield  {journal} {\bibinfo  {journal} {Journal of Statistical Mechanics: Theory and Experiment}\ }\textbf {\bibinfo {volume} {2004}},\ \bibinfo {pages} {P05003} (\bibinfo {year} {2004})}\BibitemShut {NoStop}%
\bibitem [{\citenamefont {de~Gennes}(1964)}]{degennes}%
  \BibitemOpen
  \bibfield  {author} {\bibinfo {author} {\bibfnamefont {P.~G.}\ \bibnamefont {de~Gennes}},\ }\bibfield  {title} {\bibinfo {title} {Boundary effects in superconductors},\ }\href {https://doi.org/10.1103/RevModPhys.36.225} {\bibfield  {journal} {\bibinfo  {journal} {Rev. Mod. Phys.}\ }\textbf {\bibinfo {volume} {36}},\ \bibinfo {pages} {225} (\bibinfo {year} {1964})}\BibitemShut {NoStop}%
\bibitem [{\citenamefont {Beenakker}(1992)}]{Beenakker1992}%
  \BibitemOpen
  \bibfield  {author} {\bibinfo {author} {\bibfnamefont {C.~W.~J.}\ \bibnamefont {Beenakker}},\ }\bibfield  {title} {\bibinfo {title} {Quantum transport in semiconductor-superconductor microjunctions},\ }\href {https://doi.org/10.1103/PhysRevB.46.12841} {\bibfield  {journal} {\bibinfo  {journal} {Phys. Rev. B}\ }\textbf {\bibinfo {volume} {46}},\ \bibinfo {pages} {12841} (\bibinfo {year} {1992})}\BibitemShut {NoStop}%
\bibitem [{\citenamefont {Skvortsov}\ \emph {et~al.}(2001)\citenamefont {Skvortsov}, \citenamefont {Larkin},\ and\ \citenamefont {Feigel'man}}]{Skvortsov2001}%
  \BibitemOpen
  \bibfield  {author} {\bibinfo {author} {\bibfnamefont {M.~A.}\ \bibnamefont {Skvortsov}}, \bibinfo {author} {\bibfnamefont {A.~I.}\ \bibnamefont {Larkin}},\ and\ \bibinfo {author} {\bibfnamefont {M.~V.}\ \bibnamefont {Feigel'man}},\ }\bibfield  {title} {\bibinfo {title} {Superconductive proximity effect in interacting disordered conductors},\ }\href {https://doi.org/10.1103/PhysRevB.63.134507} {\bibfield  {journal} {\bibinfo  {journal} {Phys. Rev. B}\ }\textbf {\bibinfo {volume} {63}},\ \bibinfo {pages} {134507} (\bibinfo {year} {2001})}\BibitemShut {NoStop}%
\bibitem [{\citenamefont {Dzero}\ and\ \citenamefont {Schmalian}(2005)}]{dzero2005superconductivity}%
  \BibitemOpen
  \bibfield  {author} {\bibinfo {author} {\bibfnamefont {M.}~\bibnamefont {Dzero}}\ and\ \bibinfo {author} {\bibfnamefont {J.}~\bibnamefont {Schmalian}},\ }\bibfield  {title} {\bibinfo {title} {{Superconductivity in Charge Kondo Systems}},\ }\bibfield  {journal} {\bibinfo  {journal} {Physical Review Letters}\ }\textbf {\bibinfo {volume} {94}},\ \href {https://doi.org/10.1103/physrevlett.94.157003} {10.1103/physrevlett.94.157003} (\bibinfo {year} {2005})\BibitemShut {NoStop}%
\bibitem [{\citenamefont {Matsushita}\ \emph {et~al.}(2005)\citenamefont {Matsushita}, \citenamefont {Bluhm}, \citenamefont {Geballe},\ and\ \citenamefont {Fisher}}]{matsushita2005evidence}%
  \BibitemOpen
  \bibfield  {author} {\bibinfo {author} {\bibfnamefont {Y.}~\bibnamefont {Matsushita}}, \bibinfo {author} {\bibfnamefont {H.}~\bibnamefont {Bluhm}}, \bibinfo {author} {\bibfnamefont {T.~H.}\ \bibnamefont {Geballe}},\ and\ \bibinfo {author} {\bibfnamefont {I.~R.}\ \bibnamefont {Fisher}},\ }\bibfield  {title} {\bibinfo {title} {Evidence for charge kondo effect in superconducting {Tl}-doped {PbTe}},\ }\href {https://doi.org/10.1103/PhysRevLett.94.157002} {\bibfield  {journal} {\bibinfo  {journal} {Phys. Rev. Lett.}\ }\textbf {\bibinfo {volume} {94}},\ \bibinfo {pages} {157002} (\bibinfo {year} {2005})}\BibitemShut {NoStop}%
\bibitem [{\citenamefont {Garate}(2011)}]{garate2011charge}%
  \BibitemOpen
  \bibfield  {author} {\bibinfo {author} {\bibfnamefont {I.}~\bibnamefont {Garate}},\ }\bibfield  {title} {\bibinfo {title} {{Charge-Kondo effect in mesoscopic superconductors coupled to normal metals}},\ }\href {https://link.aps.org/doi/10.1103/PhysRevB.84.085121} {\bibfield  {journal} {\bibinfo  {journal} {Physical Review B—Condensed Matter and Materials Physics}\ }\textbf {\bibinfo {volume} {84}},\ \bibinfo {pages} {085121} (\bibinfo {year} {2011})}\BibitemShut {NoStop}%
\bibitem [{\citenamefont {Zaránd}\ \emph {et~al.}(2000)\citenamefont {Zaránd}, \citenamefont {Zimányi},\ and\ \citenamefont {Wilhelm}}]{zarand_two-channel_2000}%
  \BibitemOpen
  \bibfield  {author} {\bibinfo {author} {\bibfnamefont {G.}~\bibnamefont {Zaránd}}, \bibinfo {author} {\bibfnamefont {G.~T.}\ \bibnamefont {Zimányi}},\ and\ \bibinfo {author} {\bibfnamefont {F.}~\bibnamefont {Wilhelm}},\ }\bibfield  {title} {\bibinfo {title} {Two-channel versus infinite-channel {Kondo} models for the single-electron transistor},\ }\href {https://doi.org/10.1103/PhysRevB.62.8137} {\bibfield  {journal} {\bibinfo  {journal} {Physical Review B}\ }\textbf {\bibinfo {volume} {62}},\ \bibinfo {pages} {8137} (\bibinfo {year} {2000})},\ \bibinfo {note} {publisher: American Physical Society}\BibitemShut {NoStop}%
\bibitem [{\citenamefont {Belyansky}\ \emph {et~al.}(2021)\citenamefont {Belyansky}, \citenamefont {Whitsitt}, \citenamefont {Lundgren}, \citenamefont {Wang}, \citenamefont {Vrajitoarea}, \citenamefont {Houck},\ and\ \citenamefont {Gorshkov}}]{belyansky_frustration-induced_2021}%
  \BibitemOpen
  \bibfield  {author} {\bibinfo {author} {\bibfnamefont {R.}~\bibnamefont {Belyansky}}, \bibinfo {author} {\bibfnamefont {S.}~\bibnamefont {Whitsitt}}, \bibinfo {author} {\bibfnamefont {R.}~\bibnamefont {Lundgren}}, \bibinfo {author} {\bibfnamefont {Y.}~\bibnamefont {Wang}}, \bibinfo {author} {\bibfnamefont {A.}~\bibnamefont {Vrajitoarea}}, \bibinfo {author} {\bibfnamefont {A.~A.}\ \bibnamefont {Houck}},\ and\ \bibinfo {author} {\bibfnamefont {A.~V.}\ \bibnamefont {Gorshkov}},\ }\bibfield  {title} {\bibinfo {title} {Frustration-induced anomalous transport and strong photon decay in waveguide {QED}},\ }\href {https://doi.org/10.1103/PhysRevResearch.3.L032058} {\bibfield  {journal} {\bibinfo  {journal} {Physical Review Research}\ }\textbf {\bibinfo {volume} {3}},\ \bibinfo {pages} {L032058} (\bibinfo {year} {2021})}\BibitemShut {NoStop}%
\bibitem [{\citenamefont {Novais}\ \emph {et~al.}(2005)\citenamefont {Novais}, \citenamefont {Castro~Neto}, \citenamefont {Borda}, \citenamefont {Affleck},\ and\ \citenamefont {Zarand}}]{novais_frustration_2005}%
  \BibitemOpen
  \bibfield  {author} {\bibinfo {author} {\bibfnamefont {E.}~\bibnamefont {Novais}}, \bibinfo {author} {\bibfnamefont {A.~H.}\ \bibnamefont {Castro~Neto}}, \bibinfo {author} {\bibfnamefont {L.}~\bibnamefont {Borda}}, \bibinfo {author} {\bibfnamefont {I.}~\bibnamefont {Affleck}},\ and\ \bibinfo {author} {\bibfnamefont {G.}~\bibnamefont {Zarand}},\ }\bibfield  {title} {\bibinfo {title} {Frustration of decoherence in open quantum systems},\ }\href {https://doi.org/10.1103/PhysRevB.72.014417} {\bibfield  {journal} {\bibinfo  {journal} {Physical Review B}\ }\textbf {\bibinfo {volume} {72}},\ \bibinfo {pages} {014417} (\bibinfo {year} {2005})}\BibitemShut {NoStop}%
\bibitem [{\citenamefont {Sheehy}\ and\ \citenamefont {Schmalian}(2007)}]{sheehy_quantum_2007}%
  \BibitemOpen
  \bibfield  {author} {\bibinfo {author} {\bibfnamefont {D.~E.}\ \bibnamefont {Sheehy}}\ and\ \bibinfo {author} {\bibfnamefont {J.}~\bibnamefont {Schmalian}},\ }\bibfield  {title} {\bibinfo {title} {Quantum critical scaling in graphene},\ }\href {https://doi.org/10.1103/PhysRevLett.99.226803} {\bibfield  {journal} {\bibinfo  {journal} {Phys. Rev. Lett.}\ }\textbf {\bibinfo {volume} {99}},\ \bibinfo {pages} {226803} (\bibinfo {year} {2007})}\BibitemShut {NoStop}%
\bibitem [{\citenamefont {Fisher}\ \emph {et~al.}(1989)\citenamefont {Fisher}, \citenamefont {Weichman}, \citenamefont {Grinstein},\ and\ \citenamefont {Fisher}}]{FisherFisher1990}%
  \BibitemOpen
  \bibfield  {author} {\bibinfo {author} {\bibfnamefont {M.~P.~A.}\ \bibnamefont {Fisher}}, \bibinfo {author} {\bibfnamefont {P.~B.}\ \bibnamefont {Weichman}}, \bibinfo {author} {\bibfnamefont {G.}~\bibnamefont {Grinstein}},\ and\ \bibinfo {author} {\bibfnamefont {D.~S.}\ \bibnamefont {Fisher}},\ }\bibfield  {title} {\bibinfo {title} {Boson localization and the superfluid-insulator transition},\ }\href {https://doi.org/10.1103/PhysRevB.40.546} {\bibfield  {journal} {\bibinfo  {journal} {Phys. Rev. B}\ }\textbf {\bibinfo {volume} {40}},\ \bibinfo {pages} {546} (\bibinfo {year} {1989})}\BibitemShut {NoStop}%
\bibitem [{\citenamefont {Hinlopen}\ \emph {et~al.}(2022)\citenamefont {Hinlopen}, \citenamefont {Hinlopen}, \citenamefont {Ayres},\ and\ \citenamefont {Hussey}}]{hinlopen_b2_2022}%
  \BibitemOpen
  \bibfield  {author} {\bibinfo {author} {\bibfnamefont {R.~D.~H.}\ \bibnamefont {Hinlopen}}, \bibinfo {author} {\bibfnamefont {F.~A.}\ \bibnamefont {Hinlopen}}, \bibinfo {author} {\bibfnamefont {J.}~\bibnamefont {Ayres}},\ and\ \bibinfo {author} {\bibfnamefont {N.~E.}\ \bibnamefont {Hussey}},\ }\bibfield  {title} {\bibinfo {title} {{$B^2$ to $B$-linear magnetoresistance due to impeded orbital motion }},\ }\href {https://doi.org/10.1103/PhysRevResearch.4.033195} {\bibfield  {journal} {\bibinfo  {journal} {Physical Review Research}\ }\textbf {\bibinfo {volume} {4}},\ \bibinfo {pages} {033195} (\bibinfo {year} {2022})}\BibitemShut {NoStop}%
\bibitem [{\citenamefont {Kim}\ \emph {et~al.}(2024)\citenamefont {Kim}, \citenamefont {Altman},\ and\ \citenamefont {Chatterjee}}]{kim_linear_2024}%
  \BibitemOpen
  \bibfield  {author} {\bibinfo {author} {\bibfnamefont {J.}~\bibnamefont {Kim}}, \bibinfo {author} {\bibfnamefont {E.}~\bibnamefont {Altman}},\ and\ \bibinfo {author} {\bibfnamefont {S.}~\bibnamefont {Chatterjee}},\ }\bibfield  {title} {\bibinfo {title} {Linear magnetoresistance from glassy orders},\ }\href {https://doi.org/10.1073/pnas.2405720121} {\bibfield  {journal} {\bibinfo  {journal} {Proceedings of the National Academy of Sciences}\ }\textbf {\bibinfo {volume} {121}},\ \bibinfo {pages} {e2405720121} (\bibinfo {year} {2024})}\BibitemShut {NoStop}%
\bibitem [{\citenamefont {Hewson}(1993)}]{hewson_kondo_1993}%
  \BibitemOpen
  \bibfield  {author} {\bibinfo {author} {\bibfnamefont {A.~C.}\ \bibnamefont {Hewson}},\ }\href {https://doi.org/10.1017/CBO9780511470752} {\emph {\bibinfo {title} {The {Kondo} {Problem} to {Heavy} {Fermions}}}},\ Cambridge {Studies} in {Magnetism}\ (\bibinfo  {publisher} {Cambridge University Press},\ \bibinfo {address} {Cambridge},\ \bibinfo {year} {1993})\BibitemShut {NoStop}%
\bibitem [{\citenamefont {Paul}\ and\ \citenamefont {Kotliar}(2001)}]{paul_thermoelectric_2001}%
  \BibitemOpen
  \bibfield  {author} {\bibinfo {author} {\bibfnamefont {I.}~\bibnamefont {Paul}}\ and\ \bibinfo {author} {\bibfnamefont {G.}~\bibnamefont {Kotliar}},\ }\bibfield  {title} {\bibinfo {title} {Thermoelectric behavior near the magnetic quantum critical point},\ }\href {https://doi.org/10.1103/PhysRevB.64.184414} {\bibfield  {journal} {\bibinfo  {journal} {Physical Review B}\ }\textbf {\bibinfo {volume} {64}},\ \bibinfo {pages} {184414} (\bibinfo {year} {2001})}\BibitemShut {NoStop}%
\bibitem [{\citenamefont {Georges}\ and\ \citenamefont {Mravlje}(2021)}]{georges2021skewed}%
  \BibitemOpen
  \bibfield  {author} {\bibinfo {author} {\bibfnamefont {A.}~\bibnamefont {Georges}}\ and\ \bibinfo {author} {\bibfnamefont {J.}~\bibnamefont {Mravlje}},\ }\bibfield  {title} {\bibinfo {title} {Skewed non-fermi liquids and the seebeck effect},\ }\href {https://doi.org/10.1103/PhysRevResearch.3.043132} {\bibfield  {journal} {\bibinfo  {journal} {Phys. Rev. Res.}\ }\textbf {\bibinfo {volume} {3}},\ \bibinfo {pages} {043132} (\bibinfo {year} {2021})}\BibitemShut {NoStop}%
\bibitem [{\citenamefont {Maebashi}\ and\ \citenamefont {Varma}(2022)}]{maebashi2022quantum}%
  \BibitemOpen
  \bibfield  {author} {\bibinfo {author} {\bibfnamefont {H.}~\bibnamefont {Maebashi}}\ and\ \bibinfo {author} {\bibfnamefont {C.~M.}\ \bibnamefont {Varma}},\ }\href {https://arxiv.org/abs/2207.11982} {\bibinfo {title} {Quantum-critical transport of marginal fermi-liquids}} (\bibinfo {year} {2022})\BibitemShut {NoStop}%
\bibitem [{\citenamefont {Martin}\ \emph {et~al.}(1990)\citenamefont {Martin}, \citenamefont {Fiory}, \citenamefont {Fleming}, \citenamefont {Schneemeyer},\ and\ \citenamefont {Waszczak}}]{Martin1990}%
  \BibitemOpen
  \bibfield  {author} {\bibinfo {author} {\bibfnamefont {S.}~\bibnamefont {Martin}}, \bibinfo {author} {\bibfnamefont {A.~T.}\ \bibnamefont {Fiory}}, \bibinfo {author} {\bibfnamefont {R.~M.}\ \bibnamefont {Fleming}}, \bibinfo {author} {\bibfnamefont {L.~F.}\ \bibnamefont {Schneemeyer}},\ and\ \bibinfo {author} {\bibfnamefont {J.~V.}\ \bibnamefont {Waszczak}},\ }\bibfield  {title} {\bibinfo {title} {Normal-state transport properties of $\textrm{Bi}_{2+x}\textrm{Sr}_{2-y}\textrm{CuO}_{6+\delta}$ crystals},\ }\href {https://doi.org/10.1103/PhysRevB.41.846} {\bibfield  {journal} {\bibinfo  {journal} {Phys. Rev. B}\ }\textbf {\bibinfo {volume} {41}},\ \bibinfo {pages} {846} (\bibinfo {year} {1990})}\BibitemShut {NoStop}%
\bibitem [{\citenamefont {Takagi}\ \emph {et~al.}(1992)\citenamefont {Takagi}, \citenamefont {Batlogg}, \citenamefont {Kao}, \citenamefont {Kwo}, \citenamefont {Cava}, \citenamefont {Krajewski},\ and\ \citenamefont {Peck}}]{Takagi1992}%
  \BibitemOpen
  \bibfield  {author} {\bibinfo {author} {\bibfnamefont {H.}~\bibnamefont {Takagi}}, \bibinfo {author} {\bibfnamefont {B.}~\bibnamefont {Batlogg}}, \bibinfo {author} {\bibfnamefont {H.~L.}\ \bibnamefont {Kao}}, \bibinfo {author} {\bibfnamefont {J.}~\bibnamefont {Kwo}}, \bibinfo {author} {\bibfnamefont {R.~J.}\ \bibnamefont {Cava}}, \bibinfo {author} {\bibfnamefont {J.~J.}\ \bibnamefont {Krajewski}},\ and\ \bibinfo {author} {\bibfnamefont {W.~F.}\ \bibnamefont {Peck}},\ }\bibfield  {title} {\bibinfo {title} {Systematic evolution of temperature-dependent resistivity in $\textrm{La}_{2-x}\textrm{Sr}_x\textrm{CuO}_4$},\ }\href {https://doi.org/10.1103/PhysRevLett.69.2975} {\bibfield  {journal} {\bibinfo  {journal} {Phys. Rev. Lett.}\ }\textbf {\bibinfo {volume} {69}},\ \bibinfo {pages} {2975} (\bibinfo {year} {1992})}\BibitemShut {NoStop}%
\bibitem [{\citenamefont {Giraldo-Gallo}\ \emph {et~al.}(2018{\natexlab{b}})\citenamefont {Giraldo-Gallo}, \citenamefont {Galvis}, \citenamefont {Stegen}, \citenamefont {Modic}, \citenamefont {Balakirev}, \citenamefont {Betts}, \citenamefont {Lian}, \citenamefont {Moir}, \citenamefont {Riggs}, \citenamefont {Wu}, \citenamefont {Bollinger}, \citenamefont {He}, \citenamefont {Božović}, \citenamefont {Ramshaw}, \citenamefont {McDonald}, \citenamefont {Boebinger},\ and\ \citenamefont {Shekhter}}]{Boebinger2018}%
  \BibitemOpen
  \bibfield  {author} {\bibinfo {author} {\bibfnamefont {P.}~\bibnamefont {Giraldo-Gallo}}, \bibinfo {author} {\bibfnamefont {J.~A.}\ \bibnamefont {Galvis}}, \bibinfo {author} {\bibfnamefont {Z.}~\bibnamefont {Stegen}}, \bibinfo {author} {\bibfnamefont {K.~A.}\ \bibnamefont {Modic}}, \bibinfo {author} {\bibfnamefont {F.~F.}\ \bibnamefont {Balakirev}}, \bibinfo {author} {\bibfnamefont {J.~B.}\ \bibnamefont {Betts}}, \bibinfo {author} {\bibfnamefont {X.}~\bibnamefont {Lian}}, \bibinfo {author} {\bibfnamefont {C.}~\bibnamefont {Moir}}, \bibinfo {author} {\bibfnamefont {S.~C.}\ \bibnamefont {Riggs}}, \bibinfo {author} {\bibfnamefont {J.}~\bibnamefont {Wu}}, \bibinfo {author} {\bibfnamefont {A.~T.}\ \bibnamefont {Bollinger}}, \bibinfo {author} {\bibfnamefont {X.}~\bibnamefont {He}}, \bibinfo {author} {\bibfnamefont {I.}~\bibnamefont {Božović}}, \bibinfo {author} {\bibfnamefont {B.~J.}\ \bibnamefont {Ramshaw}}, \bibinfo {author} {\bibfnamefont {R.~D.}\ \bibnamefont {McDonald}}, \bibinfo {author} {\bibfnamefont
  {G.~S.}\ \bibnamefont {Boebinger}},\ and\ \bibinfo {author} {\bibfnamefont {A.}~\bibnamefont {Shekhter}},\ }\bibfield  {title} {\bibinfo {title} {Scale-invariant magnetoresistance in a cuprate superconductor},\ }\href {https://doi.org/10.1126/science.aan3178} {\bibfield  {journal} {\bibinfo  {journal} {Science}\ }\textbf {\bibinfo {volume} {361}},\ \bibinfo {pages} {479} (\bibinfo {year} {2018}{\natexlab{b}})},\ \Eprint {https://arxiv.org/abs/https://www.science.org/doi/pdf/10.1126/science.aan3178} {https://www.science.org/doi/pdf/10.1126/science.aan3178} \BibitemShut {NoStop}%
\bibitem [{\citenamefont {Wang}\ \emph {et~al.}(2006)\citenamefont {Wang}, \citenamefont {Li},\ and\ \citenamefont {Ong}}]{wang2006nernst}%
  \BibitemOpen
  \bibfield  {author} {\bibinfo {author} {\bibfnamefont {Y.}~\bibnamefont {Wang}}, \bibinfo {author} {\bibfnamefont {L.}~\bibnamefont {Li}},\ and\ \bibinfo {author} {\bibfnamefont {N.~P.}\ \bibnamefont {Ong}},\ }\bibfield  {title} {\bibinfo {title} {Nernst effect in high-${T}_{c}$ superconductors},\ }\href {https://doi.org/10.1103/PhysRevB.73.024510} {\bibfield  {journal} {\bibinfo  {journal} {Phys. Rev. B}\ }\textbf {\bibinfo {volume} {73}},\ \bibinfo {pages} {024510} (\bibinfo {year} {2006})}\BibitemShut {NoStop}%
\bibitem [{\citenamefont {Li}\ \emph {et~al.}(2010)\citenamefont {Li}, \citenamefont {Wang}, \citenamefont {Komiya}, \citenamefont {Ono}, \citenamefont {Ando}, \citenamefont {Gu},\ and\ \citenamefont {Ong}}]{li2010diamagnetism}%
  \BibitemOpen
  \bibfield  {author} {\bibinfo {author} {\bibfnamefont {L.}~\bibnamefont {Li}}, \bibinfo {author} {\bibfnamefont {Y.}~\bibnamefont {Wang}}, \bibinfo {author} {\bibfnamefont {S.}~\bibnamefont {Komiya}}, \bibinfo {author} {\bibfnamefont {S.}~\bibnamefont {Ono}}, \bibinfo {author} {\bibfnamefont {Y.}~\bibnamefont {Ando}}, \bibinfo {author} {\bibfnamefont {G.~D.}\ \bibnamefont {Gu}},\ and\ \bibinfo {author} {\bibfnamefont {N.~P.}\ \bibnamefont {Ong}},\ }\bibfield  {title} {\bibinfo {title} {Diamagnetism and cooper pairing above ${T}_{c}$ in cuprates},\ }\href {https://doi.org/10.1103/PhysRevB.81.054510} {\bibfield  {journal} {\bibinfo  {journal} {Phys. Rev. B}\ }\textbf {\bibinfo {volume} {81}},\ \bibinfo {pages} {054510} (\bibinfo {year} {2010})}\BibitemShut {NoStop}%
\bibitem [{\citenamefont {Hsu}\ \emph {et~al.}(2021)\citenamefont {Hsu}, \citenamefont {Hartstein}, \citenamefont {Davies}, \citenamefont {Hickey}, \citenamefont {Chan}, \citenamefont {Porras}, \citenamefont {Loew}, \citenamefont {Taylor}, \citenamefont {Liu}, \citenamefont {Eaton} \emph {et~al.}}]{hsu2021unconventional}%
  \BibitemOpen
  \bibfield  {author} {\bibinfo {author} {\bibfnamefont {Y.-T.}\ \bibnamefont {Hsu}}, \bibinfo {author} {\bibfnamefont {M.}~\bibnamefont {Hartstein}}, \bibinfo {author} {\bibfnamefont {A.~J.}\ \bibnamefont {Davies}}, \bibinfo {author} {\bibfnamefont {A.~J.}\ \bibnamefont {Hickey}}, \bibinfo {author} {\bibfnamefont {M.~K.}\ \bibnamefont {Chan}}, \bibinfo {author} {\bibfnamefont {J.}~\bibnamefont {Porras}}, \bibinfo {author} {\bibfnamefont {T.}~\bibnamefont {Loew}}, \bibinfo {author} {\bibfnamefont {S.~V.}\ \bibnamefont {Taylor}}, \bibinfo {author} {\bibfnamefont {H.}~\bibnamefont {Liu}}, \bibinfo {author} {\bibfnamefont {A.~G.}\ \bibnamefont {Eaton}}, \emph {et~al.},\ }\bibfield  {title} {\bibinfo {title} {Unconventional quantum vortex matter state hosts quantum oscillations in the underdoped high-temperature cuprate superconductors},\ }\bibfield  {journal} {\bibinfo  {journal} {Proceedings of the National Academy of Sciences}\ }\href {https://doi.org/10.1073/pnas.2021216118} {10.1073/pnas.2021216118}
  (\bibinfo {year} {2021})\BibitemShut {NoStop}%
\bibitem [{\citenamefont {Michon}\ \emph {et~al.}(2019)\citenamefont {Michon}, \citenamefont {Girod}, \citenamefont {Badoux}, \citenamefont {Ka{\v{c}}mar{\v{c}}{\'\i}k}, \citenamefont {Ma}, \citenamefont {Dragomir}, \citenamefont {Dabkowska}, \citenamefont {Gaulin}, \citenamefont {Zhou}, \citenamefont {Pyon} \emph {et~al.}}]{Michon2019}%
  \BibitemOpen
  \bibfield  {author} {\bibinfo {author} {\bibfnamefont {B.}~\bibnamefont {Michon}}, \bibinfo {author} {\bibfnamefont {C.}~\bibnamefont {Girod}}, \bibinfo {author} {\bibfnamefont {S.}~\bibnamefont {Badoux}}, \bibinfo {author} {\bibfnamefont {J.}~\bibnamefont {Ka{\v{c}}mar{\v{c}}{\'\i}k}}, \bibinfo {author} {\bibfnamefont {Q.}~\bibnamefont {Ma}}, \bibinfo {author} {\bibfnamefont {M.}~\bibnamefont {Dragomir}}, \bibinfo {author} {\bibfnamefont {H.}~\bibnamefont {Dabkowska}}, \bibinfo {author} {\bibfnamefont {B.}~\bibnamefont {Gaulin}}, \bibinfo {author} {\bibfnamefont {J.-S.}\ \bibnamefont {Zhou}}, \bibinfo {author} {\bibfnamefont {S.}~\bibnamefont {Pyon}}, \emph {et~al.},\ }\bibfield  {title} {\bibinfo {title} {Thermodynamic signatures of quantum criticality in cuprate superconductors},\ }\href {https://www.nature.com/articles/s41586-019-0932-x} {\bibfield  {journal} {\bibinfo  {journal} {Nature}\ }\textbf {\bibinfo {volume} {567}},\ \bibinfo {pages} {218} (\bibinfo {year} {2019})}\BibitemShut {NoStop}%
\bibitem [{\citenamefont {Gourgout}\ \emph {et~al.}(2022)\citenamefont {Gourgout}, \citenamefont {Grissonnanche}, \citenamefont {Lalibert{\'e}}, \citenamefont {Ataei}, \citenamefont {Chen}, \citenamefont {Verret}, \citenamefont {Zhou}, \citenamefont {Mravlje}, \citenamefont {Georges}, \citenamefont {Doiron-Leyraud} \emph {et~al.}}]{gourgout2022seebeck}%
  \BibitemOpen
  \bibfield  {author} {\bibinfo {author} {\bibfnamefont {A.}~\bibnamefont {Gourgout}}, \bibinfo {author} {\bibfnamefont {G.}~\bibnamefont {Grissonnanche}}, \bibinfo {author} {\bibfnamefont {F.}~\bibnamefont {Lalibert{\'e}}}, \bibinfo {author} {\bibfnamefont {A.}~\bibnamefont {Ataei}}, \bibinfo {author} {\bibfnamefont {L.}~\bibnamefont {Chen}}, \bibinfo {author} {\bibfnamefont {S.}~\bibnamefont {Verret}}, \bibinfo {author} {\bibfnamefont {J.-S.}\ \bibnamefont {Zhou}}, \bibinfo {author} {\bibfnamefont {J.}~\bibnamefont {Mravlje}}, \bibinfo {author} {\bibfnamefont {A.}~\bibnamefont {Georges}}, \bibinfo {author} {\bibfnamefont {N.}~\bibnamefont {Doiron-Leyraud}}, \emph {et~al.},\ }\bibfield  {title} {\bibinfo {title} {Seebeck coefficient in a cuprate superconductor: particle-hole asymmetry in the strange metal phase and fermi surface transformation in the pseudogap phase},\ }\href {https://journals.aps.org/prx/abstract/10.1103/PhysRevX.12.011037} {\bibfield  {journal} {\bibinfo  {journal} {Physical Review X}\
  }\textbf {\bibinfo {volume} {12}},\ \bibinfo {pages} {011037} (\bibinfo {year} {2022})}\BibitemShut {NoStop}%
\bibitem [{\citenamefont {Schmalian}\ and\ \citenamefont {Wolynes}(2000)}]{schmalian_stripe_2000}%
  \BibitemOpen
  \bibfield  {author} {\bibinfo {author} {\bibfnamefont {J.}~\bibnamefont {Schmalian}}\ and\ \bibinfo {author} {\bibfnamefont {P.~G.}\ \bibnamefont {Wolynes}},\ }\bibfield  {title} {\bibinfo {title} {Stripe {Glasses}: {Self}-{Generated} {Randomness} in a {Uniformly} {Frustrated} {System}},\ }\href {https://doi.org/10.1103/PhysRevLett.85.836} {\bibfield  {journal} {\bibinfo  {journal} {Physical Review Letters}\ }\textbf {\bibinfo {volume} {85}},\ \bibinfo {pages} {836} (\bibinfo {year} {2000})}\BibitemShut {NoStop}%
\bibitem [{\citenamefont {Jaeger}\ \emph {et~al.}(1989)\citenamefont {Jaeger}, \citenamefont {Haviland}, \citenamefont {Orr},\ and\ \citenamefont {Goldman}}]{Jaeger1989}%
  \BibitemOpen
  \bibfield  {author} {\bibinfo {author} {\bibfnamefont {H.~M.}\ \bibnamefont {Jaeger}}, \bibinfo {author} {\bibfnamefont {D.~B.}\ \bibnamefont {Haviland}}, \bibinfo {author} {\bibfnamefont {B.~G.}\ \bibnamefont {Orr}},\ and\ \bibinfo {author} {\bibfnamefont {A.~M.}\ \bibnamefont {Goldman}},\ }\bibfield  {title} {\bibinfo {title} {Onset of superconductivity in ultrathin granular metal films},\ }\href {https://doi.org/10.1103/PhysRevB.40.182} {\bibfield  {journal} {\bibinfo  {journal} {Phys. Rev. B}\ }\textbf {\bibinfo {volume} {40}},\ \bibinfo {pages} {182} (\bibinfo {year} {1989})}\BibitemShut {NoStop}%
\bibitem [{\citenamefont {Chakravarty}\ \emph {et~al.}(1986)\citenamefont {Chakravarty}, \citenamefont {Ingold}, \citenamefont {Kivelson},\ and\ \citenamefont {Luther}}]{chakluther}%
  \BibitemOpen
  \bibfield  {author} {\bibinfo {author} {\bibfnamefont {S.}~\bibnamefont {Chakravarty}}, \bibinfo {author} {\bibfnamefont {G.-L.}\ \bibnamefont {Ingold}}, \bibinfo {author} {\bibfnamefont {S.}~\bibnamefont {Kivelson}},\ and\ \bibinfo {author} {\bibfnamefont {A.}~\bibnamefont {Luther}},\ }\bibfield  {title} {\bibinfo {title} {Onset of global phase coherence in josephson-junction arrays: A dissipative phase transition},\ }\href {https://doi.org/10.1103/PhysRevLett.56.2303} {\bibfield  {journal} {\bibinfo  {journal} {Phys. Rev. Lett.}\ }\textbf {\bibinfo {volume} {56}},\ \bibinfo {pages} {2303} (\bibinfo {year} {1986})}\BibitemShut {NoStop}%
\bibitem [{\citenamefont {Chakravarty}\ \emph {et~al.}(1988)\citenamefont {Chakravarty}, \citenamefont {Ingold}, \citenamefont {Kivelson},\ and\ \citenamefont {Zimanyi}}]{chakravartyandZimanyi}%
  \BibitemOpen
  \bibfield  {author} {\bibinfo {author} {\bibfnamefont {S.}~\bibnamefont {Chakravarty}}, \bibinfo {author} {\bibfnamefont {G.-L.}\ \bibnamefont {Ingold}}, \bibinfo {author} {\bibfnamefont {S.}~\bibnamefont {Kivelson}},\ and\ \bibinfo {author} {\bibfnamefont {G.}~\bibnamefont {Zimanyi}},\ }\bibfield  {title} {\bibinfo {title} {Quantum statistical mechanics of an array of resistively shunted josephson junctions},\ }\href {https://doi.org/10.1103/PhysRevB.37.3283} {\bibfield  {journal} {\bibinfo  {journal} {Phys. Rev. B}\ }\textbf {\bibinfo {volume} {37}},\ \bibinfo {pages} {3283} (\bibinfo {year} {1988})}\BibitemShut {NoStop}%
\bibitem [{\citenamefont {Fisher}(1987)}]{matthewfisher}%
  \BibitemOpen
  \bibfield  {author} {\bibinfo {author} {\bibfnamefont {M.~P.~A.}\ \bibnamefont {Fisher}},\ }\bibfield  {title} {\bibinfo {title} {Dissipation and quantum fluctuations in granular superconductivity},\ }\href {https://doi.org/10.1103/PhysRevB.36.1917} {\bibfield  {journal} {\bibinfo  {journal} {Phys. Rev. B}\ }\textbf {\bibinfo {volume} {36}},\ \bibinfo {pages} {1917} (\bibinfo {year} {1987})}\BibitemShut {NoStop}%
\bibitem [{\citenamefont {Sachdev}\ and\ \citenamefont {Vojta}(2000)}]{voigtaandsachdev}%
  \BibitemOpen
  \bibfield  {author} {\bibinfo {author} {\bibfnamefont {S.}~\bibnamefont {Sachdev}}\ and\ \bibinfo {author} {\bibfnamefont {M.}~\bibnamefont {Vojta}},\ }\bibfield  {title} {\bibinfo {title} {Quantum phase transitions in antiferromagnets and superfluids},\ }\href {https://doi.org/10.1016/s0921-4526(99)01726-3} {\bibfield  {journal} {\bibinfo  {journal} {Physica B: Condensed Matter}\ }\textbf {\bibinfo {volume} {280}},\ \bibinfo {pages} {333–340} (\bibinfo {year} {2000})}\BibitemShut {NoStop}%
\bibitem [{\citenamefont {Caldeira}\ and\ \citenamefont {Leggett}(1981)}]{caldeira_influence_1981}%
  \BibitemOpen
  \bibfield  {author} {\bibinfo {author} {\bibfnamefont {A.~O.}\ \bibnamefont {Caldeira}}\ and\ \bibinfo {author} {\bibfnamefont {A.~J.}\ \bibnamefont {Leggett}},\ }\bibfield  {title} {\bibinfo {title} {Influence of {Dissipation} on {Quantum} {Tunneling} in {Macroscopic} {Systems}},\ }\href {https://doi.org/10.1103/PhysRevLett.46.211} {\bibfield  {journal} {\bibinfo  {journal} {Physical Review Letters}\ }\textbf {\bibinfo {volume} {46}},\ \bibinfo {pages} {211} (\bibinfo {year} {1981})}\BibitemShut {NoStop}%
\bibitem [{\citenamefont {Tranquada}\ \emph {et~al.}(2024)\citenamefont {Tranquada}, \citenamefont {Lozano}, \citenamefont {Yao}, \citenamefont {Gu},\ and\ \citenamefont {Li}}]{tranquada_first_order_2024}%
  \BibitemOpen
  \bibfield  {author} {\bibinfo {author} {\bibfnamefont {J.~M.}\ \bibnamefont {Tranquada}}, \bibinfo {author} {\bibfnamefont {P.~M.}\ \bibnamefont {Lozano}}, \bibinfo {author} {\bibfnamefont {J.}~\bibnamefont {Yao}}, \bibinfo {author} {\bibfnamefont {G.~D.}\ \bibnamefont {Gu}},\ and\ \bibinfo {author} {\bibfnamefont {Q.}~\bibnamefont {Li}},\ }\bibfield  {title} {\bibinfo {title} {From nonmetal to strange metal at the stripe-percolation transition in $\textrm{La}_{2-x}\textrm{Sr}_x\textrm{CuO}_4$},\ }\href {https://doi.org/10.1103/PhysRevB.109.184510} {\bibfield  {journal} {\bibinfo  {journal} {Phys. Rev. B}\ }\textbf {\bibinfo {volume} {109}},\ \bibinfo {pages} {184510} (\bibinfo {year} {2024})}\BibitemShut {NoStop}%
\bibitem [{\citenamefont {Zaránd}(1995)}]{zarand_low-temperature_1995}%
  \BibitemOpen
  \bibfield  {author} {\bibinfo {author} {\bibfnamefont {G.}~\bibnamefont {Zaránd}},\ }\bibfield  {title} {\bibinfo {title} {Low-temperature behavior of a generalized two-level system: {Exact} results in the large-flavor-number limit},\ }\href {https://doi.org/10.1103/PhysRevB.51.273} {\bibfield  {journal} {\bibinfo  {journal} {Physical Review B}\ }\textbf {\bibinfo {volume} {51}},\ \bibinfo {pages} {273} (\bibinfo {year} {1995})}\BibitemShut {NoStop}%
\bibitem [{\citenamefont {von Delft}\ \emph {et~al.}(1999)\citenamefont {von Delft}, \citenamefont {Ludwig},\ and\ \citenamefont {Ambegaokar}}]{von_delft_2-channel_1999}%
  \BibitemOpen
  \bibfield  {author} {\bibinfo {author} {\bibfnamefont {J.}~\bibnamefont {von Delft}}, \bibinfo {author} {\bibfnamefont {A.~W.~W.}\ \bibnamefont {Ludwig}},\ and\ \bibinfo {author} {\bibfnamefont {V.}~\bibnamefont {Ambegaokar}},\ }\bibfield  {title} {\bibinfo {title} {The 2-{Channel} {Kondo} {Model}: {II}. {CFT} {Calculation} of {Non}-equilibrium {Conductance} through a {Nanoconstriction} {Containing} 2-{Channel} {Kondo} {Impurities}},\ }\href {https://doi.org/10.1006/aphy.1998.5897} {\bibfield  {journal} {\bibinfo  {journal} {Annals of Physics}\ }\textbf {\bibinfo {volume} {273}},\ \bibinfo {pages} {175} (\bibinfo {year} {1999})}\BibitemShut {NoStop}%
\bibitem [{\citenamefont {Castro~Neto}\ \emph {et~al.}(2003)\citenamefont {Castro~Neto}, \citenamefont {Novais}, \citenamefont {Borda}, \citenamefont {Zaránd},\ and\ \citenamefont {Affleck}}]{castro_neto_quantum_2003}%
  \BibitemOpen
  \bibfield  {author} {\bibinfo {author} {\bibfnamefont {A.~H.}\ \bibnamefont {Castro~Neto}}, \bibinfo {author} {\bibfnamefont {E.}~\bibnamefont {Novais}}, \bibinfo {author} {\bibfnamefont {L.}~\bibnamefont {Borda}}, \bibinfo {author} {\bibfnamefont {G.}~\bibnamefont {Zaránd}},\ and\ \bibinfo {author} {\bibfnamefont {I.}~\bibnamefont {Affleck}},\ }\bibfield  {title} {\bibinfo {title} {Quantum {Magnetic} {Impurities} in {Magnetically} {Ordered} {Systems}},\ }\href {https://doi.org/10.1103/PhysRevLett.91.096401} {\bibfield  {journal} {\bibinfo  {journal} {Physical Review Letters}\ }\textbf {\bibinfo {volume} {91}},\ \bibinfo {pages} {096401} (\bibinfo {year} {2003})}\BibitemShut {NoStop}%
\bibitem [{\citenamefont {Tulipman}\ and\ \citenamefont {Berg}(2023)}]{tulipman_criterion_2022}%
  \BibitemOpen
  \bibfield  {author} {\bibinfo {author} {\bibfnamefont {E.}~\bibnamefont {Tulipman}}\ and\ \bibinfo {author} {\bibfnamefont {E.}~\bibnamefont {Berg}},\ }\bibfield  {title} {{\selectlanguage {english}\bibinfo {title} {A criterion for strange metallicity in the {Lorenz} ratio}},\ }\href {https://doi.org/10.1038/s41535-023-00598-z} {\bibfield  {journal} {\bibinfo  {journal} {npj Quantum Materials}\ }\textbf {\bibinfo {volume} {8}},\ \bibinfo {pages} {1} (\bibinfo {year} {2023})}\BibitemShut {NoStop}%
\bibitem [{\citenamefont {Mahan}(2000)}]{mahan_many-particle_2000}%
  \BibitemOpen
  \bibfield  {author} {\bibinfo {author} {\bibfnamefont {G.~D.}\ \bibnamefont {Mahan}},\ }\href {https://doi.org/10.1007/978-1-4757-5714-9} {\emph {\bibinfo {title} {Many-Particle Physics}}},\ \bibinfo {edition} {3rd}\ ed.,\ Physics of Solids and Liquids\ (\bibinfo  {publisher} {Springer {US}},\ \bibinfo {year} {2000})\BibitemShut {NoStop}%
\bibitem [{\citenamefont {Nave}\ and\ \citenamefont {Lee}(2007)}]{nave_transport_2007}%
  \BibitemOpen
  \bibfield  {author} {\bibinfo {author} {\bibfnamefont {C.~P.}\ \bibnamefont {Nave}}\ and\ \bibinfo {author} {\bibfnamefont {P.~A.}\ \bibnamefont {Lee}},\ }\bibfield  {title} {\bibinfo {title} {Transport properties of a spinon {Fermi} surface coupled to a {U}(1) gauge field},\ }\href {https://doi.org/10.1103/PhysRevB.76.235124} {\bibfield  {journal} {\bibinfo  {journal} {Physical Review B}\ }\textbf {\bibinfo {volume} {76}},\ \bibinfo {pages} {235124} (\bibinfo {year} {2007})}\BibitemShut {NoStop}%
\bibitem [{\citenamefont {Thill}\ and\ \citenamefont {Huse}(1995)}]{Thill1995}%
  \BibitemOpen
  \bibfield  {author} {\bibinfo {author} {\bibfnamefont {M.}~\bibnamefont {Thill}}\ and\ \bibinfo {author} {\bibfnamefont {D.}~\bibnamefont {Huse}},\ }\bibfield  {title} {\bibinfo {title} {Equilibrium behaviour of quantum ising spin glass},\ }\href {https://www.sciencedirect.com/science/article/pii/037843719400247Q} {\bibfield  {journal} {\bibinfo  {journal} {Physica A: Statistical Mechanics and its Applications}\ }\textbf {\bibinfo {volume} {214}},\ \bibinfo {pages} {321} (\bibinfo {year} {1995})}\BibitemShut {NoStop}%
\end{thebibliography}%

\onecolumngrid

\appendix

\section{Relation to the theory of anomalous metals}
\label{appendix:anomalous_metals}

The nature and consequences of the superconducting quantum fluctuations
associated with small superconducting puddles in a metallic host 
has
been a topic of extended interest  \cite{Jaeger1989,chakluther,chakravartyandZimanyi,matthewfisher,voigtaandsachdev,feigelman_weak_2002,Spivak2001,Spivak2008}. Much of this analysis was carried out in the context of the theoretical
puzzles raised by the experimental observations
of an ``anomalous metal''~\cite{Kapitulnik_colloquium_2019} proximate to a superconducting quantum
critical point, primarily in systems whose constituents are relatively
conventional metals and superconductors. In common with our analysis,
these studies are distinct from studies of purely bosonic models that
have also been studied in similar contexts, in the sense that the
long-time dynamics of each individual puddle is ultimately controlled
by some sort of dissipative coupling to the surrounding fermionic
(metallic) heat bath, often referred to as Caldeira-Leggett dynamics~\cite{caldeira_influence_1981}.

We now discuss the main differences in the underlying assumptions
concerning the microscopic physics of the problems studied
previously and the one we consider in this work:
\begin{enumerate}
\item \textit{The existence of a well-defined charging energy:} In conventional superconductors,
the condition that the SC puddles must have a radius that is larger
or of order the coherence length  $\xi_0$ implies that the number of electrons and hence,
the number of Cooper pairs fluctuates rapidly. Then the quantization
of the charge on the puddle, the charging energy, and the displacement (background)
charge, 
both of which play a key role in our analysis, are either not important
or not well-defined. 

This is different in the present theory, in which we assume the
existence of a well-defined charging energy, which imposes an effective
constraint on the (integer) number of Cooper pairs on a puddle. Specifically, in contrast to cases where significant coupling between the SC and metal exponentially reduces the charging energy, here, we assume that the renomalization of the charging energy in the interacting problem is not too strong. This is possible when the Andreev conductance of from a generic puddle to the surrounding metal (which in our analysis is proportional to $\alpha_\perp$) is not too large. This condition 
is plausible in materials, such as the cuprates, where
$k_{F}\xi_0$ is not too large. Additionally, strong correlation effects
could plausibly suppress the effective transmission amplitudes of
electrons between different patches of material, even without invoking
an artificial insulating barrier surrounding the superconducting puddles~\cite{tranquada_first_order_2024}.

\item \textit{The importance of Josephson coupling between puddles:} In
the standard (classical) theory of fluctuation conductivity, there
are two distinct leading-order effects: Aslamasov-Larkin (AL) terms
contribute an additive contribution to the conductivity and reflect
the fact that current can be carried without dissipation over distances
of order the superconducting correlation length by the superconducting
fluctuations themselves. In contrast, Maki-Thompson (MT) terms give
an additive contribution to the resistivity and reflect the scattering
of current-carrying quasi-particles from the superconducting fluctuations.
For the most part, studies motivated by the observation of anomalous
metals focused on quantum terms analogous to the AL terms. Given
that the resistivity in the anomalous metal regime is always found
to be much smaller than the normal state resistivity (sometimes by
as much as four orders of magnitude), it is apparent that the most
significant effect of superconducting fluctuations is that they provide
an efficient, parallel mechanism for charge transport. Moreover, since
the SC state is a perfect conductor, it is reasonable to expect the
conductivity to diverge on close approach to a metal to superconductor
QCP. Thus, studies motivated by the anomalous metal notionally focused
on the way coherence between neighboring puddles grows with decreasing
$T$ or $B$, or upon increasing the concentration of puddles.

By contrast, what we have focused on are single puddle effects, and
have explicitly assumed that correlations between neighboring puddles
are negligible (or irrelevant, in the RG sense, in the metallic phase). We thus focus on MT-type effects, in which the (Andreev)
scattering of quasi-particles of the metallic host off of the superconducting
fluctuations on a single puddle make an additive contribution to the
resistivity. Furthermore, the existence of a well-defined charging energy implies that the superconducting susceptibility of a puddle is not exponentially large (as $\chi \sim 1/E_c$). Hence, while a sufficiently large Josephson coupling will eventually result in a transition to a SC phase, the assumption that the charging energy is well-defined also implies that this critical value of the Josephson coupling is not exponentially small. That is, the condition for mean-field order $\chi_i \sum_i J_{ij} \sim 1$
allows for a wider metallic regime for a given density of puddles.


\end{enumerate}

\section{Comparison of different microscopic realizations of puddles}

\label{appendix: puddles}

 \begin{figure}[b]
\centering
\includegraphics[width=0.6\textwidth]{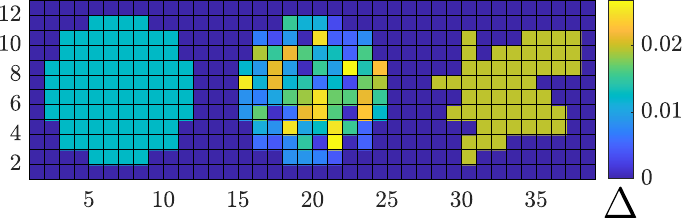} 
\caption{Comparison of a circular uniform (left), circular with disorder (center) and percolating cluster (right) type puddles used in the following calculations. In all cases the magnitude is normalized such that $\sum_r \Delta_r=\pi R^2$ (i.e. the $0$th Fourier component is identical). As can be seen, our estimation of $R$ for the third case slightly overestimates the radius.}
\label{fig:puddles pics}
\end{figure}
In the main text of this work, for the sake of simplicity we mainly consider uniform circular puddles. In this appendix, we present calculations of quantities discussed above for more generic puddles, comparing 4 microscopic models for the shapes of the puddles: 
(1) Uniform circular puddles, as discussed in the main text,
(2) Circular puddles with additional disorder realized via fluctuations of the value of the gap on each site. Specifically, we choose $\Delta_r = \Delta\times\Theta(R-r)\times \zeta_r$ where $0<\zeta_r<1$ is uniformly distributed, (3) Uniform circular puddles with a $d$-wave form factor, and
(4) Connected clusters arising from a percolation process. In all cases, we scale the magnitude of $\Delta$ so that $\sum_r \Delta_r = \pi R^2$.
A comparison of puddles of each type is shown in Fig.~\ref{fig:puddles pics}.

\begin{figure}[H]
\centering
\includegraphics[width=0.9\textwidth]{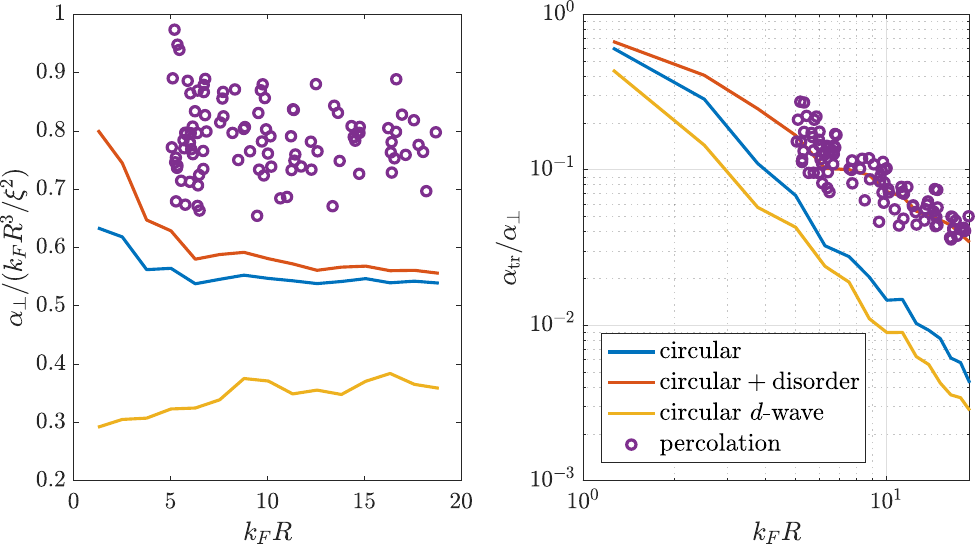} 
\caption{Comparison of $\alpha_\perp(R)$ (left) and $\alpha_{\rm tr}/\alpha_\perp$ (right) for the $4$ types of puddles considered above (in the 1st order born approximation, i.e. assuming $R\ll\xi_0$). For all types of puddles we see the expected scaling $\alpha_\perp \sim k_FR^3/\xi_0^2$ 
. As mentioned in the main text, the ratio $\alpha_{\rm tr}/\alpha_\perp$ for uniform puddles is proportional to $1/R^2$. Adding gap disorder on short length scales increases it significantly and leads to it scaling like $1/R$ instead.}
\label{fig:alpha puddles}
\end{figure}

We compare the $4$ different models of puddles with respect to their scaling of $\ell_{1ch}\propto \log(E_c/T_{1ch})$ with $R$, defined in Appendix.~\ref{appendix: kondo} (Fig.~\ref{fig:l puddles}) and $\alpha_\perp,\alpha_{\rm tr}$
computed within the Born approximation with $R$, defined in Eq.~\eqref{alpha tr}(Fig.~\ref{fig:alpha puddles}). Note that for puddles of type 4 we define $R$ as the maximal distance between a point in the puddle from its center of mass, which slightly overestimates the value of $R$ one would expect.

 \begin{figure}[b]
\centering
\includegraphics[width=0.7\textwidth]{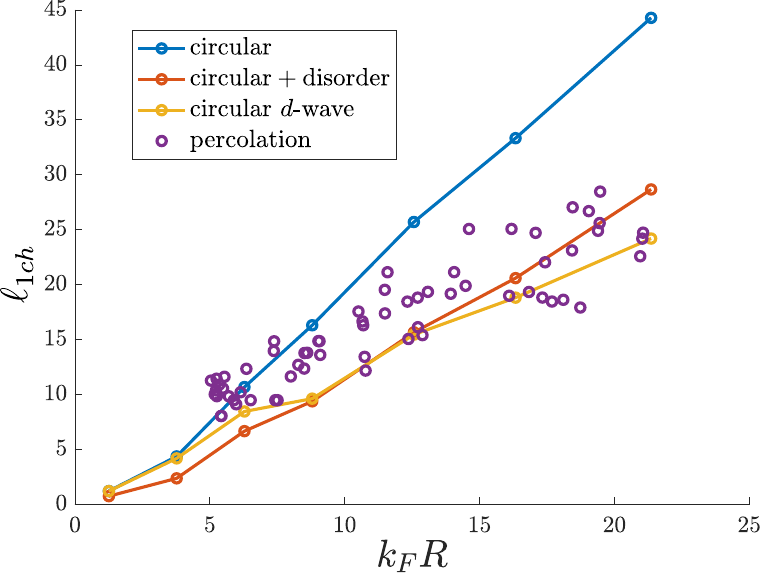} 
\caption{Comparison of $\ell_{1ch}$ for the $4$ types of puddles considered above. The uniform circular puddles have a higher $\ell_{1ch}$ than those with disorder, since the disorder further smears out the distribution of eigenvalues of the coupling matrices, and thus makes it easier for the highest eigenvalue to separate from the rest. All types of puddles show a linear increase with puddle size $R$, as expected.}
\label{fig:l puddles}
\end{figure}

The difference in the scaling properties of $\alpha(R),\alpha_{tr}(R)$ in the uniform and disordered cases can easily be understood by a simple calculation. Assume $\Delta_r\propto \Theta(R-r)\times \zeta_r$ with $\zeta_r$ disorder with correlation length $l_{\rm dis}$, i.e. $\left<\zeta_r\right>=0,\left<\zeta_r\zeta_0\right>=A(l_{\rm dis})\exp(-r^2/2l_{\rm dis}^2)$. Note that uniform circular puddles correspond to $l_{\rm dis}\gg R,A(l_{\rm dis})=const$, while spatially uncorrelated disorder (on the atomic scale) correspond to the limit $l_{\rm dis}\ll R, A(l_{\rm dis})\propto a^2/l_{\rm dis}^2$. The Fourier transform of $\Delta_r$ is given by a ($2$-dimensional) convolution:
\begin{align}
    \Delta_k &\propto  \left(\frac{R}{a}\right)^2 \int_q f(Rq)\zeta(k-q) \nonumber\\
    \left<\zeta(k)\zeta(-k')\right> &\propto l_{\rm dis}^2 A(l_{\rm dis}) \exp(-k^2l_{\rm dis}^2/2) \delta^{(2)}(k-k')
\end{align}
where $R^2f(kR)$ is the Fourier transform of the step function $\Theta(R-r)$, and decays as $|f(x\gg1)|\sim 1/x^2$. One can then simply calculate $\alpha$:
\begin{align}
    \alpha  & \propto \left(\frac{R}{a}\right)^4 \int_{k,k'\in FS} \int_{q,q'} f(Rq)f(-Rq')\left<\zeta(k-k'-q)\zeta(k'-k+q')\right> \nonumber\\ &\propto  \left(\frac{R}{a}\right)^4l_{\rm dis}^2A(l_{\rm dis})\int_{\delta k=k-k'} \int_q f(Rq)^2 
    \exp\left(-(\delta k-q)^2l_{\rm dis}^2/2\right)
\end{align}
where we have replaced the double integral over $k,k'$ on the Fermi surface by an integral over momentum differences on the Fermi surface $\delta k$, and ignored factors of $k_Fa$ (and overall $\rho_F^2\Delta^2$ etc).
For a uniform puddle we can take $l_{\rm dis}\to\infty$ and $l_{\rm dis}^2$ times the gaussian may be replaced with a delta-function, giving $\alpha\propto R^4 \times R^{-1}$ where $1/R$ comes from the domain of the 1d integral over $\delta k$. On the other hand, for uncorrelated disorder the integral over $q$ will be restricted to the domain $|q|<1/R$, giving $\alpha\propto R^4\times R^{-2}\times l_{\rm dis}^{-1}$. Overall we find 
\begin{equation}
    \alpha \propto \begin{cases} \left(\frac{R}{a}\right)^3, & l_{\rm dis} \gg R \\ \frac{R^2}{al_{\rm dis}}, & l_{\rm dis} \ll R \end{cases}
\end{equation}
Similarly, $\alpha_{tr}$ can be evaluated by adding a factor of $\delta k^2$ to the integral (coming from $1+\cos(\theta_k-\theta_{k'})$), which then gives
\begin{equation}
    \alpha_{tr} \propto \begin{cases} \frac{R}{a}, & l_{\rm dis} \gg R \\ \frac{aR^2}{l_{\rm dis}^3}, & l_{\rm dis} \ll R \end{cases}
\end{equation}
and 
\begin{equation}
    \alpha_{tr}/\alpha \propto \frac{a^2}{\min(R^2,l_{\rm dis}^2)}
\end{equation}
(for $l_{\rm dis}\geq a$). However, in our calculation of puddles of type (2), the disorder has expectation value $1$, which can be achieved by $\zeta_r\to 1+\zeta_r$. Then (plugging $l_{\rm dis}=a$) one simply adds up the two contributions (there are no cross terms since $\left<\zeta_r\right>=0$):
\begin{align}
    \alpha &\propto \left(\frac{R}{a}\right)^3 +\left(\frac{R}{a}\right)^2 \sim R^3 \nonumber \\
    \alpha_{tr} &\propto \frac{R}{a} +\left(\frac{R}{a}\right)^2 \sim R^2 
\end{align}
i.e. the uniform component dominates $\alpha$ while the disordered component dominates $\alpha_{tr}$. This leads to $\alpha_{tr}/\alpha \sim 1/R$, as seen in the numerical calculation above.

\section{Suppression of Kondo screening}

\label{appendix: kondo}
In the main text, we consider the TLS-electron interaction term:

\begin{align}
    H_{\rm int} &= \sum_{\vec{k},\vec{k}'} \left(g_\perp(\vec{k},\vec{k}') \sigma^{-} c^\dagger_{\vec{k},\uparrow}c^\dagger_{\vec{k}',\downarrow} +{\rm h.c.} \right)  \nonumber\\
    &+ \sum_{\vec{k},\vec{k}',s} g_z(\vec{k},\vec{k}')\sigma^z c^\dagger_{\vec{k},s}c_{\vec{k}',s},
\end{align}
where $\{\sigma^z, \sigma^\pm\}$ are Pauli operators acting on the two charge states of the puddle. 
Following Ref.~\cite{zarand_two-channel_2000}, we apply a particle-hole transformation on the spin-$\downarrow$ electrons
\begin{equation}
    c^\dagger_\downarrow \to c_\downarrow 
\end{equation}
We replace the FS coordinates $\vec{k},\vec{k}'$ by general indices $i,j$ and define the matrices:   
\begin{align}
J_{x} & =\begin{pmatrix}0 & g_\perp\\
g_\perp^{\dagger} & 0
\end{pmatrix} \nonumber \\
J_{y} & =\begin{pmatrix}0 & ig_\perp\\
-ig_\perp^\dagger & 0
\end{pmatrix} \nonumber \\
J_{z} & =\begin{pmatrix}g_z & 0\\
0 & -g_z
\end{pmatrix}.
\label{J mats g perp z}
\end{align}
This allows us to write the interaction in a generic multi-channel Kondo form:
\begin{equation}
H_{\rm Kondo}=\sum_{a=x,y,z}\sum_{i,j} (J_a)_{ij} c^\dagger_i c_j \sigma^a .
\label{H gen kondo}
\end{equation}
The TLS couples to the electrons via
the three matrices $(J_a)_{ij}$, which we will now consider to be generic. These matrices encode all the information on the structure of the TLS-electron interaction in flavor space. Within this notation, the flavors $i,j=1,\cdots,N$ also include spin flavor. For example, the familiar isotropic single channel Kondo model will be given by $N=2$ (with flavors $i,j = \uparrow,\downarrow$) and $J_a=\lambda\tau_a$ where $\tau_a$ are Pauli matrices; the anisotropic Kondo model by $J_a=\lambda^a\tau_a$, and and the $n$-channel Kondo model is given by $J_a=(\lambda_1 \tau_a) \oplus (\lambda_2 \tau_a) \oplus \cdots \oplus (\lambda_n \tau_a)$ with $N=2n$. Note that if the flavor corresponds to the electron spin, this will be the spin-Kondo model, but it does not necessarily have to be the case, for example in the puddle case the flavor is associated with charge degrees of freedom.

The RG equations for this Hamiltonian have been derived to $2$-loop order (i.e. $J^3$) in \cite{zarand_low-temperature_1995} (and are discussed in greater detail in \cite{von_delft_2-channel_1999}). Note that there, an \textit{additional} flavor index with $N_f$ flavors and exact degeneracy was introduced, so that the system flowed to the $N_f$-channel fixed point at $J\sim 1/N_f$, which for large enough $N_f$ lies in the perturbative regime, thus controlling the calculation. However, since we will be interested in the question \textit{how long does the system spend (in RG time) in the weak coupling regime} throughout the flow, the equations to this order are sufficient for us even without invoking this additional flavor. In matrix form, the RG equations are given by:
\begin{equation}   
\frac{dJ_{a}}{d\ell}=-i\frac{\rho_F}{\pi}\epsilon_{abc}J_{b}J_{c}-\frac{\rho_F^2}{2\pi^2}\left(J_{a}\text{tr}\left(J_{b}J_{b}\right)-J_{b}\text{tr}\left(J_{a}J_{b}\right)\right)
\label{rg g mats}
\end{equation}
where $\epsilon_{abc}$ is the Levi-Civita tensor, and summation over repeated indices is implied (note the difference of $2\pi$ in the convention for the density of states relative to  \cite{zarand_low-temperature_1995}).
As pointed out in \cite{zarand_low-temperature_1995}, the only stable fixed points of these equations are those for which 
\begin{align}
    J_a = \lambda \tau_a \oplus 0_{N-2\times N-2},
    \label{1c g}
\end{align}
with $2\pi\rho_F \lambda=1$,
corresponding to single-channel fixed points.  Note that to this order of the equations, the fixed point is ``artificially" set to finite coupling.
Instead of relying on it, we know that once the ``single-channel" matrix structure Eq.~\eqref{1c g} is reached, the system will flow to the strong-coupling Kondo fixed point within a flow time $\delta \ell_{K}\propto 1/ \lambda_{1ch}$ where $\lambda_{1ch}$ is the value of the interaction once the matrices approximately reach the form Eq.~\eqref{1c g}. Thus, we will take on a $2$-step RG approach, and analyze how much RG time is required for the system to go from given initial conditions to the structure Eq.~\eqref{1c g}. Remarkably, we will find that for quite generic initial conditions this will scale like $\ell_{1ch}\sim N$, so that the scale at which Kondo behaviour appearis  $T_{1ch} \sim E_c\exp(-\ell_{1ch})$ will be exponentially suppressed by the ``size" of the matrix $N$. A similar conclusion was reached for a related model in Ref. \cite{zarand_two-channel_2000}.

\subsection{Kondo screening in a random-matrix model}
It is simpler to understand this behavior in a model in which the bare matrices $J_a(\ell=0)$ are taken from a distribution of random matrices with probability measure:
\begin{equation}
P\left(J_{x},J_y,J_z\right)\sim\exp\left(-\sum_{a=x,y,z}\frac{N^{2}}{2\mathcal{J}_{a}^{2}}\text{tr}\left(J_{a}^{2}\right)\right).
\label{rand mat dist}
\end{equation}
Demanding \textit{exact} (as opposed to statistical) symmetries, e.g., ${\rm U}(1)$ as in our theory or ${\rm SU(2)}$, introduce correlations between the three coupling matrices; see below. In addition, $\mathcal{J}_{a}$ are the widths of the distributions of the matrix elements of the three matrices $J_a$.
Note that this corresponds to the couplings in the random matrix model given in Appendix \ref{appendix: random mat model}, albeit with the factors of $1/N$ absorbed in $J$. It is useful to define the scalars:
\begin{align}
\alpha_{ab} & =\frac{\rho_F^2}{2\pi^2}\text{tr}\left(J_{a}J_{b}\right)\nonumber \\
\beta & =-i\left(\frac{\rho_F}{2\pi}\right)^3\epsilon_{abc}\text{tr}\left(J_{a}J_{b}J_{c}\right).
\end{align}
For initial conditions taken from the distribution Eq.~\eqref{rand mat dist} we have $\alpha_a\equiv\alpha_{aa}\sim\mathcal{O}\left(1\right),\alpha_{a\neq b}\sim\mathcal{O}\left(\frac{1}{N}\right),\beta\sim\mathcal{O}\left(\frac{1}{N^{3/2}}\right)$,
while for comparison, in the single-channel Kondo case $\alpha_{aa},\beta\sim\mathcal{O}\left(1\right),\alpha_{a\neq b}=0$.
Starting with arbitrary initial conditions, we can look at the flow equations
for the above quantities (which are derived straightforwardly from Eq.~\eqref{rg g mats}):
\begin{align}
\frac{d\alpha_{ab}}{d\ell} & =4\beta\delta_{ab}-2\left(\alpha_{ab}\alpha_{cc}-\alpha_{ac}\alpha_{bc}\right) \nonumber \\
\frac{d\beta}{d\ell} & =\frac{3}{8}\left(\frac{\rho_F}{\pi}\right)^4\text{tr}\left(J_{a}J_{a}J_{b}J_{b}-J_{a}J_{b}J_{a}J_{b}\right)-\frac{3}{2}\alpha_{aa}\beta.
\label{set of RG flow eqs}
\end{align}
 Observe that the set in Eqs.~\eqref{set of RG flow eqs} are not generically closed. Nevertheless,
in the initial stages of the flow, to leading order
in $\tfrac{1}{N}$ the only interesting dynamics are captured by the diagonal $\alpha_{aa}$ which constitute a closed set of equations (here the index $a$ is not
summed over):
\begin{equation}
\frac{d\alpha_{a}}{d\ell}=-2\left(\sum_{c\neq a}\alpha_{c}\right)\alpha_{a}+\mathcal{O}\left(\frac{1}{N^{2}}\right).
\label{rg alpha large N}
\end{equation}
This is the ``frustrated" regime, where the couplings $\alpha_{a}$ all decrease. 
Eventually, this behavior will change due to the growth of $\beta$. The increase of $\beta$ is governed by the term proportional to $\sum_{a,b}\text{tr}\left(J_{a}J_{b}J_{b}J_{a}-J_{a}J_{b}J_{a}J_{b}\right)$, which is proportional to $1/N$ as will show next, while $\beta\sim N^{-3/2}$ and is therefore subleading. The leading contribution is determined by the underlying symmetry of the couplings. For example, in the ${\rm SU}(2)$ symmetric case, the couplings can be written as $J_a = g\otimes \sigma_a$ and the ${\rm U}(1)$ symmetric case is given by the form Eq.~\eqref{J mats g perp z} with $g,g_\perp,g_z$ taken from the distribution Eq.~\eqref{rand mat dist} ($g,g_z$ being hermitian); and the generic case corresponds to completely uncorrelated distributions for the three components. We find that the contributions to $d\beta/d\ell$ in Eq. (\ref{set of RG flow eqs}) are given to leading order in $1/N$ by (with symmetry $S={\rm SU(2),U(1),generic}$) by
\begin{eqnarray}
   \frac{\mathcal{B}_{S}}{N} \equiv \frac{3}{8}\left(\frac{\rho_F}{\pi}\right)^4 \sum_{a,b}\text{tr}\left(J_{a}J_{b}J_{b}J_{a}-J_{a}J_{b}J_{a}J_{b}\right)  \approx \begin{cases}
\frac{9}{N}\alpha^{2} & \text{SU}\left(2\right)\\
\frac{3}{N}\left(\alpha_{\perp}^{2}+\alpha_{\perp}\alpha_{z}\right) & \text{U}\left(1\right)\\
\frac{3}{2N}\sum_{a\ne b}\alpha_{a}\alpha_{b} & \text{generic}
\end{cases} 
\end{eqnarray}
By inserting this in the flow equation for $\beta$ we see that
\begin{eqnarray}
    \frac{d\beta}{d\ell} =\frac{\mathcal{B}_{S}}{N} + \mathcal{O}(N^{-3/2})
\end{eqnarray}
Thus we expect that after a scale $\ell$ that satisfies $\ell^{-1} \sim\frac{\mathcal{B}_{S}}{N}$
the quantity $\beta$ will become  $\mathcal{O}(1)$ and the flow will change
its nature to the single-channel Kondo behavior, with $\alpha_{a}$ increasing. We can work this out more precisely. Consider for simplicity a case with SU(2) symmetry, i.e., $\alpha_{a}=\alpha$ (either statistical or exact symmetry), so that the equations simplify. Other cases show the same scaling of $T_{1ch}$ with modified coefficients. 

To leading order in $1/N,\alpha$, the RG equations are given by
\begin{align}
\frac{d\alpha}{d\ell} & =-4\alpha^{2}+4\beta \nonumber \\
\frac{d\beta}{d\ell} & =\frac{C}{N}\alpha^{2}-\frac{9}{2}\alpha\beta+\mathcal{O}\left(\frac{1}{N^{2}}\right) \nonumber 
\end{align}
(as derived above, $C=9$ for exact SU($2$) or $9/2$ for a statistical SU($2$) symmetry). 
As long as $\beta\ll\alpha^{2}$, we can solve for $\alpha\left(\ell\right)$
easily:
\begin{equation}
\alpha\left(\ell\right)=\frac{\alpha_{0}}{1+2\alpha_{0}\ell}
\end{equation}
Let us define $x=1+2\alpha_{0}\ell$, then:
\begin{align}
\frac{d\beta}{dx} =\frac{C\alpha_{0}}{N}\frac{1}{x^{2}}-\frac{9}{2}\frac{\beta}{x}
\Rightarrow\beta(x)=\frac{2C \alpha_0\left(x^{7/2}-1\right)}{7N x^{9/2}}
\end{align}
where $\beta(0)\sim\mathcal{O}\left(\frac{1}{N^{3/2}}\right)$ has
been neglected. Then the flow of $\alpha$ changes its nature at $x_{1ch}$ where $d \alpha/d\ell \left(x_{1ch}\right)=0$, which is given by:
\begin{align}
2C\left(x^{7/2}-1\right)=7N x^{5/2}\alpha_0
\end{align}
and solved approximately by 
\begin{equation}
x_{1ch}\approx\frac{7}{2C}N\alpha_{0}\Rightarrow\ell_{1ch}\approx\frac{7}{4C}N.
\end{equation}
Thus, the upturn associated with Kondo behaviour will appear only below the energy scale $T_{1ch}/E_c\sim e^{-\ell_{1ch}} \sim e^{-N}$.
At this point of the flow $\alpha_{1ch}=\frac{2C}{7N}$ so assuming
that this is dominated by one channel with eigenvalue $J_{max}\sim N^{-1/2}$
the Kondo temperature will be $T_{K}/T_{1ch} \sim  e^{-1/\sqrt{N}}$.

\begin{figure}[H]
\centering
\includegraphics[width=0.6\textwidth]{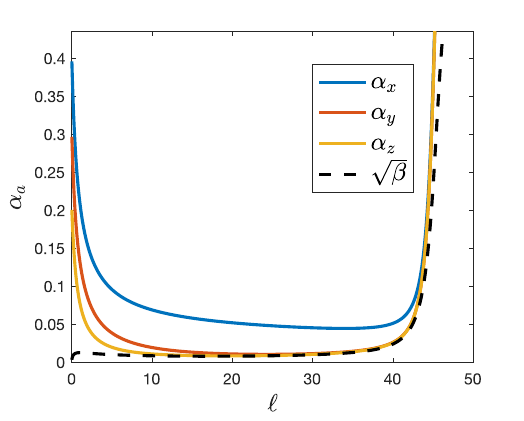} 
\caption{Flow of the couplings $\alpha_{a}$ and $\sqrt{\beta}$ for $N=200$ and $\vec{\alpha}_0=(0.4,0.3,0.2)$. $\vec{\alpha}$ flow according to Eq.~\eqref{rg alpha large N} until $\beta\sim\alpha^2$, at which point the flow changes to the single-channel Kondo behavior.
}
\label{fig:rg_rmt}
\end{figure}

All of this can also be seen 
by 
direct numerical integration of the 
the RG equations Eq.~\eqref{rg g mats}.
For example, in Fig.~\ref{fig:rg_rmt} we plot $\alpha_{a}(\ell),\beta(\ell)$ for $N=200,\vec{\alpha}(\ell=0)=\left(0.4,0.3,0.2\right)$.
As one can see, for many decades of the flow $\beta$
is negligible and $\alpha_{a}$ decay as described by Eq.~\eqref{rg alpha large N}, with $\alpha_{y},\alpha_z$ decaying to $0$ and $\alpha_{x}$ saturating (since it is the dominant coupling). 
After a long time $\ell_{1ch}\sim N$ this behavior breaks down and $\beta$ rapidly increases, changing the behavior to the usual Kondo scaling, with all couplings 
marginally relevant.

This behavior can also be understood from the random matrix model in Appendix \ref{appendix: random mat model}. There, in the limit $N\to\infty$ the TLS only feels particle-hole excitations from the electrons, and the RG equations Eq.~\eqref{rg alpha large N} arise via the mapping to a multi-bath spin boson model (with the particle-hole excitations acting as the source for the bosonic bath). This limit has been studied extensively for general cases in previous works \cite{BashanTunable,tulipman2024solvable}.

\subsection{Kondo scale in our model}


Let us now study the RG flow Eq.~\eqref{rg g mats} with initial conditions corresponding to the superconducting puddle problem, using the structure of the $J$ matrices given in Eq.~\eqref{J mats g perp z}.
Note that this structure is enforced by the $U(1)$ charge conservation symmetry of the problem, and will be maintained
throughout the flow. One can thus rewrite the RG equations in terms of
$g_\perp,g_z$ instead of $J_{x,y,z}$.


Remarkably, the suppression of $T_{1ch}$ in these systems does not rely on any specific structure of the couplings (such as a random-matrix assumption), but rather arises due to the dense spectrum of eigenvalues of the coupling matrices, leading to strong frustration. We repeat the above calculations for Andreev scattering from circular puddles of different sizes $R$, with $g_\perp(\ell=0)$ corresponding to Eq.~\eqref{g perp} and $g_z(\ell=0)=0$. In order to isolate the effect of the puddle size, we scale $g_\perp(0)\to J_0\times g_\perp(0)/\sqrt{\alpha_\perp(0)}$ so that all calculations have the same $\alpha_\perp(0)\propto(\rho_FJ_0)^2$ and thus collapse onto a single curve in the initial stage of the flow. The resulting graphs of $\alpha_\perp(\ell)$ and $\ell_{1ch}(R)\approx \log(E_c/T_{1ch})$ are presented in Fig.~\ref{fig:rg and susc} in the main text. Clearly, all graphs collapse onto the curve given by Eq.~\eqref{alpha of ell} up to some flow time $\ell_{1ch}$ (defined in these calculations by the location of the minimum in $\alpha_\perp(\ell)$) which scales linearly with $k_FR$.

\begin{figure}[h!]
\centering
\includegraphics[width=0.7\textwidth]{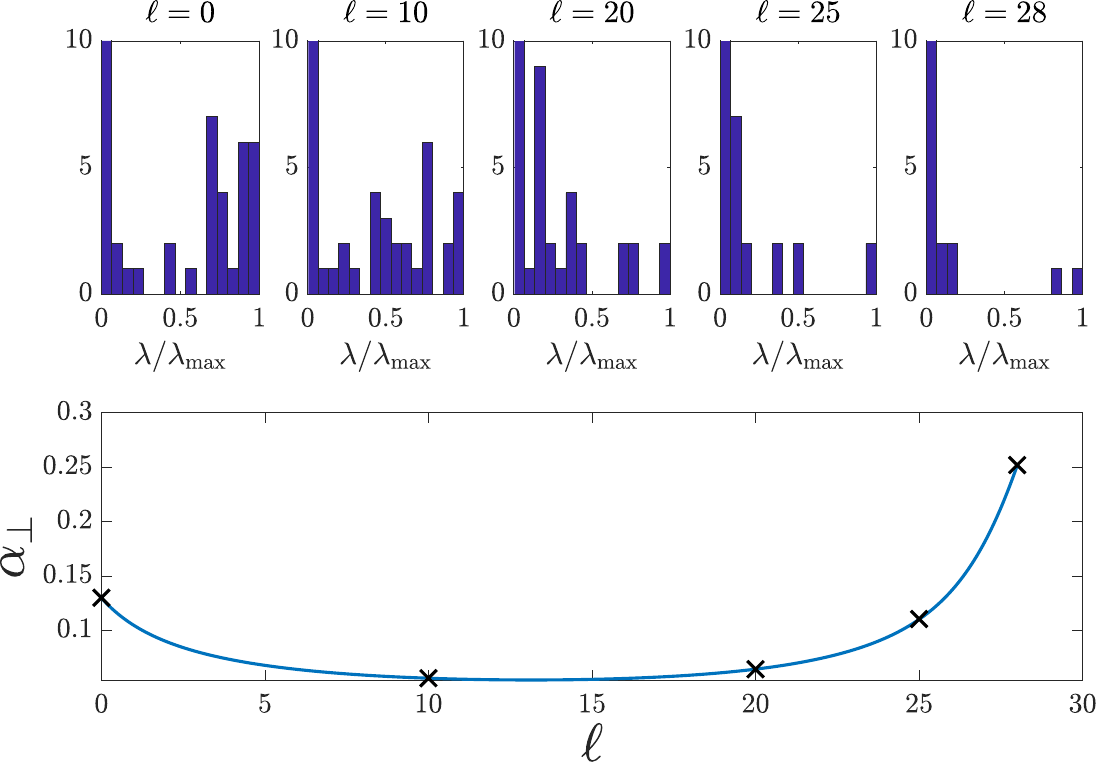} 
\caption{Upper row: Histograms of the (positive half of the) spectrum of $J_x$ throughout the RG flow for a circular puddle with $k_FR=7.5$ at different times of the RG flow, marked in $\times$'s on the graph of $\alpha_\perp(\ell)$ in the bottom row.
As explained in the text, initially the eigenvalues are bunched near the maximal value, and are gradually pushed away from the maximal eigenvalue throughout the flow.}
\label{fig:histograms}
\end{figure}

This can be understood as follows. In the limit $k_FR\to0$ the matrix $g_\perp$ is a constant with all-to-all scattering on the Fermi surface (since the typical momentum transfer $q_0\sim1/R$ is larger than the whole Fermi surface); the eigenvector  $\sim(1,1,\cdots,1)$ is the only one with nonzero eigenvalue and the TLS is immediately Kondo-screened in the $s$-wave channel. Conversely, as $k_FR\to\infty$ the matrix $g_\perp(0)$ approaches the identity matrix (since the interaction range in $k$-space decays like $1/R$), and its nonzero eigenvalues are bunched densely near the maximal eigenvalue (in the limit $k_FR\to \infty$ the system will be at an ``infinite-channel" Kondo critical point \cite{zarand_two-channel_2000}). For any finite but not too small $R$ the initial spectrum is dense without a well-separated maximal eigenvalue, and throughout the flow the largest eigenvalue slowly separates from the rest while all others decay to $0$, as can be seen in the histograms of the eigenvalues of $\tilde{g}_x$ throughout the flow, shown in Fig.~\ref{fig:histograms}.

\section{Random $g$-matrix model}
\label{appendix: random mat model}

Here we consider a random matrix variant of the low-energy Hamiltonian that can be viewed as a coarse-grained description of the TLS Hamiltonian in Eq.~(\ref{TLS Ham}) in the main text. We will show that this model shares important properties with the more physically motivated puddle models, where the coupling matrices $g_\perp$, $g_z$ are derived microscopically, as described in Appendix \ref{appendix: puddles}.
Specifically, the one-loop approximation for the puddle models becomes exact in the random matrix model, and the suppression of the Kondo temperature can be obtained analytically and follows the same scaling as in the puddle models. This exemplifies the fact that the puddle models naturally behave similarly to the random matrix model even when the number of channels (the dimension of the coupling matrices) is not very large. 

The random matrix model is constructed as follows. Every unit cell contains $N$ electron channels and $M$ puddles, $N,M\gg1$ (with fixed ratio $M/N\sim \mathcal{O}(1)$), and we allow for arbitrary electron-puddle interactions within each unit cell. Scaling the couplings accordingly, we arrive at
 \begin{equation}
H_{\rm TLS}(\vec{r})= -\sum_{i} \frac{1}{2} h_{\vec{r},i} \sigma_{\vec{r},i}^{z}+ \frac{1}{N} \sum_{ijl} \left( g_{\perp}^{\vec{r},ijl}\sigma_{\vec{r},i}^{+}c_{\vec{r},j}c_{\vec{r},l}+{\rm h.c.}+  g_z^{\vec{r},ijl}\sigma^z_{\vec{r},i}c_{\vec{r},j}^{\dagger}c_{\vec{r},l} \right).
\end{equation}
The Hamiltonian of the full, coarse-grained system is given by 
\begin{equation}
H=\sum_{\k;l}(\varepsilon_{\k} - \mu)c^\dagger_{\k,l}c_{\k,l}+\sum_{\vec{r}} H_{\rm TLS}(\vec{r}).
\end{equation}
Here the electronic dispersion $\varepsilon_\k$ and the chemical potential $\mu$ are $l$-independent. We consider random, spatially and flavor-uncorrelated complex couplings, namely $g_\perp=g_1+ig_2$, with $g_1,g_2,g_z$ being Gaussian random variables with zero mean and variances $g_{1}^2,g_{2}^2,g_z^2$ respectively, and we define $g_\perp^2=g_1^2+g_2^2$. 

To proceed, assuming a trivial structure in replica space, we express the disorder-averaged partition function $\overline{Z}=\int D\{g\} P(\{g\}) Z[\{g\}]$ (where $P$ is the normal probability distribution for each $g_{ijl}$) with an effective action for the electron and TLS correlators. We follow the steps of Refs.~\cite{BashanTunable,tulipman2024solvable} (were a similar derivation is presented in greater detail): We integrate over the interaction disorder $g$, assuming the system is self-averaging (i.e., using the replica diagonal Ansatz); and then introduce the bilocal fields
\begin{equation}
    G_{\vec{k}}(\tau,\tau') = -\frac{1}{N} \sum_{i} c_{\vec{k},i}(\tau) \bar{c}_{\vec{k},i}(\tau'),\quad \chi_{ab}(\tau,\tau') = \frac{1}{M} \sum_{l} \sigma_a(\tau) \sigma_b(\tau'), {\rm where }~a,b\in \{x,y,z\},
\end{equation}
via the Lagrange multiplier fields $\Sigma,\Pi$, respectively. Then, integrating over the electrons yields the effective action
\begin{align}
      S_{\rm eff}&=-N\text{Tr}\ln\left(G_{0}^{-1}-\Sigma\right) - N\int_{\tau,\tau'}\sum_{\boldsymbol{r},\boldsymbol{r}'}\sum_{\sigma}G_{\boldsymbol{r},\boldsymbol{r}'}\left(\tau',\tau\right)\Sigma_{\boldsymbol{r},\boldsymbol{r}'}\left(\tau,\tau'\right) + \frac{M}{2}\int_{\tau,\tau'}\sum_{\boldsymbol{r}}\sum_{ab}\chi_{ab,\boldsymbol{r}}\left(\tau,\tau'\right)\Pi_{ab,\boldsymbol{r}}\left(\tau',\tau\right)\nonumber
      \\\nonumber &-\frac{M}{2}\int_{\tau,\tau'}\sum_{\boldsymbol{r}}2 g_\perp^2 \bigg[\left(G_{\boldsymbol{r}}\left(\tau',\tau\right)^2+G_{\boldsymbol{r}}\left(\tau,\tau'\right)^2\right)\left(\chi_{xx,\boldsymbol{r}}\left(\tau,\tau'\right)+\chi_{yy,\boldsymbol{r}}\left(\tau,\tau'\right)\right) \\\nonumber & + \left(G_{\boldsymbol{r}}\left(\tau',\tau\right)^2-G_{\boldsymbol{r}}\left(\tau,\tau'\right)^2\right)\left(i\chi_{xy,\boldsymbol{r}}\left(\tau,\tau'\right)-i\chi_{yx,\boldsymbol{r}}\left(\tau,\tau'\right)\right)\bigg]
      \\& -\frac{M}{2}\int_{\tau,\tau'}\sum_{\boldsymbol{r}} g_z^2 \left(G_{\boldsymbol{r}}\left(\tau',\tau\right)G_{\boldsymbol{r}}\left(\tau,\tau'\right)\right)\chi_{zz,\boldsymbol{r}}\left(\tau,\tau'\right) \nonumber
      \\& + \sum_{\boldsymbol{r}}\sum_{l=1}^{M}S_{\text{Berry}}\left[\boldsymbol{\sigma}_{l,\boldsymbol{r}}\right] - \int_{\tau}\sum_{\boldsymbol{r}}\sum_{l=1}^{M}{h}_{l,\boldsymbol{r}} {\sigma^z}_{l,\boldsymbol{r}} - \frac{1}{2}\int_{\tau,\tau'}\sum_{\boldsymbol{r}}\sum_{ab}\Pi_{ab,\boldsymbol{r}}\left(\tau',\tau\right)\sum_{l=1}^{M}\sigma^a_{l,\boldsymbol{r}}\left(\tau\right)\sigma^b_{l,\boldsymbol{r}}\left(\tau'\right).
\end{align}
Here $S_{\rm Berry}[\boldsymbol{\sigma}_{l,\boldsymbol{r}}]$ denotes the Berry phase of the TLSs in the spin-coherent state path integral representation of the partition function. The saddle point equations, obtained by varying $S_{\rm eff}$ with respect to $\Pi,\chi,\Sigma,G$, respectively, are given by 
\begin{align}
    \chi_{ab,\boldsymbol{r}}(\tau,\tau') &= \frac{1}{M}\sum_{i=1}^{M} \sigma^a_{i,\boldsymbol{r}}(\tau) \sigma^b_{i,\boldsymbol{r}}(\tau') \\
    \Pi_{ab}(\tau) &= \delta_{ab}(\delta_{ax}+\delta_{ay})\Pi_{+}(\tau) + i\epsilon_{abz}\Pi_{-}(\tau)+\delta_{ab}\delta_{az}\Pi_{z}(\tau)\\
    \Pi_{\pm}(\tau) &= 2g_\perp^2\left(G(\tau)^2\pm G(-\tau)^2\right) \\ 
    \Pi_z(\tau) &= -g_z^2 G(\tau)G(-\tau)\\G_{\boldsymbol{k}}\left(i\omega\right)&=\frac{1}{i\omega-\varepsilon_{\boldsymbol{k}}-\Sigma\left(i\omega\right)}, \\
    \Sigma_{\boldsymbol{r}} (\tau)&=-2\frac{M}{N}g^2  G_{\boldsymbol{r}}(-\tau) \left(\chi_{xx,\boldsymbol{r}}(\tau)+i\chi_{xy,\boldsymbol{r}}(\tau)\right)
\end{align}
In the last equation, we used the fact that due to $U(1)$ symmetry, 
\begin{align}
    \chi_{xx}(\tau)&=\chi_{yy}(\tau)=\chi_{xx}(-\tau),\\ \chi_{xy}(\tau)&=-\chi_{yx}(\tau)=\chi_{xy}(-\tau).
\end{align}
In addition, $G(\tau) \equiv \int_{\vec{k}} G_{\vec{k}}(\tau)$ denotes the local Green's function, and note that the self-energy $\Sigma$ is completely local in the random matrix model. Assuming particle-hole symmetry, i.e., $G(\tau) = G(-\tau)$, further simplifies the saddle point equations. Small breaking of the particle-hole symmetry essentially translates to an additional field in the $z$ direction acting on the puddles, shifting the distribution of $h_{\vec{r},i}$. This does not change the physics qualitatively, and henceforth we focus on the particle-hole symmetric scenario. 

The saddle point equations can be solved provided we know the \textit{average} TLS susceptibility $\chi_{ab}$ (with respect to $h$). To do so, we identify the spin-boson action related to the TLSs and the electronic bath $\Pi$:
\begin{equation}
    S_{\rm TLS} = \sum_{\boldsymbol{r}}\sum_{l=1}^{M}S_{\text{Berry}}\left[\boldsymbol{\sigma}_{l,\boldsymbol{r}}\right] - \int_{\tau}\sum_{\boldsymbol{r}}\sum_{l=1}^{M}{h}_{l,\boldsymbol{r}} {\sigma^z}_{l,\boldsymbol{r}} - \frac{1}{2}\int_{\tau,\tau'}\sum_{\boldsymbol{r}}\sum_{ab}\Pi_{ab,\boldsymbol{r}}\left(\tau',\tau\right)\sum_{l=1}^{M}\sigma^a_{l,\boldsymbol{r}}\left(\tau\right)\sigma^b_{l,\boldsymbol{r}}\left(\tau'\right).
\end{equation}
As stated in the main text, this action describes a collection of decoupled spin-boson problems with Ohmic baths $\Pi_a(\omega)\propto |\omega|$ in the $U(1)$ symmetric point. The resulting RG equations for the couplings $\alpha_{\perp,z}$ were derived in \cite{castro_neto_quantum_2003,novais_frustration_2005} and are exactly those given in Eq.~\eqref{RG spin-boson U1}. The TLS susceptibility for a given $h$ can be obtained by perturbative techniques \cite{belyansky_frustration-induced_2021}, and is given in Eq.~\eqref{chi tls rpa}. The averaging procedure is identical to the one described in App.~\ref{appendix: more details}, where the corresponding dimensionless parameter $\alpha_a$ is defined here by $\alpha_a \equiv \frac{\rho_F^2}{\pi^2} {\rm tr}(|g_a|^2)$. The approximate one-loop self-energy in Eq.~\eqref{1 loop se} becomes the exact result in the random-matrix model. 

\section{More details on averaging of $\chi'',\Sigma''$, single-particle properties and transport}
\label{appendix: more details}

\subsection{Averaging over background charge $h$}

\label{appendix: averaged susc}
In this appendix we analyze the flow equations to determine the scaling form of the susceptibility Eq.~\eqref{scaling susc}.  The flow
equation for the longitudinal field $h$ of Eq.~\eqref{RG spin-boson U1} is
\begin{equation}
\frac{d{h}(\ell)}{d\ell}=(1-2\alpha_{\perp}(\ell)){h}(\ell),
\end{equation}
were $\alpha_\perp(\ell)$ is given by Eq.~\eqref{alpha of ell}. As in the main text, we omit the argument $(0)$ from bare coupling values. The solution is
\begin{equation}
{h}\left(\ell\right)=\frac{e^\ell {h}}{1+2\alpha_{\perp}\ell},
\end{equation}
where we restrict ourselves to the case $\alpha_z=0$ as it is negligible in the regime relevant here.
To determine the transverse susceptibility we add the source term $\int_{\tau}\left(h_{+}\left(\tau\right)\sigma_{+}\left(\tau\right)+h_{-}\left(\tau\right)\sigma_{-}\left(\tau\right)\right)$
to the action. The RG equations for the 
transverse fields are 
\begin{equation}
\frac{d{h}_{\pm}(\ell)}{d\ell}=(1-\alpha_{\perp}(\ell)){h}_{\pm}(\ell)
\end{equation}
with the solution 
\begin{equation}
{h}_{\pm}\left(l\right)=\frac{e^\ell {h}_{\pm}}{\left(1+2\alpha_{\perp}\ell\right)^{1/2}}.
\end{equation}
Finally, the
free energy obeys engineering scaling 
\begin{equation}
F\left({h},{h}_{\pm},\alpha_{\perp},T\right)=e^\ell F\left({h}\left(\ell\right),{h}_{\pm}\left(\ell\right),\alpha_{\perp}\left(\ell\right),e^\ell T\right).
\end{equation}
To determine the susceptibility we take the second derivative of the free energy 
 with respect to the transverse fields
\begin{eqnarray}
\chi\left({h},\alpha_{\perp},\omega,T\right) & \equiv & -\left.\frac{\delta^2 F\left({h},{h}_{\pm},\alpha_{\perp}\right)}{\delta{h}_{+}\left(\omega\right)\delta{h}_{-}\left(\omega\right)}\right|_{{h}_{\pm}=0}\nonumber \\
 & = & \frac{dh_{+}}{dh_{+}\left(\ell\right)}\frac{dh_{-}}{dh_{-}\left(\ell\right)}e^\ell \chi\left({h}\left(\ell\right),\alpha_{\perp}\left(\ell\right),e^\ell\omega,e^\ell T\right)\nonumber \\
 & = & \frac{e^{-\ell} \chi\left({h}(\ell),\alpha_{\perp}(\ell),e^\ell\omega,e^\ell T\right)}{1+2\alpha_{\perp}\ell}.
\end{eqnarray}

When averaging over $h$ with a uniform distribution we can change variables:
\begin{equation}
dh=\frac{dh}{dh\left(\ell\right)}dh\left(\ell\right)=\left(1+2\alpha_{\perp}\ell\right)dh\left(\ell\right).
\end{equation}
We arrive at the important finding that the integral for the {\it averaged}
susceptibility can be evaluated either using the bare theory or, equivalently,
using the low-energy theory:
\begin{equation}
\left<\chi\right>_{\bar{n}}\equiv \int_{-E_c}^{E_c} \chi(h,\alpha_\perp,\omega,T,\ell=0) \frac{dh}{2E_c} = \int_{-E_c}^{E_c} \chi(h(\ell),\alpha_\perp(\ell),e^\ell\omega,e^\ell T) \frac{dh(\ell)}{2E_c}. \label{scaling of averaged susc}
\end{equation}

It is now straightforward to calculate the averaged susceptibility. For concreteness, we may use the explicit expression for the free TLS susceptibility in Eq.~\eqref{weak coupling susc} of the main text. After performing the RG flow down to $\ell_*$, analytically continuing to real frequencies and taking the imaginary part, we find
\begin{equation}
    \chi''(h(\ell_*),\alpha_\perp(\ell_*),\omega,T)= -\pi\tanh\left(\frac{\beta h(\ell_*)}{2}\right) \delta(h(\ell_*)+\omega),
\end{equation}
which is valid for $\alpha(\ell_*)\ll1$,
and inserting this into Eq.~\eqref{scaling of averaged susc} leads to Eq.~\eqref{sus averaged over h} of the main text.
Note that the cancellation of the RG logarithms is special to the two-point function $\chi$, and happens since $\sigma^z \sim \sigma^+\sigma^-$, and thus to leading order $d\log(h)/d\ell = d\log(h_+ h_-)/d\ell$ near the weak coupling fixed point. Although this relation may break down for the two-point function due to higher order corrections in $\alpha$, these differences are subleading in $1/\ell$ and thus do not contribute to the leading behavior of $\chi$ at low energies.
Furthermore, this cancellation does not hold for higher-order correlation functions. Specifically, the scaling relation for the averaged $2k$-point function $\chi^{(2k)}=\left<\mathcal{T} \sigma^- \sigma^+ \cdots \sigma^- \sigma^+\right>$ reads
\begin{equation}
\left<\chi^{(2k)}\right>_{\bar{n}} = \frac{1}{2E_c}\int_{-E_c}^{E_c} \frac{\chi^{(2k)}(h(\ell))}{\left(1+2\alpha_\perp\ell\right)^{k-1}} dh(\ell). 
\end{equation}

We return to our discussion on the averaged susceptibility, Eq.~\eqref{sus averaged over h}. Notice that the form of Eq.~\eqref{sus averaged over h} does not depend on the exact functional form of $\chi''$ for a given $h$. Instead, it relies on the existence of a scaling form for $\chi$ at low energies. For example, for $T=0$, one has $\chi=\chi_{xx}+i\chi_{xy}$ where $\chi_{ab}=\left<\sigma^a\sigma^b\right>$, and $\chi''_{xx}(
\omega)$ ($\chi_{xy}(
\omega)$) is a symmetric (antisymmetric) function of $h$ (since the sign of $h$ can be absorbed into a redefinition $\sigma^y\to-\sigma^y$). Thus in the averaging Eq.~\eqref{scaling of averaged susc} only $\chi_{xx}$ will contribute. Defining a dimensionless scaling function $f$ via $\chi''_{xx}(\omega,h(\ell_*))=\frac{1}{h(\ell_*)}f(\omega/h(\ell_*))$, we see that 
\begin{align}
    \int_{-E_c}^{E_c}\chi''(\omega,h(\ell_*)) \frac{dh(\ell_*)}{2E_c} = \frac{{\rm sign (\omega)}}{E_c}\int_{|\omega|/E_c}^\infty f(x)\frac{dx}{x} = \frac{{\rm sign (\omega)}}{E_c} \left(\int_{0}^\infty f(x)\frac{dx}{x} -\mathcal{O}\left(\frac{|\omega|}{E_c}\right)\right)
    \label{susc scaling av}
\end{align}
where we have changed variables to $x = \omega/h(\ell_*)$. The lower integration limit can be continued to $0$ since for small frequencies $f(x\ll1)\propto x$, and the upper limit converges due to the sum rule $\int_0^\infty \chi''_{xx}(\omega) d\omega = 1/2$. 
Thus, for energies lower than the upper cutoff, the exact form of $\chi''$ could only slightly change the numerical prefactor. 
For example, this calculation can be performed exactly using the full TLS correlators, obtained via an RPA-like resummation of bubble diagrams for the TLS, which broadens the delta-function peak. The expression (after performing RG flow down to $\ell_*$) is given by \cite{belyansky_frustration-induced_2021}
\begin{equation}
    \chi_{xx}''(h(\ell_*),\alpha
    (\ell_*),\omega)= \frac{\pi\alpha
    (\ell_*)\omega\left(h(\ell_*)^2+(1+\pi^2\alpha
    (\ell_*)^2)\omega^2\right)}{\left(h(\ell_*)^2-(1+\pi^2\alpha
    (\ell_*)^2)\omega^2\right)^2+(2\pi\alpha
    (\ell_*) h(\ell_*)\omega)^2}.
    \label{chi tls rpa}
\end{equation}
The integral over $h$ reads
\begin{equation}
\left<\chi''(\omega)\right>_{\bar{n}} = \frac{1}{2E_c}\left(\arctan\left(\frac{E_c-\omega}{\pi\alpha(\ell_*)\omega}\right)+\arctan\left(\frac{E_c+\omega}{\pi\alpha(\ell_*)\omega}\right)\right) = \frac{\pi}{2E_c}{\rm sign}(\omega) -  \frac{\pi\alpha(\ell_*)\omega}{E_c^2} +\cdots .
\end{equation}
which agrees with Eq.~\eqref{sus averaged over h} in the main text to leading order in $\omega$.
Interestingly, in this case, the prefactor of the leading term is the same as one gets from the $\alpha=0$ expression, although there is no a priori reason for this.
 Going beyond the approximation used to derive Eq.~\eqref{chi tls rpa}, one would generally expect that the integral in Eq.~\eqref{susc scaling av} would depend on $\alpha$, i.e. $\int_0^\infty f(x)dx/x=\frac{\pi}{2}\tilde{f}(\alpha(\ell_*))$ with $\tilde{f}(0)=1$. This will generically lead to subleading in $1/\log(\omega)$ corrections of the form:
 \begin{equation}
 \left<\chi''(\omega)\right>_{\bar{n}} =  \frac{\pi}{2E_c}{\rm sign}(\omega) \tilde{f}\left(\frac{\alpha}{1+2\alpha\log(E_c/|\omega|)}\right) \approx  \frac{\pi}{2E_c}{\rm sign}(\omega) \left(1+\tilde{f}'(0)\frac{\alpha}{1+2\alpha\log(E_c/|\omega|)}+\cdots\right).
 \end{equation}
such that the leading term is the same as the one derived from the weak coupling approximations.

\subsection{Self energy and transport time}

\label{appendix: transport}

Here, we briefly outline the derivation of the self-energy and transport time of a single puddle of size $R$, averaged over the background charges. For example, at one-loop order, the self-energy on the Fermi surface is given by 
 \begin{align}
\Sigma''_{\vec{k}}( \omega)&= \int \frac{d\Omega}{2\pi}\sum_{\vec{k}'} |g_\perp(\vec{k},\vec{k}')|^2 G''_{\vec{k}'}(\omega+\Omega)\chi''(\Omega) \left(\coth\left(\frac{\Omega}{2T} \right) - \tanh\left(\frac{\omega+\Omega}{2T} \right)\right)\nonumber\\
& \approx \rho_F \int \frac{d\Omega}{2\pi}\int d\varepsilon_{\vec{k'}}\int \frac{ dk'_{\parallel} }{2\pi} |g_\perp(\vec{k},\vec{k}')|^2 G''_{\vec{k}'}(\omega+\Omega)\chi''(\Omega) \left(\coth\left(\frac{\Omega}{2T} \right) - \tanh\left(\frac{\omega+\Omega}{2T} \right)\right) \nonumber\\
& \approx  - \pi E_F \alpha_\perp \eta_{\vec{k}} \int\frac{d\Omega}{2\pi}\chi''(\Omega)\left(\coth\left(\frac{\Omega}{2T} \right) - \tanh\left(\frac{\omega+\Omega}{2T} \right)\right) 
\label{se one loop derivation}
\end{align}
where, as in the main text, $\eta_{\vec{k}} \propto \int \frac{ dk'_{\parallel} }{2\pi} |g_\perp(\vec{k},\vec{k}')|^2 / \alpha_{\perp}$ is the angular dependence of the self-energy which is determined by the characteristics of the puddles, e.g., their superconducting order parameter, shape, size, and disorder strength. The real part of the retarded self-energy is obtained by Kramers-Kronig. At $T=0$ $\Sigma'_{\k}(\omega) = \tfrac{\pi}{2}\Sigma''_{\k}(\omega) {\rm sgn}(\omega)\ln\left(\tfrac{E_c}{\omega}\right)$.

Note that the transport time and single particle lifetime are generically different, since scattering off of TLSs may not degrade the electrical current if it is dominated by Andreev backscattering. This difference does not affect the $T$-scaling, but only enters into the numerical prefactor. To see this explicitly, we analyze the reduction to the transport coefficient below for approximately isotropic scattering, i.e., either $s$-wave, or $d$-wave with sufficiently disordered puddles. Considerable anisotropy further reduces the coefficient but the exact weighting expression is more involved.

We follow Ref. \cite{tulipman_criterion_2022} (particularly the supplementary material)
to apply the Prange-Kadanoff reduction scheme and derive a quantum
Boltzmann equation (QBE) for our model. Using the fact that the spectral
function of the electrons is sharply peaked on the Fermi surface we
define a generalized distribution function 
\begin{align}
f\left(\hat{\boldsymbol{k}},\omega,\boldsymbol{r},t\right)\equiv-i\int\frac{d\varepsilon}{2\pi}G^{<}\left({\boldsymbol{k}},\omega,\boldsymbol{r},t\right)
\end{align}
and similarly $1-f=-i\int_{\varepsilon}G^{>}$. Here we are employing the conventional Keldysh notation where $G^{<(>)}$ denotes the lesser (greater) component of the Green's function. The restriction of the
electrons to the Fermi surface is valid provided that $\left|\Sigma^{''}\left(\omega\lesssim T\right)\right|\ll v_{F}q_{*}\left(T\right)$
for all scattering mechanisms (i.e. disorder and electron-puddle scattering).
Here $q_{*}\left(T\right)$ is the typical momentum transfer in collisions.
Since for electron-puddle scattering it holds that $q_{*}\sim1/R$
this condition is satisfied as long as the puddle is not too large as $v_{F}/R\sim E_{F}/k_FR$
and as long as $n_{\text{pud}}\ll1/R$. 

The QBE is given by 
\begin{align}
\mathcal{D}\left[f\left(\hat{\boldsymbol{k}},\omega,\boldsymbol{r},t\right)\right]=\mathcal{I}_{\text{coll}}\left[f\left(\hat{\boldsymbol{k}},\omega,\boldsymbol{r},t\right)\right]
\end{align}
 where the differential operator is given by 
\begin{align}
\mathcal{D}=\left(1-\partial_{\omega}\Sigma^{'}\right)\partial_{t}+\partial_{t}\Sigma^{'}\partial_{\omega}-\nabla_{\boldsymbol{r}}\Sigma^{'}\cdot\nabla_{\boldsymbol{k}_{F}}+\nabla_{\boldsymbol{k}_{F}}\left(\varepsilon_{\boldsymbol{k}}+\Sigma^{'}\right)\cdot\nabla_{\boldsymbol{r}}
\end{align}
and the collision integral is given by 

\begin{align}
\mathcal{I}_{\text{coll}}=\int_{\varepsilon_{\boldsymbol{k}}}\Sigma^{>}\left(\boldsymbol{k},\omega\right)G^{<}\left(\boldsymbol{k},\omega\right)-G^{>}\left(\boldsymbol{k},\omega\right)\Sigma^{<}\left(\boldsymbol{k},\omega\right).
\end{align}
Here we have that
\begin{align}
\Sigma^{<}\left(\boldsymbol{k},\omega\right) & =\int_{\boldsymbol{k}',\nu}\left|g_\perp\left(\boldsymbol{k},\boldsymbol{k}'\right)\right|^{2}\chi''\left(\nu\right)\left\{ \left(n_{0}\left(\nu\right)+1\right)G^{<}\left(-\boldsymbol{k}',\omega+\nu\right)+n_{0}\left(\nu\right)G^{<}\left(-\boldsymbol{k}',\omega-\nu\right)\right\} \\
\Sigma^{>}\left(\boldsymbol{k},\omega\right) & =\int_{\boldsymbol{k}',\nu}\left|g_\perp\left(\boldsymbol{k},\boldsymbol{k}'\right)\right|^{2}\chi''\left(\nu\right)\left\{ n_{0}\left(\nu\right)G^{>}\left(-\boldsymbol{k}',\omega+\nu\right)+\left(n_{0}\left(\nu\right)+1\right)G^{>}\left(-\boldsymbol{k}',\omega-\nu\right)\right\} 
\end{align}
where the $-\boldsymbol{k}'$ rather than $\boldsymbol{k}'$ is due
to the Andreev form of the interaction. The explicit dependence on
the sign $\pm\boldsymbol{k}$ cannot be disregarded as in the self-energy
since we are not working in equilibrium. Inserting the definitions
above, after carrying the integration of $\varepsilon_{\boldsymbol{k}}$
we obtain
We will use the QBE to extract the ratio $\lambda_{\text{tr}}/\lambda$
(or $\alpha_{\text{tr}}/\alpha$). Let us first consider $g\left(\boldsymbol{k},\boldsymbol{k}'\right)=const$. In this case
there are no vertex corrections. The collision integral is conveniently
given by 
\begin{align}
\mathcal{I}_{\text{coll}}\left[f\right]=2\Sigma''\left(\omega\right)\delta f
\end{align}
where we introduced the deviation from equilibrium of the generalized
distribution function as 
\begin{align}
f=f_{0}+\delta f
\end{align}
and $f_{0}$ is the Fermi-Dirac distribution in thermal equilibrium.
In the presence of an applied electric field the QBE is given by 
\begin{align}
v_{F}\hat{\boldsymbol{k}}\cdot\boldsymbol{E}\partial_{\omega}f_{0}=2\Sigma''\left(\omega\right)\delta f.
\end{align}
Inserting $\delta f=k_{F}\hat{\boldsymbol{k}}\cdot\boldsymbol{\delta f}\left(\omega\right)$
we have 
\begin{align}
\delta f_{i}\left(\omega\right)=\frac{v_{F}}{k_{F}}\frac{\partial_{\omega}f_{0}}{2\Sigma''\left(\omega\right)}E_{i}\quad i=x,y
\end{align}
Since the electrical current is given by $\boldsymbol{J}_{\text{el}}=-\rho_{F}\int_{\hat{\boldsymbol{k}},\omega}v_{F}\hat{\boldsymbol{k}}\delta f\left(\hat{\boldsymbol{k}},\omega\right)$
we have 
\begin{align}
\sigma_{\text{el}}=\frac{\rho_{F}v_{F}^{2}}{16T}\int_{\omega}\frac{\text{sech}^{2}\left(\frac{\omega}{2T}\right)}{\left|\Sigma''\left(\omega\right)\right|}
\label{conductivity}
\end{align}
i.e. $\alpha_{\text{tr}}=\alpha$ as expected in the case with no
vertex corrections (Eq.~\eqref{conductivity} agrees with the expressions in \cite{BashanTunable,tulipman2024solvable}, evaluated using the Kubo formula in the absence of vertex corrections). 

Next we proceed to evaluate the reduction of $\alpha_{\text{tr}}/\alpha$
due to the non-trivial form factor $g\left(\boldsymbol{k},\boldsymbol{k}'\right)$.
To this end we will use the variational formulation of the QBE \cite{mahan_many-particle_2000,nave_transport_2007,tulipman_criterion_2022}. We
parametrize the deviation from equilibrium as $f=f_{0}+\phi\partial_{\omega}f_{0}$.
The variational dc resistivity is given by 
\begin{align}
\rho_{\text{var}}\left[\phi\right]=\rho_{0}\left[\phi\right]+\rho_{\text{pud}}\left[\phi\right]
\label{variational rho}
\end{align}
with $\rho_{0}$ being the residual resistivity and 
\begin{align}
\rho_{\text{pud}}\left[\phi\right]=\frac{\left\langle \phi,\mathcal{P}_{\text{pud}}\phi\right\rangle }{\left|\left\langle \phi,X\right\rangle \right|^{2}}
\end{align}
where 
\begin{align*}
\left\langle \phi,\mathcal{P}_{\text{pud}}\phi\right\rangle  & =2\nu_{0}\beta\int_{\omega,\hat{\boldsymbol{k}},\omega',\hat{\boldsymbol{k}'},\nu}\left|g_\perp\left(\boldsymbol{k},\boldsymbol{k}'\right)\right|^{2}\chi''\left(\nu\right)f_{0}\left(\omega\right)\left(1-f\left(\omega'\right)\right)n_{0}\left(\nu\right)\\
 & \times\left(\phi\left(\hat{\boldsymbol{k}},\omega\right)-\phi\left(-\hat{\boldsymbol{k}'},\omega'\right)\right)^{2}\delta\left(\omega'-\omega-\nu\right)
\end{align*}
and 
\begin{align}
\left|\left\langle \phi,X\right\rangle \right|^{2}=\left|\rho_{F}\int_{\hat{\boldsymbol{k}},\omega}v_{F}\hat{\boldsymbol{k}}\phi\left(\hat{\boldsymbol{k}},\omega\right)\partial_{\omega}f_{0}\left(\omega\right)\right|^{2}.
\end{align}
The minimizer of Eq.~\eqref{variational rho} corresponds to the physical resistivity, i.e., $\rho_{\text{phys}}\leq\rho_{\text{var}}$.
For impurity dominated scattering the physical solution is $\phi=-\eta k_{F}\hat{\boldsymbol{k}}\cdot\boldsymbol{n}$
where $\boldsymbol{n}$ is a unit vector in the direction of the electric
field $\boldsymbol{E}$. This Ansatz is valid in cases where scattring
is approximately isotropic. 

Notice that 
\begin{align}
\left\langle \phi,\mathcal{P}_{\text{pud}}\phi\right\rangle =\underbrace{\int_{\hat{\boldsymbol{k}},\hat{\boldsymbol{k}'}}\left|g_\perp\left(\boldsymbol{k},\boldsymbol{k}'\right)\right|^{2}\left(\phi\left(\hat{\boldsymbol{k}}\right)-\phi\left(-\hat{\boldsymbol{k}'}\right)\right)^{2}}_{\mathcal{R}}\times\left(\text{frequency integrals}\right).
\label{variational scaling exp}
\end{align}
The frequency integrals correspond to the previously obtained expression for the resistivity (or conductivity) with a constant form factor. We have introduced the ratio $\mathcal{R}$ that determines the suppression of the transport coefficient due to vertex corrections. 
Inserting $\phi=-\eta k_{F}\hat{\boldsymbol{k}}\cdot\boldsymbol{n}$:
\begin{align}
\mathcal{R} & =k_{F}^{2}\eta^{2}\int_{\hat{\boldsymbol{k}},\hat{\boldsymbol{k}'}}\left|g_\perp\left(\boldsymbol{k},\boldsymbol{k}'\right)\right|^{2}\left(\left(\hat{\boldsymbol{k}}+\hat{\boldsymbol{k}'}\right)\cdot\boldsymbol{n}\right)^{2}\\
 & =\frac{k_{F}^{2}\eta^{2}}{2}\int_{\hat{\boldsymbol{k}},\hat{\boldsymbol{k}'}}\left|g_\perp\left(\boldsymbol{k},\boldsymbol{k}'\right)\right|^{2}\left|\hat{\boldsymbol{k}}+\hat{\boldsymbol{k}'}\right|^{2}\\
 & =k_{F}^{2}\eta^{2}\int_{\hat{\boldsymbol{k}},\hat{\boldsymbol{k}'}}\left|g_\perp\left(\boldsymbol{k},\boldsymbol{k}'\right)\right|^{2}\left(1+\cos\theta_{\boldsymbol{k},\boldsymbol{k}'}\right).
\end{align}
where in the first equality we averaged over $\boldsymbol{n}$.
Since $\alpha_\perp=\alpha_{\text{tr}}$ for $|g_\perp\left(\boldsymbol{k},\boldsymbol{k}'\right)|^2={\rm const}$, we define $g_\perp^2 = \int_{\hat{\boldsymbol{k}},\hat{\boldsymbol{k}'}} \left|g_\perp\left(\boldsymbol{k},\boldsymbol{k}'\right)\right|^{2} $ and using the above we have that 
\begin{align}
\frac{\alpha_{\text{tr}}}{\alpha_\perp}=\frac{1}{g_\perp^{2}}\int_{\hat{\boldsymbol{k}},\hat{\boldsymbol{k}'}}\left|g_\perp\left(k_{F}\left(\hat{\boldsymbol{k}}+\hat{\boldsymbol{k}'}\right)\right)\right|^{2}\left(1+\cos\left(\theta_{\boldsymbol{k}}-\theta_{\boldsymbol{k}'}\right)\right) \label{alpha tr}
\end{align}
as given in the main text. While the exact form of Eq.~\eqref{alpha tr} depends on the Ansatz for $\phi$, the $T$-linear scaling of the resistivity is independent of this Ansatz. Specifically, our Ansatz is appropriate when the form factor in the self-energy $\eta_\k$ is roughly constant and would need to be modified if the single-particle scattering rate has significant anisotropy on the Fermi surface. 
Such anisotropy may arise if the superconducting order parameter has $d-$wave symmetry, as discussed below Eq. (\ref{1 loop se}).
However, since the $T$-scaling of the resistivity is determined by the frequency integrals in Eq.~\eqref{variational scaling exp}, which are left unchanged, the scaling will remain $T$-linear. To this end, the effect of anisotropy is to further reduce the magnitude of $\alpha_{\rm tr}/\alpha$ due to ``short-circuiting'' of the hot directions on the Fermi surface. 

\subsection{Averaging over puddle size $R$}
\label{appendix: R average}

We consider the averaging over the different puddle sizes. Imagining that puddles form due to a percolation-like process, it is reasonable to assume that the puddle distribution is exponential in the area of the puddles, for puddles larger than the coherence length $\xi_0$. We hence take the distribution of $R$ to be $P(R)=\mathcal{N}^{-1} \Theta(R-\xi_0)R\exp(-R^2/2w^2)$ where the width $w$ is determined by the disorder. Importantly, since the susceptibility of large puddles is exponential in $R$ due to the renormalized charging energy, $1/E_c(R)\propto e^{k_F R}$, if the distribution of $R$ has significant weight for large puddles, one could encounter quantum-Griffiths behavior \cite{Thill1995}, where large puddles dominate the averaged susceptibility. Since our choice of $P(R)$ decays sufficiently fast, quantum-Griffiths effects are absent. The averaging over $R$ can be done numerically using the expressions for $\alpha_\perp(R)$ and $E_c(R)$ (given in the main text in the paragraphs following Eq.~\eqref{g perp} and Eq.~\eqref{TLS Ham}).
For example, to evaluate the self-energy (for simplicity we set $T=0$) we need to consider the following expression
\begin{align}
    \bigg<\alpha_\perp(R) \left<\chi''(\omega,R)\right>_{\bar{n}}\bigg>_R  &= \frac{\pi}{2}{\rm sign}(\omega) \int_0^\infty  \frac{\alpha_\perp(R) }{E_c(R)} \Theta(E_c(R)-|\omega|)P(R) dR. \nonumber \\
    &
      \equiv \frac{\pi\mathcal{C}(\xi_0,w,r_c(\omega))}{2U}{\rm sign}(\omega) ,
    \label{size ave sus}
\end{align}
where we define the dimensionless function 
\begin{equation}
    \mathcal{C}(\xi_0,w,r_c) \equiv \mathcal{N}^{-1} \int_{\xi_0}^{r_c} dr r^3 \frac{\alpha_\perp(r)}{z(\alpha_\perp(r))} \exp \left(-r^2 / 2w^2 \right).
\end{equation}
Here $r_c \sim \log(U/|\omega|)$ is the solution of $E_c(r_c)=|\omega|$. For small enough frequencies we may let $r_c \to \infty$ such that the prefactor in the self energy $\mathcal{C}(\xi_0,w,r_c\to\infty)$ is independent of frequency, leading to a form of the self energy similar to that given in the main text \eqref{1 loop se}.
This approximation is valid below the energy scale $\omega_*(\xi_0,w)$ at which the tail of the integral above $r_c$ is significant, i.e. which satisfies $\mathcal{C}(r_c\to\infty) = a\mathcal{C}(r_c(\omega_*))$ with $a>1$ some arbitrary number (for example, applying this logic to the case $\mathcal{C}(r_c)\sim\tanh(r_c/r_c(\omega_*))$ would give $a=1/\tanh(1)\approx1.31$).
Similarly, one can define $\mathcal{C}_{\rm tr}$ as the coefficient in the transport scattering rate (or resistivity) by replacing the $\alpha_\perp$ in the numerator of the above expression by $\alpha_{\rm tr}$.

The approximation in the main text treats the distribution as sharply peaked around an average value, $\bar{R}$. Although this approximation does not fit very well to the above calculations (i.e. there is not a wide regime of parameters where $\omega_*=E_c(\bar{R}),\mathcal{C}=\alpha_\perp(\bar{R})/E_c(\bar{R}),\mathcal{C}_{\rm tr}=\alpha_{\rm tr}(\bar{R})/E_c(\bar{R})$ for some $\bar{R}$), it is not a crucial point, and the parameters $\omega_*,\mathcal{C}
,\mathcal{C}_{\rm tr}$ can be treated as independent phenomenological variables of the model, with the important constraint $\mathcal{C}_{\rm tr}<\mathcal{C}$. In other words, the approximation captures the qualitative behavior, but not the exact numerical values of the above.

\subsection{Distribution of $E_c$} 
\label{Appendix: Ec distribution}

In the text, we treat the renormalized charging energy $E_c$ as a function of the typical Andreev conductance between the metal and the puddle. However, since the charging energy is exponential in the Andreev conductance, fluctuations in the conductance might affect the charging energy significantly. To see this, we use the exact expression for the renormalized charging energy as a function of the Andreev transmission coefficients of each channel, $0\leq \mathcal{T}_l<1$ \cite{feigelman_weak_2002}:
\begin{align*}
\frac{E_{c}\left(\{\mathcal{T}_l\}\right)}{E_{c,0}}&=\prod_l\left(1-\mathcal{T}_{l}\right)^{1/2}=\exp\left(-S\right)\\
S & \equiv-\frac{1}{2}\sum_{l}\log\left(1-\mathcal{T}_{l}\right)
\end{align*}
with additional logarithmic corrections in $\alpha$ to $S$ not considered here.
To proceed we may consider a distribution of the transmission coefficients $\mathcal{T}_l$ of channels
$l=1,\cdots,N$. In the limit of $N\gg 1$, by virtue of the central limit theorem (CLT), we may express $S$ as
\begin{align*}
S & =s_{0}N+\sqrt{N}\sigma x\\
x & \sim\mathcal{N}\left(0,1\right)\\
s_{0} & =\left<-\frac{1}{2}\log\left(1-\mathcal{T}\right)\right>_{\mathcal{T}}\\
\sigma^{2} & =\left<\left(-\frac{1}{2}\log\left(1-\mathcal{T}\right)-s_{0}\right)^{2}\right>_{\mathcal{T}},
\end{align*}
where $x$ is a normally distributed random variable with zero mean and a variance of 1.
The fluctuations in the renormalized charging energy are then log-normal
distributed. We define the median renormalized charging energy $\bar{E}_c=E_{c,0}e^{-s_0N}$, and obtain that
\begin{align*}
P\left(E_c\right) & =P\left(x\right)dx/dE_c\propto\frac{\bar{E}_c}{E_c}\exp\left(-\frac{\log\left(\bar{E}_c/E_c\right)^{2}}{2N\sigma^{2}}\right)=\exp\left(\log\left(\bar{E}_c/E_c\right)\left(1-\frac{\log\left(\bar{E}_c/E_c\right)}{2N\sigma^{2}}\right)\right).
\end{align*}
This distribution has three regimes:
\[
P\left(E_c\right)\sim\begin{cases}
\exp\left(-\frac{\log^{2}\left(\bar{E}_c/E_c\right)}{2N\sigma^{2}}\right) & E_c\ll r^{-1}\bar{E}_c \quad{\rm 
 or }\quad  E_c\gg r\bar{E}_c\\
\frac{1}{E_c} & r^{-1}\bar{E}_c \ll E_c\ll r\bar{E}_c
\end{cases}
\]
where $r=\exp\left(2N\sigma^{2}\right)$. Note that for this result
to be valid in the two outer regimes
one must have $2N\sigma^{2}\ll\sigma N$ i.e. $\sigma\ll1$, otherwise the
CLT is not applicable for the corresponding values of $S$. If $\sigma\sim1$
then the behavior of $P\left(E_c\right)$ in these limits will not
be universal. 

If $r\gg1$ then most puddles lie in this intermediate regime with $P(E_c)\sim 1/E_c$. For the resistivity to be linear in $T$, most puddles should have $E_c\gg T$. Thus, the regime with linear in $T$ resistivity is actually $T\ll r^{-1}\bar{E}_c$. If $N\sigma^2\lesssim 1$ (i.e. $r\sim \mathcal{O}(1)$) this aligns with the previous estimation of the upper cutoff $\bar{E}_c$. Otherwise, in order for the $T$ linear regime to be well defined, $r^{-1} \bar{E}_c$ must be well separated from $T_{1ch}$ of the puddles with the largest charging energies. If $T_{1ch}\sim E_ce^{-cN}$ this requires that $4\sigma^2 < c$ for large $N$.

\end{document}